\documentclass[12pt]{article}

\usepackage[style=nature,backend=biber,natbib=true,maxbibnames=5,minbibnames=1,defernumbers=true]{biblatex} %
\DeclareBibliographyCategory{mainrefs} %
\addtocategory{mainrefs}{lazer2018science,hanley2024machine,dufour2024ammeba,cook2017neutralizing,deverna2024factchecking,pfander2025spotting,psychological2022psychological,spitale2023ai,goldstein2024persuasive,porter2021global,martel2024factchecker,pennycook2021shifting,vanderlinden2017inoculating,roozenbeek2022psychological,hoes2024prominent,modirrousta2023gamified,cialdini2021influence,roozenbeek2019fake,lewandowsky2021countering,pennycook2024inoculation,dasanmartino2019finegrained,piskorski2023semeval,carrascofarre2022fingerprints,guess2020digital,berger2023debunking,costello2024durably,kozyreva2024toolbox,hasanain2024arpro,glaser2017discovery,corbin2014basics,milgrom1986relying,kamenica2011bayesian,loewenstein2001risk,slovic2007affect,douglas2017psychology,pennycook2020fighting,scherer2020susceptibility,simchon2026inoculation,pennycook2019lazy,bai2025llm,matz2024potential,salvi2025conversational,toubia2025database,wang2017liar,yang2022rawfc}
\usepackage{enumitem}
\usepackage{graphicx}
\usepackage{amsmath}
\usepackage{bbm}
\usepackage{geometry}
\usepackage{blkarray}
\PassOptionsToPackage{table}{xcolor}%
\usepackage{tikz}
\usepackage[english]{babel}
\usepackage{mathrsfs}

\usepackage[T1]{fontenc}
\usepackage{newtxtext}
\usepackage{amsthm}
\usepackage{amssymb}
\usepackage{stackrel}
\usepackage[singlespacing]{setspace}
\usepackage{pgfplotstable}
\usepackage{makecell}
\usepackage{subcaption}
\usepackage{float}
\usepackage{threeparttable}
\usepackage{amsfonts}
\usepackage[section]{placeins}
\usepackage{hyperref} %
\hypersetup{hidelinks} %
\usepackage{tocloft} %
\usepackage{bm}
\usepackage{algorithm}
\usepackage{algpseudocode}
\usepackage{tabularx}
\usepackage{booktabs}
\usepackage{longtable}
\usepackage[table]{xcolor}
\usepackage{multirow}

\newcommand{\secref}[1]{\emph{#1}} %
\let\citep\supercite
\DeclareCaptionLabelSeparator{pipe}{\ $|$\ } %
\begin{document}
\addtocontents{toc}{\protect\setcounter{tocdepth}{-1}} %
\setcounter{secnumdepth}{0} %

\title{A misinformation playbook sharpens judgement without breeding cynicism\thanks{Corresponding authors: Tong Guo (tong.guo@duke.edu) and Qiaozhu Mei (qmei@umich.edu). Affiliations: Guo, Fuqua School of Business, Duke University; Ye, Carlson School of Management, University of Minnesota; Xie, School of Information, University of Michigan; Mei, School of Information and Department of Electrical Engineering and Computer Science, University of Michigan. We acknowledge feedback and comments from seminar participants at the GenAI Lab at TUM School of Management, Duke University, The Hong Kong PolyU Business School, The PKU Guanghua School of Management, the Marketing Science Conference 2025 \& 2026 and the 13th Triennial Invitational Choice Symposium. We are grateful to Vladimir Khismatullin and Xin Zheng for their excellent research assistance. This research was conducted without external funding. The authors declare no competing interests. In preparing this paper, the authors utilized generative AI models as tools in their main analysis and to copyedit.}}
\author{Tong Guo \and Teng Ye \and Yutong Xie \and Qiaozhu Mei}
\date{October 15, 2026}
\setcounter{footnote}{0}
\maketitle

\begin{abstract}
\noindent Misinformation harms public health and democracy, and generative artificial intelligence now writes falsehoods as persuasive as people's. Defences range from fact-checking individual claims to training people to resist falsehoods. Yet many claims are too new to check, and interventions that teach doubt can reduce belief in truth too. We propose a defence that targets not the facts of each claim but the persuasion strategies that recur across them. Here we show that learning misinformation's persuasion playbook, mapped from 214,646 fact-checked claims in 51 languages, can sharpen human and machine judgement of claims they know little about, without making either more cynical about true claims. The playbook spans social influence, information manipulation and emotional appeals, combined around a core of fear, conspiracy and fabrication. Briefed with it, plus a reminder that persuasion is not proof of falsehood, a language model better distinguished true from machine-generated false claims about events after its training data. In a pre-registered experiment, the same briefing improved people's discernment of claims they reported as unfamiliar and left their belief in true claims unchanged. Averaged over all claims, the gain was not statistically significant. Because it targets what recurs, this defence need not wait for claim-specific fact-checks.
\end{abstract}

\newpage

Misinformation increasingly reaches people before anyone has checked it. Generative artificial intelligence (AI) adds to the challenge: it writes falsehoods as persuasive as those written by people\citep{spitale2023ai,goldstein2024persuasive}, and machine-generated content is a growing share of misinformation online\citep{hanley2024machine,dufour2024ammeba}. Fact-checks are effective, but they usually arrive after a claim has spread and address one claim at a time\citep{porter2021global,martel2024factchecker}. Large language models (LLMs) used to screen content face the same gap: asked about events after their training data, they have no claim-specific knowledge to rely on. Simply encouraging doubt is not the solution. People already doubt true news more than they believe false news\citep{pfander2025spotting}, and scepticism applied to true and false claims alike can reduce belief in accurate information without improving discernment\citep{hoes2024prominent,modirrousta2023gamified,deverna2024factchecking}. How, then, can people and machines judge claims that have yet to be checked?

We argue that part of the answer lies less in what a false claim says than in how it tries to make us believe it. Facts differ from one claim to another, but strategies used to sell them stay familiar: borrowed authority, appeals to fear\citep{carrascofarre2022fingerprints} and hints that ``they don't want you to know'' repeat across crises, countries and languages. Two lines of research already look beyond the facts of individual claims. Detectors of propaganda and persuasion techniques in news text have been used to flag questionable content automatically\citep{dasanmartino2019finegrained,piskorski2023semeval}. Among the many interventions against misinformation\citep{kozyreva2024toolbox}, accuracy prompts remind people to consider whether content is true\citep{pennycook2021shifting}, and inoculation, or prebunking, teaches people to recognize certain manipulation techniques rather than refute particular claims\citep{vanderlinden2017inoculating,cook2017neutralizing,roozenbeek2019fake,roozenbeek2022psychological,lewandowsky2021countering}; both have reduced susceptibility to misinformation in experiments.

There is reason for doubt, however. Inoculation with manipulation techniques has improved people's ability to tell true from false headlines in some evaluations but not others. A reanalysis of gamified inoculation found more scepticism without better discrimination\citep{modirrousta2023gamified}, and inoculation against emotionally manipulative language was shown to improve discernment only when combined with an accuracy prompt\citep{pennycook2024inoculation}. The techniques taught are a small set chosen by researchers rather than measured empirically\citep{dasanmartino2019finegrained,piskorski2023semeval}; most prior work has relied on researcher-authored materials or on samples confined to a few topics, one language and one short time window\citep{psychological2022psychological}. How techniques combine has been reported as raw pair counts in propaganda-annotated news\citep{hasanain2024arpro}, but not against a prevalence-adjusted baseline in fact-checked misinformation.

Here we address these limits. Rather than hand-picking which techniques to teach, we map misinformation's persuasion `playbook' (the strategies it uses to make claims believable, how often each is used and which tend to appear together) from a large, multilingual collection of professionally fact-checked claims. We then ask whether teaching it can sharpen the judgement of LLMs and people about claims they know little about, without breeding doubt about true claims. Because the same knowledge could also help produce more persuasive falsehoods, we also test it in the hands of an attacker.

\section{Mapping the persuasion playbook}\label{sec:results}\label{sec:approach}\label{subsec:prevalence}

We draw on the ClaimReview database maintained by Duke Reporters' Lab. We analyse the 214,646 claims fact-checked between January 2019 and April 2024 by 236 non-partisan organizations in 51 languages, most often English (31.6\%), Arabic (13.7\%) and Spanish (10.6\%). This scope is much broader than that of single-organization benchmarks such as LIAR\citep{wang2017liar} and RAWFC\citep{yang2022rawfc} (\emph{Methods}). We continue to track the data through August 2025 for the experiments, which test AI-generated claims anchored to events after the evaluator's October 2023 training cutoff (\emph{Methods}).

\begin{figure}[htp]
    \centering
    \begin{subfigure}[b]{\linewidth}
    \includegraphics[width=\linewidth]{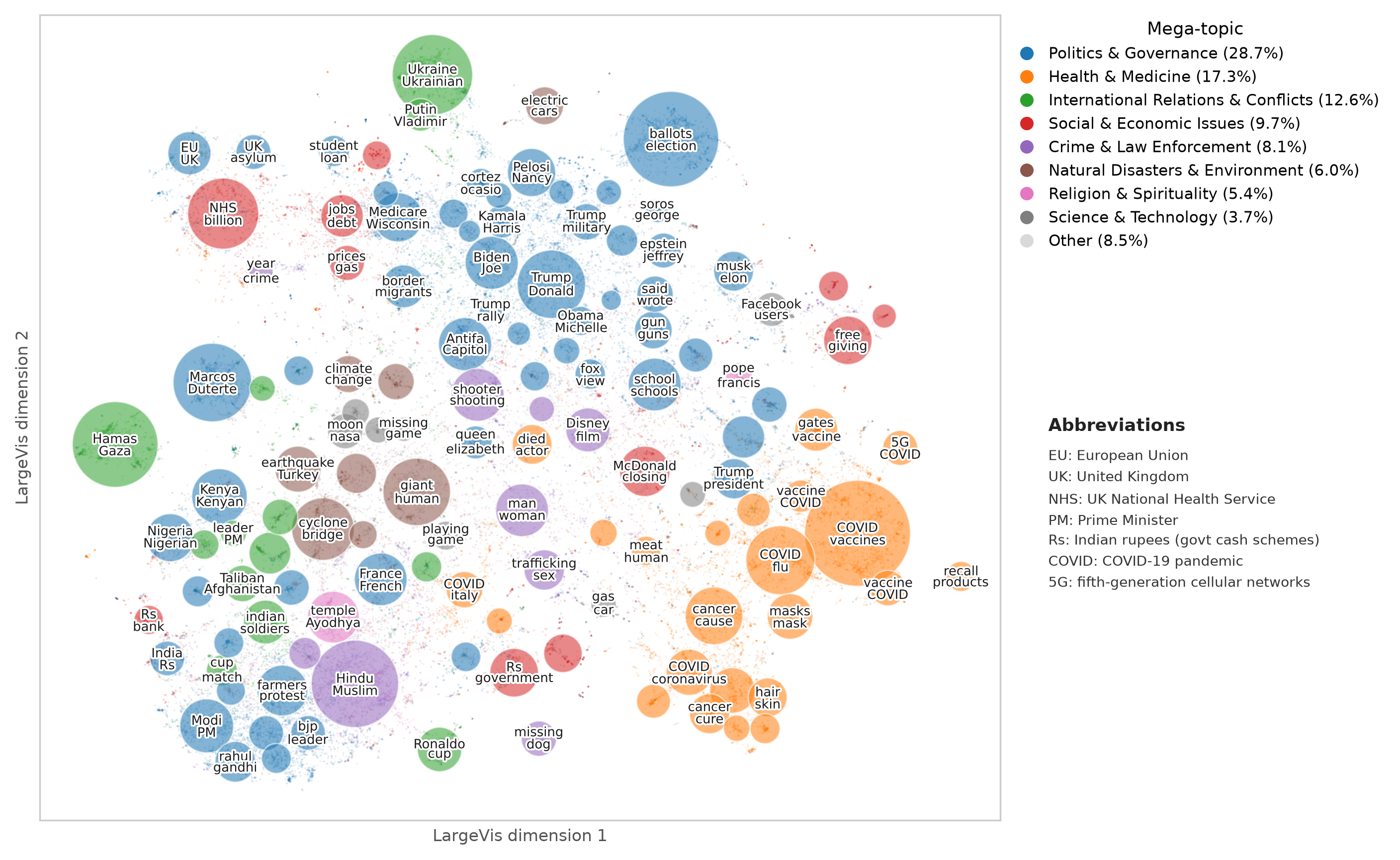}
    \caption{}\label{fig:topiccloud}
    \end{subfigure}
    \begin{subfigure}[b]{\linewidth}
        \centering
        \includegraphics[width=0.75\linewidth]{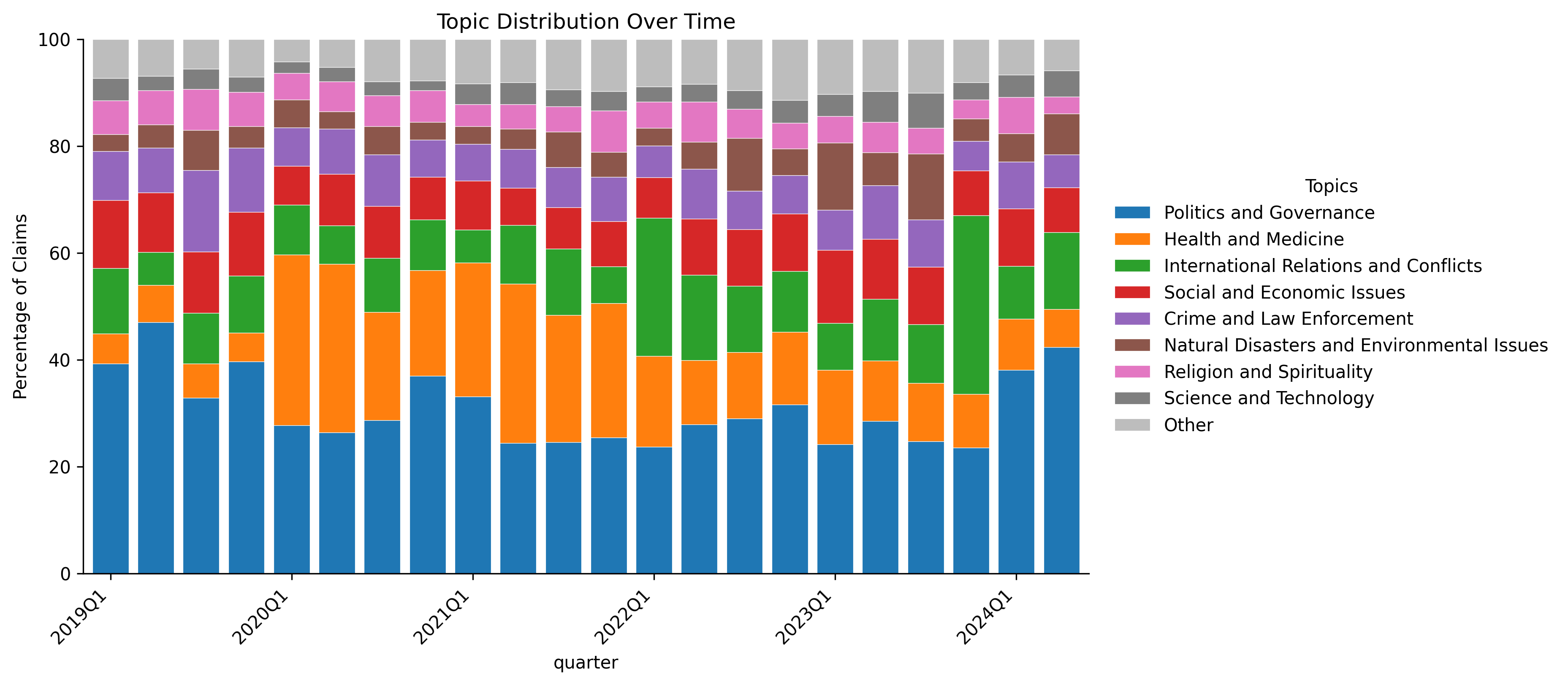}
        \caption{}\label{fig:topictime}
    \end{subfigure}
    \caption{\textbf{Topics of the 214{,}646 fact-checked claims.} \textbf{a}, Micro- and mega-topics of the English-language claims ($n = 66{,}130$): each bubble is a micro-topic cluster sized by claim volume and placed by semantic similarity; legend percentages give each mega-topic's share of all 214{,}646 claims. \textbf{b}, Monthly mega-topic distribution, January 2019 to April 2024.}
    \label{fig:topics}
\end{figure}

LLM-assisted topic modelling of the full 214,646-claim corpus recovers eight mega-topics (\emph{Methods}). \emph{Politics \& Governance} is the largest, at nearly 30\% of claims, followed by \emph{Health \& Medicine} (17.3\%) and \emph{International Relations \& Conflicts} (12.6\%). The remaining five, \emph{Social \& Economic Issues}, \emph{Crime \& Law Enforcement}, \emph{Natural Disasters \& Environmental Issues}, \emph{Religion \& Spirituality} and \emph{Science \& Technology}, are smaller but each well represented (see the legend of Fig.~\ref{fig:topiccloud}). Health claims surged during the COVID-19 pandemic, and political claims peaked around major elections (Fig.~\ref{fig:topictime}).

To show the micro-topics within these mega-topics, Fig.~\ref{fig:topiccloud} maps the English-language claims, with each bubble a micro-topic cluster sized by claim volume and placed by semantic similarity. US political and geopolitical claims, regional politics across South and Southeast Asia and Africa, and health and science claims centred on COVID-19 form the largest clusters, with economic, environmental, religious, and sports \& entertainment topics scattered throughout.

We next ask two descriptive questions about these claims. First, which persuasion strategies does misinformation use most often? Second, which strategies appear together? The experiments that follow then test whether a briefing built from the answers changes how evaluators judge new claims (estimands in Supplementary Table~\ref{tab:estimands}).

We derived the strategies inductively (\emph{Methods}). An LLM (GPT-4o) first coded the persuasion tactics in every claim in its original language, and two researchers then consolidated the resulting 28,000 tactic codes through grounded-theory coding\citep{glaser2017discovery,corbin2014basics} into 48 strategies (Supplementary Table~\ref{tab:48strat}). We organize these into 13 groups under three frameworks (Supplementary Table~\ref{tab:membership}). The first comprises six of Cialdini's seven principles of social influence\citep{cialdini2021influence}. The second comprises three information-manipulation mechanisms that degrade the epistemic quality of information itself, namely \emph{hiding}, \emph{manipulating} and \emph{fabricating} information\citep{milgrom1986relying,kamenica2011bayesian}. The third comprises four affective strategies that appeal primarily to emotion\citep{loewenstein2001risk,slovic2007affect,douglas2017psychology}. Together, these 13 groups cover over 99.9\% of claims.

\begin{figure}[p]
    \centering
    \begin{subfigure}[b]{0.49\linewidth}
    \centering
    \includegraphics[width=\linewidth]{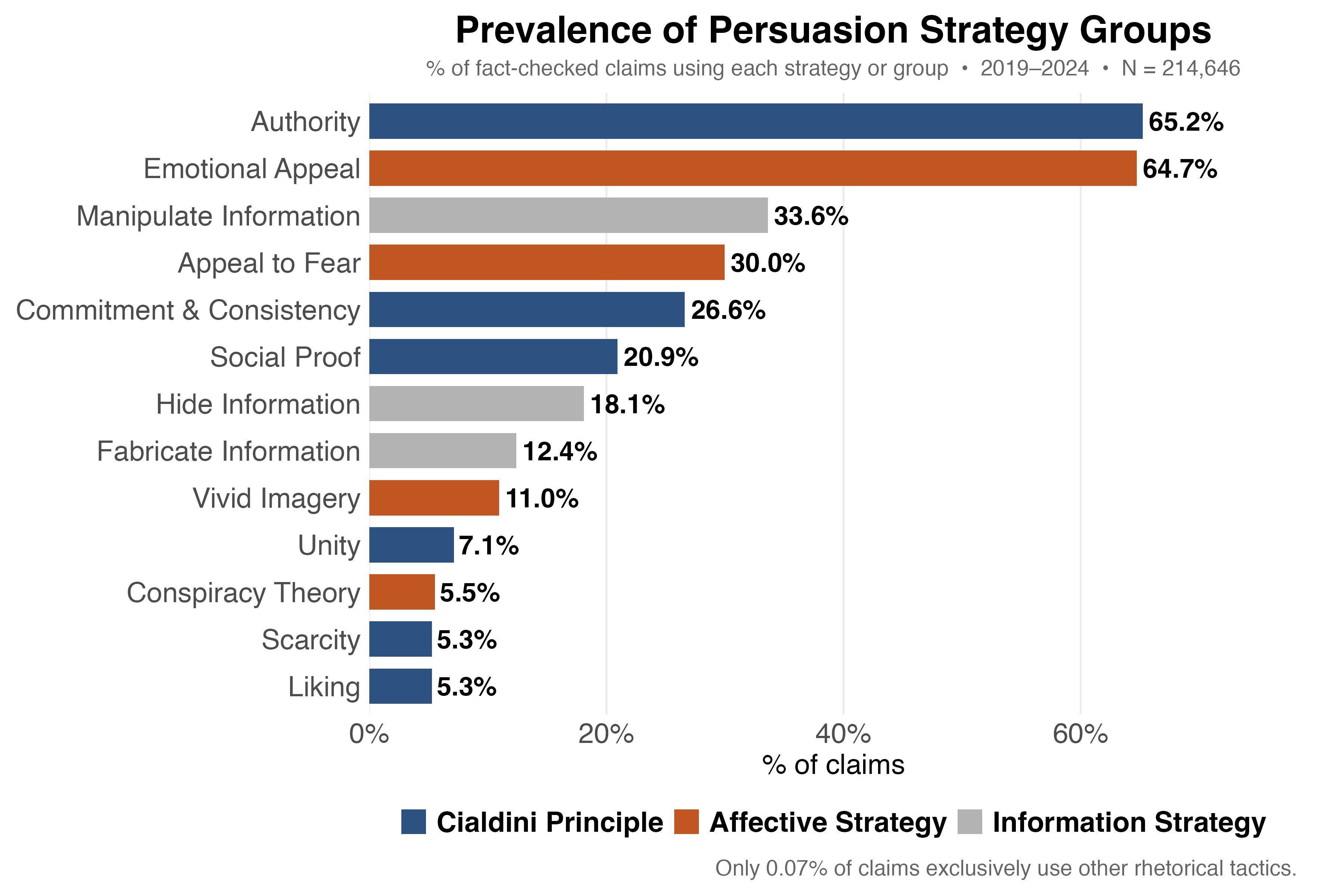}
    \caption{}\label{fig:cialdini_pct}
    \end{subfigure}\hfill
    \begin{subfigure}[b]{0.46\linewidth}
    \centering
    \includegraphics[width=\linewidth]{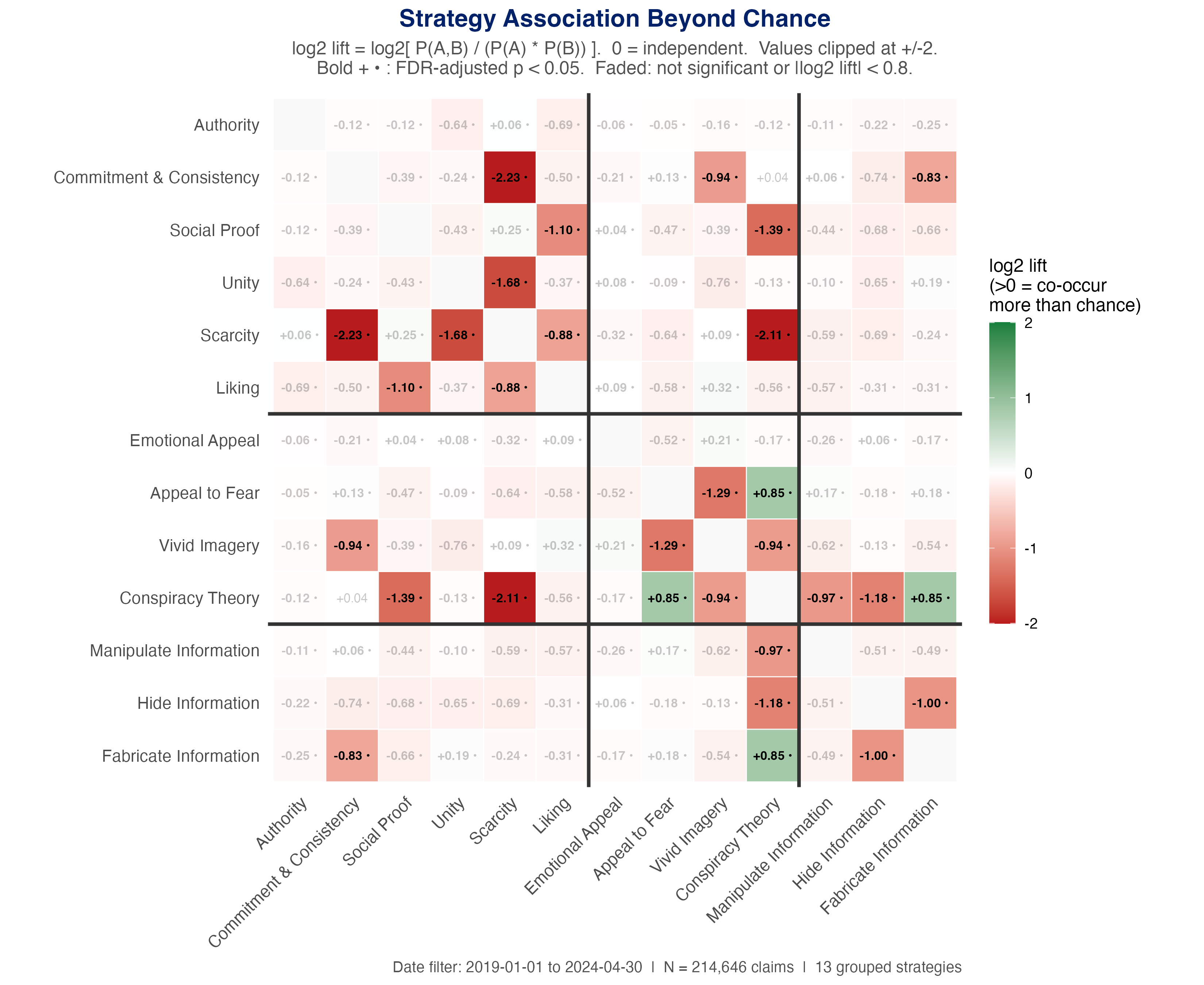}
    \caption{}\label{fig:cooc_lift}
    \end{subfigure}
    \begin{subfigure}[b]{\linewidth}
    \centering
    \includegraphics[width=\linewidth,height=0.42\textheight,keepaspectratio]{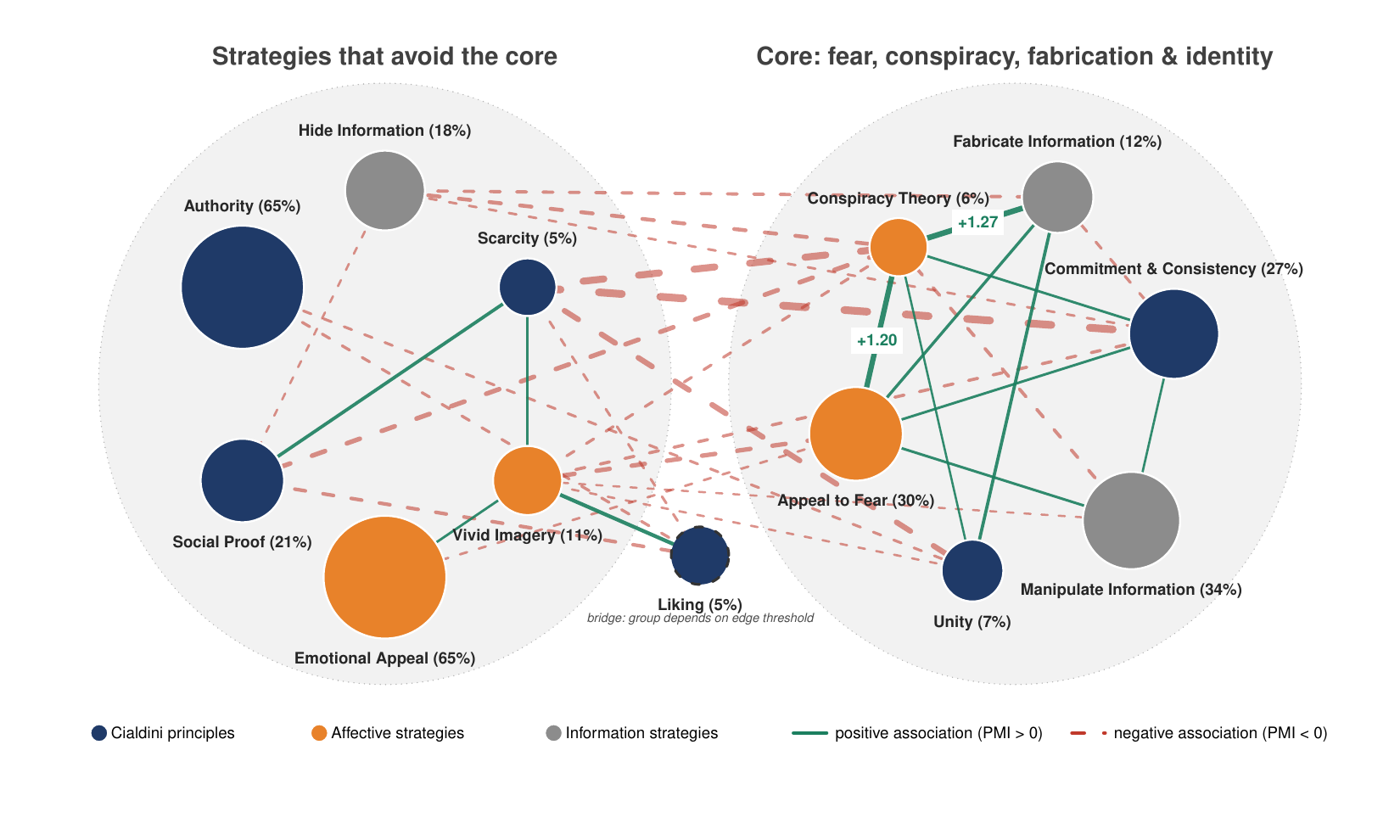}
    \caption{}\label{fig:network}
    \end{subfigure}
    \caption{\textbf{Prevalence and co-occurrence of the 13 strategy groups.} \textbf{a}, Share of claims using each group (any constituent strategy), coloured by framework (blue, Cialdini principles; orange, affective; grey, information manipulation); shares sum to more than 100\% because a claim can use up to five strategies. \textbf{b}, Pairwise co-occurrence relative to independence (pointwise mutual information, clipped at $\pm 2$), the matrix on which the briefings draw. \textbf{c}, Signed network relative to a maximum-entropy baseline that fixes each claim's number of strategies and each strategy's prevalence. Solid green edges, positive associations; dashed red edges, negative associations (BH-adjusted $q < 0.001$, $|\log_2 \text{lift}| \geq 0.3$; width proportional to $|\log_2 \text{lift}|$); node size, prevalence. Shaded regions show the two-group partition: the group containing \emph{Conspiracy Theory} is cohesive, the other is held together mainly by avoiding it, and \emph{Liking} (dashed outline) changes group across thresholds (Supplementary Information section~\ref{appendix:network}). $N = 214{,}646$ claims, January 2019 to April 2024.}
    \label{fig:cooc}
\end{figure}

\emph{Authority} and \emph{Emotional Appeal} are the most pervasive groups, each appearing in approximately 65\% of claims (Fig.~\ref{fig:cialdini_pct}; a group is counted if any of its strategies appears, Supplementary Tables~\ref{tab:48strat} and~\ref{tab:membership}). Among the other Cialdini principles, \emph{Commitment \& Consistency} (26.6\%) and \emph{Social Proof} (20.9\%) follow, whereas \emph{Unity}, \emph{Scarcity} and \emph{Liking} each appear in roughly 5--7\% of claims. Among the other affective strategies, \emph{Appeal to Fear} is the most common (30\%), followed by \emph{Vivid Imagery} and \emph{Conspiracy Theory}. Among information strategies, \emph{Manipulate Information} (33.6\%) is far more common than outright fabrication or hiding information. This pattern is consistent with misinformation distorting information more often than inventing it.

A claim rarely relies on a single persuasion strategy. Fig.~\ref{fig:cooc_lift} compares how often each pair of strategy groups appears together with how often it would if the groups occurred independently (pointwise mutual information). Because each claim carries only a few labels, however, strategies can also appear to avoid one another simply because they compete for those labels. For the network in Fig.~\ref{fig:network}, we compare each pair against a stricter maximum-entropy baseline that holds fixed the number of strategies in each claim and the prevalence of each strategy (\emph{Methods}). We describe pairs as positively or negatively associated when they co-occur more or less often than expected under this baseline. %

Three patterns emerge (Fig.~\ref{fig:network}; Supplementary Information section~\ref{appendix:network}). First, \emph{Authority} and \emph{Emotional Appeal} appear together in many claims, but no more often than either baseline predicts. Second, five strategies form a tightly linked core. \emph{Conspiracy Theory} is positively associated with \emph{Fabricate Information} and \emph{Appeal to Fear}, joined by appeals to group identity (\emph{Unity}) and to prior commitments (\emph{Commitment \& Consistency}). This core holds under every edge threshold and partition we examined, and most other strategies avoid it. Consistent with this, 71\% of the network's triangles are structurally balanced, against 51\% when edge signs are shuffled ($p = .019$). In particular, \emph{Conspiracy Theory} co-occurs less often than expected with social proof and vivid imagery, consistent with conspiracies operating as self-contained, evidence-light narratives\citep{douglas2017psychology}. The core is a general threat-narrative recipe, of which conspiracy theories are the most concentrated form: the core persists when conspiracy-labelled claims are excluded (Supplementary Information section~\ref{appendix:network}). The division of the remaining strategies is less stable (Supplementary Information section~\ref{appendix:network}). Third, strategies within the same theoretical framework tend to replace rather than reinforce one another, so the network cuts across our three frameworks. These patterns supply the content of the briefings we test next.

The core of the playbook is shared across languages. In each of the 15 languages with at least 3{,}000 claims (93\% of the corpus), the co-occurrence network correlates closely with the pooled network (median $r = 0.83$), and the positive associations of \emph{Conspiracy Theory} with \emph{Appeal to Fear} and with \emph{Fabricate Information} hold in all 15 (Supplementary Information section~\ref{appendix:by_lg}). How often each strategy appears still varies: \emph{Emotional Appeal} appears in 80\% of Arabic claims against 62\% of English and 60\% of Spanish ones, and \emph{Commitment \& Consistency} in 32\% of English claims against 14\% of Arabic ones. Some pairings vary too: \emph{Unity} and \emph{Conspiracy Theory} appear together more than twice as often as expected in Arabic claims, but slightly less often than expected in Spanish ones.

The strategy mix also varies systematically with a claim's topic, language and dissemination channel. For example, conditional on topic, language and time, social-media claims are more likely to be coded with affective, social and identity-based strategies, whereas information-manipulation strategies are more common in long-form channels. These associations are descriptive and do not necessarily reflect deliberate targeting by producers (Supplementary Information section~\ref{appendix:variation}).

\section{Briefing a language-model evaluator}\label{subsec:evaluator_prebunk}\label{sec:prebunking}\label{sec:application}

We now ask whether briefing evaluators with the playbook changes how they judge new claims about which they know little. We first tested a language-model evaluator (GPT-4o-mini), prompted to answer as simulated US adults whose profiles were built from their answers to more than 500 survey questions\citep{toubia2025database}, on claims about events after its October 2023 training cutoff. We used 240 false claims generated by GPT-4o and 40 true claims about the same events that we compiled from news outlets and official sources, similar in length to the false claims. Because every false claim was AI-generated, the experiment measures protection against AI-generated misinformation. Strategy knowledge might also simply make an evaluator suspicious of rhetoric. We therefore separated it from a short reminder that the presence of a persuasion strategy does not by itself make a claim false (hereafter, the decoupling instruction), because persuasion techniques characterize \emph{how} content is communicated, not \emph{whether} it is true\citep{dasanmartino2019finegrained}.

Each evaluation proceeded in three steps (\emph{Methods}). First, the model received one of five briefings: an accuracy prompt (C1)\citep{pennycook2021shifting}, strategy knowledge (C2), strategy knowledge plus the decoupling instruction (C3, the full briefing), the decoupling instruction alone (C0b), or no briefing (C0). Second, it received a single claim, with no other context, and the profile of one of 50 personas drawn at random from a pool of 2,058\citep{toubia2025database}. Third, it rated, as that persona, how believable the claim was on a 1--7 scale. Under the two strategy briefings (C2 and C3), it first named the strategies it detected. Every persona rated all 280 claims under every briefing, each in a separate model call that carried no memory of any other evaluation. We measured discernment as the gap between a persona's mean believability ratings of true and of false claims (hereafter, the discernment gap; the pre-registered primary outcome of the human experiment), and report the area under the receiver operating characteristic curve (AUC), which depends only on how claims are ranked, as a check.

\begin{figure}[!tbp]
    \centering
    \includegraphics[width=0.75\linewidth]{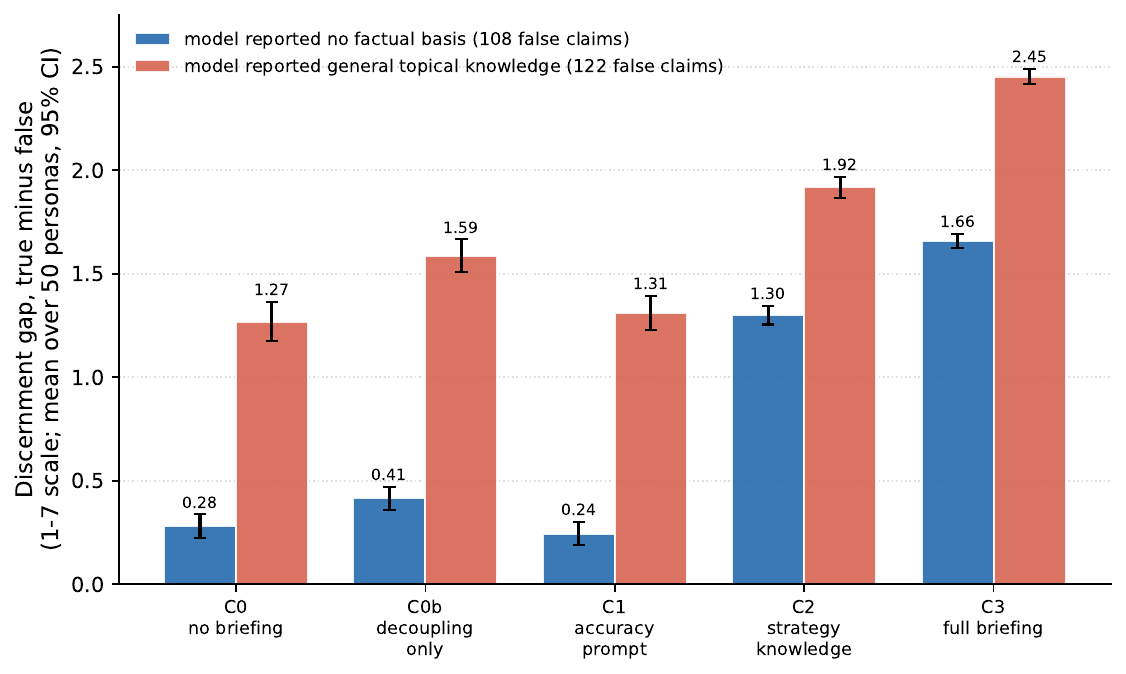}
    \caption{\textbf{Discernment of post-cutoff claims by a language-model evaluator.} Discernment gap (mean rating of true claims minus mean rating of false claims, 1--7 scale) under each briefing, separately for false claims on which the model reported no factual basis (blue) or general topical knowledge (red); means over 50 personas with 95\% confidence intervals. C0, no briefing; C0b, decoupling only; C1, accuracy prompt; C2, strategy knowledge; C3, full briefing. The AUC version is Supplementary Fig.~\ref{fig:post_cutoff_moderator_auc}.}
    \label{fig:post_cutoff_moderator}
\end{figure}

We find strategy knowledge effectively improves discernment among LLM personas. Without a briefing, the model rated true claims $0.77$ points more believable than false ones (95\% CI $[0.71, 0.84]$; this no-briefing floor was collected in a separate run with the same claims and personas, \emph{Methods}). The full briefing widened this gap to $2.06$ points ($[2.03, 2.09]$), an increase of $1.29$ points ($[1.23, 1.34]$; $p = 3 \times 10^{-44}$), whereas the accuracy prompt left it unchanged ($0.77$; $p = .92$; Fig.~\ref{fig:post_cutoff_moderator}). Both components of the full briefing contributed to the effect. Strategy knowledge alone widened the gap to $1.61$ points and the decoupling instruction alone to $1.00$ point, and combining them widened it by a further $0.45$ points over strategy knowledge alone ($[0.41, 0.49]$; $p = 2 \times 10^{-27}$; Supplementary Table~\ref{tab:discrimination_full}). AUC, which ignores how far apart the ratings lie, ranked the briefings similarly, except that the two strategy briefings were tied ($0.818$ and $0.812$, versus $0.682$ without a briefing and $0.666$ under the accuracy prompt). Importantly, the full briefing widened the gap from both sides, raising mean believability of true claims (from $4.03$ to $4.69$) while lowering that of false claims (from $3.25$ to $2.63$).

We next examined whether the gain varied with the knowledge the model reported having. We first probed the model's knowledge basis for each claim (\emph{Methods}) and classified the claims into two main groups (10 further false claims fell into two minor categories and are omitted from this comparison; Supplementary Information section~\ref{appendix:basis_probe}). For claims where it reported no reliable factual basis ($n = 108$ false claims), the gap was $0.28$ points without a briefing, and the full briefing widened it by $1.38$ points ($[1.33, 1.42]$). For claims where the model reported relevant general knowledge of the topic ($n = 122$), the gap without a briefing was larger, at $1.27$ points. Here, the full briefing widened it by $1.19$ points ($[1.11, 1.26]$; Fig.~\ref{fig:post_cutoff_moderator}). The gain was thus $0.19$ points larger where the model reported no factual basis ($[0.13, 0.26]$; $p = 4 \times 10^{-7}$). Comparing the full briefing with the decoupling instruction alone isolates the contribution of strategy knowledge, and this contribution was $0.38$ points larger on claims for which the model reported no factual basis than on claims for which it reported general knowledge ($1.24$ versus $0.87$; $[0.32, 0.43]$; $p = 1 \times 10^{-18}$). AUC showed the same pattern (Supplementary Fig.~\ref{fig:post_cutoff_moderator_auc}). Thus, strategy knowledge added most where the evaluator reported the least factual basis, consistent with the view that such briefings help most where evaluators have least prior knowledge to draw on.

To test whether this result depends on one model, we repeated the experiment with a second model family (Gemini 2.5 Flash) using the same 50 personas. Each model was scored on the claims for which it reported no factual basis. The full briefing again produced the widest gap in both models (Gemini 2.5 Flash, $1.96$ points versus $0.88$ without a briefing; GPT-4o-mini, $1.66$ versus $0.28$; Supplementary Table~\ref{tab:crossmodel} and Supplementary Fig.~\ref{fig:crossmodel_gap}). The accuracy prompt left GPT-4o-mini's gap unchanged and widened Gemini 2.5 Flash's modestly ($+0.28$ points, $[0.21, 0.35]$), far less than the full briefing ($+1.08$, $[1.01, 1.15]$). Interestingly, the models differed when strategy knowledge was given without the decoupling instruction. For Gemini 2.5 Flash, this briefing narrowed the gap below its no-briefing level ($0.44$ versus $0.88$; $-0.44$, $[-0.53, -0.36]$), because believability of true claims fell ($5.26$ to $4.82$) while that of false claims did not. %
With the decoupling instruction added, the same knowledge raised belief in true claims and lowered it in false ones. AUC gave the same ranking of briefings (Supplementary Information section~\ref{appendix:crossmodel}).

\section{Briefing human evaluators}\label{subsec:human_prebunk}

We turned next to people, asking whether the same strategy knowledge helps them tell true from false claims. In a pre-registered experiment, 749 US adults recruited on Prolific were randomly assigned to one of the same five briefings. In the two strategy arms, participants read about the 13 strategy groups, how common each is and which tend to appear together, and practised identifying them on three claims with feedback. Each participant then rated 20 claims (7 true, 13 false) drawn from the 280 claims used with the language model, on the same 1--7 scale, explained each rating in free text and reported how familiar they had been with each claim's topic (\emph{Methods}). The primary outcome was the same discernment gap.

\begin{figure}[htp]
    \centering
    \includegraphics[width=\linewidth]{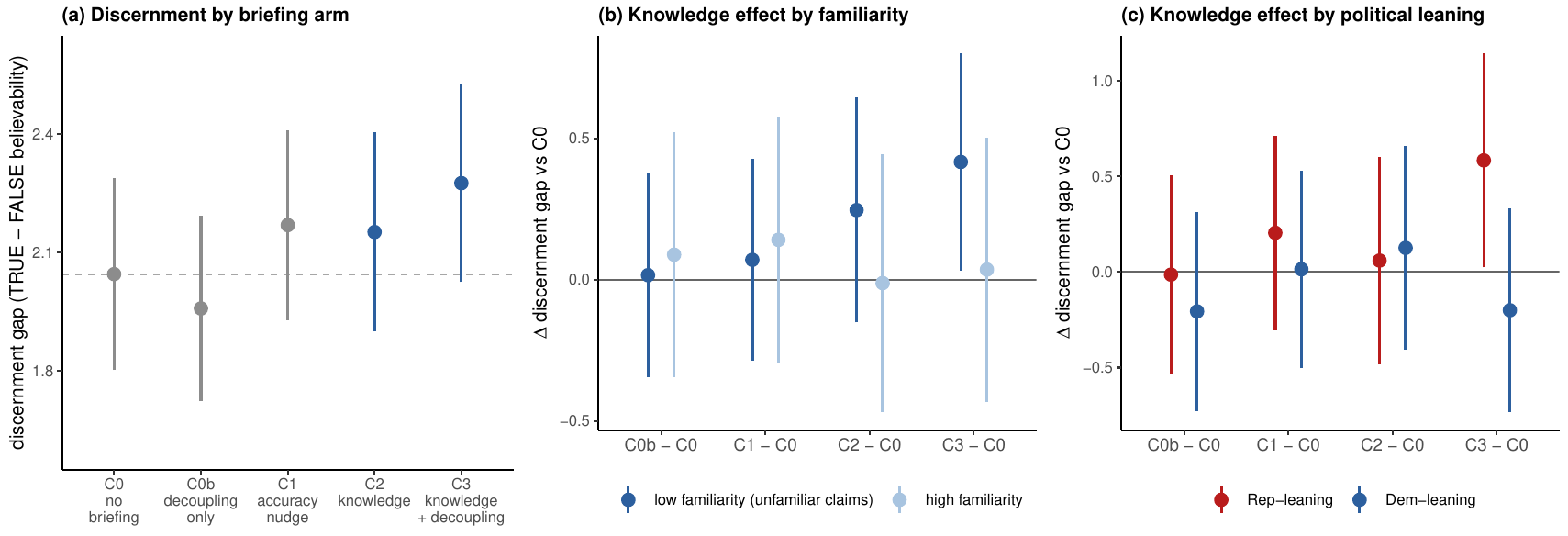}
    \caption{\textbf{Human prebunking experiment.} Pre-registered; $N = 749$ (155, 155, 154, 140 and 145 participants under C0, C0b, C1, C2 and C3). \textbf{a}, Discernment gap (mean believability of true minus false claims, 1--7 scale) by briefing; means with 95\% confidence intervals; dashed line, C0 mean. The confirmatory contrasts (C3 versus C0 and versus C0b) are not significant after correction for multiple comparisons. \textbf{b}, Change in the gap relative to C0 by the participant's familiarity with each claim's topic (low, $\leq 2$; high, $\geq 3$ on a 1--5 scale); pre-registered test of dependence on familiarity, $p = .0004$. \textbf{c}, The same contrasts within party (exploratory; difference across parties not significant; Independents omitted). The decomposition into belief in true and false claims is in Supplementary Fig.~\ref{fig:studyB_decomposition}.}
    \label{fig:studyB_human}
\end{figure}

We did not find that the full briefing significantly improved average discernment. The discernment gap was $0.23$ points higher under the full briefing than under no briefing ($d = 0.15$, $p = .39$) and $0.32$ points higher than under the decoupling instruction alone ($d = 0.21$, $p = .21$; both adjusted for multiple comparisons). Both estimates are positive but are not statistically significant (Fig.~\ref{fig:studyB_human}a). The secondary per-participant AUC contrasts were nominally significant (versus no briefing, $\Delta = +0.042$, $p = .047$; versus the decoupling instruction alone, $\Delta = +0.044$, $p = .039$) (Supplementary Information section~\ref{appendix:studyB_results}). Put simply, the full briefing raised the chance of ranking a true claim above a false one from 76\% to 80\%, even though the average discernment gap did not differ significantly across briefing arms.

Did the full briefing help more with claims that participants found unfamiliar? In a pre-registered secondary test, the effect of the briefings on discernment depended on familiarity ($p = .0004$). The full briefing's advantage over no briefing was $+0.42$ (95\% CI $[+0.03, +0.80]$) on unfamiliar claims and $+0.04$ ($[-0.43, +0.50]$) on familiar ones (Fig.~\ref{fig:studyB_human}b), whereas the accuracy prompt showed no such moderation ($p = .32$). Participants also used what they were taught: under the full briefing, their explanations for false claims named a taught strategy three times as often as under no briefing ($4.1\%$ versus $1.3\%$, $p = 2 \times 10^{-9}$). As with the language model, strategy knowledge helped most where evaluators had the least prior knowledge to draw on (Supplementary Information section~\ref{appendix:studyB_heterogeneity}).%

As an exploratory contrast (95\% confidence intervals in brackets), we also compared the full briefing with the accuracy prompt, an established intervention\citep{pennycook2021shifting}. We find the accuracy prompt did not improve discernment over no briefing, either overall ($+0.12$ [$-0.22$, $0.46$]) or on unfamiliar claims ($+0.07$ [$-0.29$, $0.43$]). The full briefing did not differ significantly from the accuracy prompt overall ($+0.11$ [$-0.24$, $0.45$]). However, on unfamiliar claims its discernment gap was $0.35$ larger ([$0.00$, $0.69$], $p = .050$; Supplementary Information section~\ref{appendix:studyB_heterogeneity}).

Neither strategy briefing, with or without the decoupling instruction, bred cynicism. Belief in true claims was virtually identical across all five briefings (5.08 to 5.09), and each briefing was equivalent to no briefing within the bound pre-registered for this test ($\pm 0.3d$, applied here to every arm; $p \leq .007$; Supplementary Table~\ref{tab:studyB_tests}). Belief in true claims did not fall on familiar or unfamiliar claims; on unfamiliar claims, the full briefing's estimate was imprecise and, if anything, positive ($+0.16$ [$-0.12$, $0.45$]; exploratory; Supplementary Information section~\ref{appendix:studyB_results}). Instead, the full briefing lowered belief in false claims ($-0.23$ [$-0.45$, $-0.01$], $p = .042$; exploratory). The full briefing thus made people more sceptical of false claims without making them more sceptical of true ones.

Finally, as a pre-registered exploratory analysis, we examined whether the gains differed by political leaning. They did not differ significantly by party ($p = .58$). In exploratory, uncorrected estimates, Republican-leaning participants had the lowest discernment of unfamiliar claims without a briefing (gap $1.20$, versus $2.02$ for Democratic leaners) and the largest overall point estimate ($+0.58$ against no briefing; $+0.60$ on unfamiliar claims). Estimates for Democratic leaners were near zero (Fig.~\ref{fig:studyB_human}c), and familiarity was similar across parties (Supplementary Information section~\ref{appendix:studyB_heterogeneity}).

In sum, the gain was largest where factual knowledge was limited for both evaluators, on claims for which the LLM evaluators reported no factual basis and on unfamiliar claims for people. Measured on the same 1--7 scale, the full briefing widened the GPT-4o-mini evaluator's true-false discernment gap by $1.29$ points but people's by only $0.23$ points on average ($0.42$ on unfamiliar claims). %

\section{The adversarial stress test}\label{sec:adversarial}\label{subsec:strategy-guided-prediction}

Knowledge that helps evaluators recognize persuasion strategies could also be used to make falsehoods more persuasive. We therefore stress-tested the playbook against this misuse: does strategy knowledge make AI-generated misinformation more convincing, and does briefing the evaluator offset any such gain? We ran this test with simulated evaluators only. It examines AI-generated claims judged by AI evaluators, the setting in which such content is produced and screened at scale, and it requires rating 2,000 claims under every briefing, more deliberately persuasive false content than we judged appropriate to show participants. Its results therefore concern machine evaluators, not human persuasion. We generated 2,000 false claims about 20 post-cutoff events under five generator conditions (A1--A5) that mirror the evaluator briefings, from zero-shot generation (A1) to generation guided by the playbook's prevalence and co-occurrence patterns (A3--A5), and had simulated personas rate them (\emph{Methods}).

Strategy knowledge did make the generated claims more believable. We evaluated the generated claims with a panel of 100 personas drawn from the same pool as in the discernment experiment\citep{toubia2025database}. Under no briefing, mean believability rose with each added level of strategy knowledge, from $2.68$ under zero-shot generation to $2.94$ under guidance from corpus-wide co-occurrence patterns on the 1--7 scale, and topic-specific co-occurrence guidance added no further gain ($2.93$). Even so, the most believable generations remained less believable than the fact-checked claims that actually circulated about the same events, against which each generated claim was matched by semantic similarity ($3.13$; the full 2,058-persona evaluation gives the same ordering, Supplementary Information section~\ref{subsec:persona_eval}). Generations guided by the playbook's prevalence and co-occurrence patterns also matched the strategy profiles of fact-checked claims about the same events modestly more closely than zero-shot generations did (Supplementary Information section~\ref{appendix:strategy_alignment}).

The full briefing more than offset this gain. The same 100-persona panel rated all 2,000 claims under each evaluator briefing, and we computed each rating's shift ($\Delta$) from the same persona's unbriefed rating of the same claim (Supplementary Fig.~\ref{fig:adversarial_prebunk}). Every briefing lowered believability. The full briefing produced the largest reduction ($-0.60$) and the decoupling instruction alone the smallest ($-0.07$). For the generator's most believable condition, guidance from corpus-wide co-occurrence patterns, the full briefing erased the $+0.26$ knowledge gain and brought believability $0.32$ below the zero-shot baseline (Supplementary Information section~\ref{appendix:adversarial_full}). Because this test uses false claims only, it measures suppression of belief, not discernment. A Gemini 2.5 Flash evaluator replicated the reversal (Supplementary Information section~\ref{appendix:adversarial_crossmodel}), reducing concern that it depends on a shared generator--evaluator family.

\section{Discussion}
\label{sec:conclusion}
A common response to misinformation that outpaces fact-checking is generalized doubt. We present evidence that, in the tested settings, learning misinformation's persuasion playbook can instead sharpen judgement of unfamiliar claims without lowering belief in true ones. Three features distinguish the approach (Table~\ref{tab:defences}). First, it can be delivered without prior knowledge of the claim being judged, so it applies where fact-checks and the evaluator's own knowledge of the claims run out. Second, the playbook is derived empirically from fact-checked claims in 51 languages, and it shows that strategies can work together across theoretical categories. Third, the gain in discernment was achieved without reducing the credibility of true claims. %

Why did strategy knowledge help most when evaluators knew least? One plausible explanation is that strategy cues substitute for missing facts. An evaluator who knows the facts can assess a claim directly; one who does not may rely more on how it is presented. The playbook tells such evaluators which forms of persuasion strategies misinformation tends to use and which ones tend to appear together. The discernment gain was indeed largest on claims for which the model reported no factual basis and on claims that people rated as unfamiliar. The accuracy prompt, which asks evaluators to use what they already know but offers no new cues, did little on these claims: nothing for GPT-4o-mini or for people, and only a small gain for Gemini 2.5 Flash. The decoupling instruction probably helps by stopping evaluators from treating persuasive wording alone as a sign of falsehood. The playbook may also help most where there is most room to improve; participants with low baseline discernment showed larger point estimates, although these subgroup estimates are exploratory. Two limits apply. We did not test evaluators who knew the facts of the claims, and familiarity was self-reported after the briefing, so it marks limited prior knowledge rather than a manipulated one.

Our results favour teaching the playbook together with the principle that a persuasion strategy is not proof of falsehood. When Gemini 2.5 Flash received the strategy knowledge without that principle, its belief in true claims fell while its belief in false claims did not (\secref{Briefing a language-model evaluator}). GPT-4o-mini showed no such cost, and in people the pre-registered analogue was not detected. This finding is consistent with evidence that inoculation against emotional manipulation improved human discernment only when paired with an accuracy prompt\citep{pennycook2024inoculation}.

Fact-checking corrects specific misbeliefs, but only for the claims it checks\citep{porter2021global,martel2024factchecker,berger2023debunking}. Persuasion strategies recur across topics, languages and time, raising the possibility that learning to recognize them could help evaluators assess claims beyond those used in training\citep{roozenbeek2022psychological,lewandowsky2021countering,vanderlinden2017inoculating,pennycook2019lazy,berger2023debunking}. Strategy-level inoculation can therefore complement fact-checking, particularly for new and fast-moving claims, where benefits concentrated. The playbook might also help fact-checkers anticipate problematic narratives likely to emerge around scheduled events.

We built and tested the playbook as a defence, but were mindful that it can also serve an attacker. Strategy-guided prompting made AI-generated misinformation more convincing to LLM personas (\secref{The adversarial stress test}), in line with evidence that AI-generated messages persuade people\citep{bai2025llm,matz2024potential,salvi2025conversational}. The tactics in the playbook are already widely used, but organizing them together may make them easier to deploy. The defensive value of the playbook does not remove this risk. For this reason, the labelled corpus, generation prompts and 2{,}000-claim generation set are available only on reasonable request, with each request evaluated individually (\emph{Data availability}).

Taken together, teaching misinformation's persuasion playbook sharpened AI discernment and, for people, discernment of claims they knew little about, without breeding cynicism. Our study did not isolate the causal effects of any of these persuasion strategies on belief or sharing, which would be a valuable next step. Furthermore, all false claims were AI-generated, and those in the discernment experiments were anchored to events that postdated the evaluator's training data and for which the model reported no factual basis, so that they would be new to the evaluators. The discernment gain we measure therefore applies to AI-generated misinformation, and whether it extends to human-authored misinformation remains to be tested. In our experiment, the human effects were small and conditional on self-reported familiarity. Completion was also lower in the strategy briefings (85\% versus 92\%), so those who finished them may have been more motivated (bounds that allow for this selective dropout remain positive but imprecise; Supplementary Information section~\ref{appendix:studyB_attrition}). Shorter briefings might reduce dropout, but they could also teach less, since much of the briefing's value lies in its detail about how strategies are used and combined. It is also worth testing whether a larger trial would detect an average effect, and how long the discernment gain would last. Nonetheless, the playbook offers a promising defence against AI-generated misinformation that moves faster than fact-checking can follow.

\printbibliography[category=mainrefs,title={References}] %

\begin{table}[h!]
\centering
\scriptsize
\setlength{\tabcolsep}{3pt}
\caption{\textbf{The playbook briefing compared with existing defences against misinformation.} Entries are qualitative design features and findings documented in the cited evaluations (or here), not head-to-head efficacy evidence; only the accuracy prompt was tested directly against the playbook.}
\label{tab:defences}
\begin{tabular}{>{\raggedright\arraybackslash}p{2.6cm}>{\raggedright\arraybackslash}p{2.0cm}>{\raggedright\arraybackslash}p{2.3cm}>{\raggedright\arraybackslash}p{2.6cm}>{\raggedright\arraybackslash}p{2.1cm}>{\raggedright\arraybackslash}p{1.9cm}}
\toprule
\textbf{Defence} & \textbf{Claim-specific verdict or factual knowledge?} & \textbf{Applies to unseen claims?} & \textbf{Belief in true content} & \textbf{Delivery} & \textbf{Evaluators tested} \\
\midrule
Fact-checks and labels\citep{porter2021global,martel2024factchecker} & Yes: a verdict per claim & Directly covers checked claims & Scepticism increased in cited evaluation\citep{hoes2024prominent} & Professional check per claim, after spread & Human; verdicts usable by AI \\
\addlinespace[2pt]
Accuracy prompt\citep{pennycook2021shifting} & Prompts accuracy consideration; can draw on available knowledge & No detected gain on unfamiliar claims here (this study) & Unchanged in people; slightly lower in both models (this study) & One question & Both; no gain for GPT-4o-mini, small gain for Gemini 2.5 Flash (this study) \\
\addlinespace[2pt]
Technique inoculation, games and videos\citep{roozenbeek2019fake,roozenbeek2022psychological} & No & Designed to transfer to the techniques taught & Mixed: discrimination gains in a meta-analysis\citep{simchon2026inoculation}; more scepticism without discrimination gain in a reanalysis of games\citep{modirrousta2023gamified}; gain only in combination with an accuracy prompt in one trial\citep{pennycook2024inoculation} & Short game or video & Human \\
\addlinespace[2pt]
Media-literacy tips\citep{guess2020digital} & No & Designed to apply generally (generic heuristics) & Scepticism increased in cited evaluation\citep{hoes2024prominent} & List of tips & Human \\
\addlinespace[2pt]
Dialogue with AI\citep{costello2024durably} & Model supplies belief-specific information & Personalized to one belief per dialogue & Not reported & Personalized multi-turn conversation & AI delivers to humans \\
\addlinespace[2pt]
Playbook briefing (this study) & No & Tested on post-cutoff AI-generated claims; human gains concentrated on unfamiliar claims, average contrasts n.s. & Equivalent within bound in people; rose in both models (combined briefing) & Written briefing or prompt & Both, tested here \\
\bottomrule
\end{tabular}
\end{table}

\clearpage
\section{Methods}\label{sec:methods}

\subsection{The ClaimReview corpus}\label{sec:data}

Our data come from the ClaimReview project, coordinated by Duke Reporters' Lab and supported by the Google News Initiative. ClaimReview is a structured schema that fact-checking organizations use to tag their fact-check articles so that search engines, applications and social media platforms can index them; before 2025 it served as the primary fact-checking layer for Google Search, Microsoft Bing and Facebook (\url{https://www.factcheckinsights.org/guide}).\footnote{For how Duke Reporters' Lab identifies eligible fact-checkers, see \url{https://reporterslab.org/how-we-identify-fact-checkers}.} Our coding pull contains 236,954 fact-checked claim records. Of these, 214,646 have an effective fact-check date (the review date, with the claim date as fallback) between January 2019 and April 2024; they were contributed in 51 languages by 236 non-partisan organizations on all six inhabited continents and form the sample for all persuasion-pattern analyses. Major contributors include PolitiFact, Snopes.com and FactCheck.org, alongside hundreds of other organizations adhering to the International Fact-Checking Network's Code of Principles. This coding sample provides the strategy briefings. A second pull covering November 2023 through August 2025 added 26,992 records, bringing the full collection to 251,326 unique claims. We use the post-November 2023 records to extract seeding events used in the prebunking experiments. %

Each record contains the text of the claim under scrutiny, the URL of the original source, the fact-checker's accuracy rating, the fact-checker's identity and organizational affiliation, and the date of the fact-check. Because the data are collected by professional journalists rather than crowdworkers or algorithms, the corpus is an expert-annotated record of newsworthy fact-checked claims. More than 75\% of reviewed claims are rated false or misleading, consistent with fact-checkers spending their limited time on content that poses genuine informational risk. See the distribution of verdicts in Supplementary Information section~\ref{appendix:corpus_overview}. About 30\% of claims trace back to social media platforms, judged by their original source URL.

ClaimReview covers far more topics, languages and fact-checking organizations than existing misinformation benchmarks. Datasets such as LIAR (PolitiFact political claims only)\citep{wang2017liar} and RAWFC (raw reports sourced from Snopes)\citep{yang2022rawfc} each draw on a single organization; our data span eight mega-topics and 51 languages, with no researcher-imposed topical pre-selection. English accounts for the largest share of the corpus (Supplementary Information section~\ref{appendix:corpus_overview}), followed by Arabic, Spanish, Hindi, Portuguese and dozens of other languages.

\subsection{LLM-assisted topic modelling}\label{subsec:topics}

To recover topics in an interpretable form, we ran a three-step topic-modelling procedure assisted by large language models (LLMs; \texttt{gpt-4o-2024-08-06}, temperature 0) that aggregates micro-topics into eight mega-topics. In its published evaluation, this approach outperformed traditional alternatives such as latent Dirichlet allocation (LDA) and BERTopic\citep{pham2023topicgpt}. Supplementary Information section~\ref{appendix:megatopics} details the procedure and supporting evidence. The eight mega-topics are \emph{Politics \& Governance}, \emph{Health \& Medicine}, \emph{International Relations \& Conflicts}, \emph{Social \& Economic Issues}, \emph{Crime \& Law Enforcement}, \emph{Natural Disasters \& Environmental Issues}, \emph{Religion \& Spirituality}, and \emph{Science \& Technology}.

\subsection{Strategy annotation and theoretical coding}\label{subsec:gtm}

We identified the persuasive tactics in each claim inductively, following the Grounded Theory Method (GTM)\citep{glaser2017discovery,urquhart2022grounded,corbin2014basics}. Adapting Gao et al.\citep{gao2024collabcoder}, we used GPT-4o (\texttt{gpt-4o-2024-08-06}, temperature 0) for zero-shot open coding\citep{gilardi2023chatgpt}: the model read every claim in its original language and extracted up to five tactics per claim with detailed rationales, and the extracted tactic codes and final strategy labels were standardized in English. Open coding produced about 28,000 distinct first-order tactic codes across over 922,000 instances. Two researchers then consolidated these through constant comparison (axial coding) into a final taxonomy of 48 persuasion strategies, spanning more than 9,600 unique constituent tactics that appear over 833,000 times across the dataset. Supplementary Table~\ref{tab:48strat} lists all 48 strategies with definitions and prevalence.

In the theoretical coding phase, we organize the 48 strategies under three complementary frameworks (Cialdini principle of social influence; information framework; affective framework). This follows from the well-documented dual structure of persuasion mechanisms \citep{dellavigna2010persuasion}: an information-based channel operating through Bayesian updating on signals, and a preference- or affect-based channel operating through heuristics and emotional arousal. Six of Cialdini's seven principles of social influence \citep{cialdini2021influence} (\emph{Authority}, \emph{Commitment \& Consistency}, \emph{Liking}, \emph{Scarcity}, \emph{Social Proof}, \emph{Unity}) account for 21 of the 48 codes (mappings in Supplementary Table~\ref{tab:membership}). Three information-manipulation mechanisms that degrade the epistemic quality of information itself (\emph{hiding}, \emph{manipulating}, and \emph{fabricating} information) account for 12 \citep{milgrom1986relying,kamenica2011bayesian}. Four affective strategies that operate through emotional arousal rather than cognitive reasoning (\emph{emotional appeal}, \emph{appeal to fear}, \emph{vivid imagery}, \emph{conspiracy theory}) complete the framework \citep{loewenstein2001risk,slovic2007affect,douglas2017psychology}. Over 99.9\% of claims employ strategies from at least one of the three frameworks; the remainder are grouped under ``other rhetorical tactics'' (Supplementary Table~\ref{tab:membership}).

Our 48-strategy taxonomy connects to the fine-grained propaganda- and persuasion-technique taxonomies developed in natural language processing\citep{dasanmartino2019finegrained,piskorski2023semeval}; Supplementary Table~\ref{tab:semeval_map} maps the two label spaces. The main differences are how our taxonomy was developed and what it covers. We followed a bottom-up approach to identify strategies from the 51-language empirical sample in ClaimReview, and used it to estimate how often strategies appear and occur together across the corpus. The existing propaganda taxonomies have no direct equivalents for several tactics in our taxonomy, such as conspiracy theory, vivid imagery and misleading statistics. This reflects their focus on rhetoric in general news content, whereas our taxonomy focuses on fact-checked falsehoods (Supplementary Information section~\ref{appendix:semeval_map}).

We calculated the corpus-wide prevalence of each strategy group. For each pair of strategy groups we computed pointwise mutual information (PMI): the $\log_2$ lift over independence (Fig.~\ref{fig:cooc_lift}). To adjust for the fact that strategies compete for a limited number of slots (each claim carries up to five labels), we also compared each strategy pair against a maximum-entropy null. The null keeps each claim's label count as observed and fills those slots at random, with each strategy weighted so that it appears as often as it does in the data in expectation. Pairs that depart from this null (BH-adjusted $q < 0.001$ and $|\log_2 \text{lift}| \geq 0.3$) enter a signed network as positive or negative edges, and we partitioned the network by maximizing signed modularity\citep{gomez2009analysis}. We checked bloc membership across eight edge-retention rules ($q \in \{0.01, 0.001\}$, $|\log_2 \text{lift}| \in \{0.1, 0.2, 0.3, 0.4\}$) and report any strategy that falls in each bloc under half of the rules as a bridge. Null-model details, per-strategy bloc stability and alternative partitions are in Supplementary Information section~\ref{appendix:network}. Finally, we compared strategy use on and off social media. We coded a claim as originating on social media when its original source URL pointed to a social media platform, and regressed this indicator on the 13 strategy-group indicators in a linear probability model with mega-topic, language and year-month fixed effects and heteroskedasticity-robust standard errors. See Supplementary Information section~\ref{appendix:by_channel} for details. %

\subsection{LLM evaluator prebunking experiment}

\emph{Design.} We prebunk two lightweight LLM evaluators, GPT-4o-mini (\texttt{gpt-4o-mini-2024-07-18}, temperature 0) and, for replication in a second model family of comparable size and cost, Gemini 2.5 Flash (\texttt{gemini-2.5-flash}), with four evaluator-side briefings that separate strategy knowledge (acquired from \secref{Mapping the persuasion playbook}) from decoupling, along with a no-briefing baseline (C0):
\begin{itemize}
    \item[\textbf{C1}] \emph{accuracy prompt}: ``to the best of your knowledge, is this claim accurate?'', with no strategy content\citep{pennycook2021shifting}.
    \item[\textbf{C2}] \emph{strategy knowledge}: the 13 strategy groups, their corpus prevalence and substantively meaningful co-occurrence pairs ($\lvert\log_2\text{lift}\rvert \geq 0.5$; Fig.~\ref{fig:cooc_lift}).
    \item[\textbf{C3}] \emph{full briefing}: C2 augmented with an explicit instruction that ``the presence of a persuasion strategy describes \emph{how} a claim is communicated, not \emph{whether} it is true.''
    \item[\textbf{C0b}] \emph{decoupling only}: the same decoupling instruction without any strategy knowledge. %
\end{itemize}
In C2 and C3 the evaluator also names the strategies it detects before scoring. In C3 and C0b the decoupling instruction also asked the model to assess factual plausibility independently of style and to reserve scepticism for claims that are both strategy-laden and implausible. Under each briefing condition, the model reports a believability score for each claim on 1--7 scale. In later analysis, we report the per-persona discernment gap (mean believability of true minus false claims), matching the pre-registered primary outcome of the human experiment. Because the same 50 personas rate all claims under every briefing, we compared briefings with paired $t$-tests across personas. We further report per-persona ROC-AUC which is invariant to how the model uses the rating scale (Supplementary Table~\ref{tab:discrimination_full}). Supplementary Information section~\ref{appendix:conditions} provides the full specifications and evaluation protocol.

\emph{Claim stimuli and personas.} The claim set comprises 240 AI-generated false claims (generated with \texttt{gpt-4o-2024-08-06}, temperature 0.6) anchored to eight seeding events that occurred after the evaluator's October 2023 training cutoff, plus 40 true claims about the same events (human-curated from published news sources). These true and false claims are length-matched (median 17 vs.\ 19 words). Stimuli generation details, event descriptions and sources are in Supplementary Information sections~\ref{appendix:conditions}--\ref{appendix:true-claims}. Post-cutoff timing alone does not guarantee that the model lacks knowledge on the claims, so we probe the evaluator model in a context-free API call to elicit the model's admitted basis for each judgement (Supplementary Information section~\ref{appendix:basis_probe}). The probe identifies 108 of the 240 false claims as having no basis (the evaluator reported no reliable factual basis) and 122 as drawing on general knowledge (it reported relevant general topical knowledge). The contrast between these two strata helps test the role of verification capacity on the effectiveness of the strategy knowledge briefing. Fifty personas with different demographic and political leanings were drawn at random from a 2,058-persona pool of LLM-simulated personas, built from a publicly released dataset of consenting US survey respondents\citep{toubia2025database}. Each persona evaluated all 280 claims under the four briefings and, in a separate run, under no briefing. Within-persona-level tests quantify sensitivity across prompt-conditioned evaluations of a single model, not sampling variability across claims, events, generators or model families.

\emph{Cross-model replication.} We replicated the five conditions (C0,C1,C2,C3,C0b) with Gemini 2.5 Flash, holding claims, personas, and prompts fixed. We also re-ran the knowledge basis probe for that model, and scored each model on its own no-basis claims (108 false claims for GPT-4o-mini, 105 for Gemini 2.5 Flash) against the 40 shared true claims. See Supplementary Information section~\ref{appendix:crossmodel} for details.

\subsection{Human prebunking experiment}

\emph{Inclusion and ethics.} The study was conducted from 7 July 2026 to 8 July 2026, approved by the Duke Campus Institutional Review Board (protocol 2025-0404). All participants gave informed consent. At the end of the study, participants were debriefed with the true or false status of every claim they had rated. The study was pre-registered on AsPredicted (\url{https://aspredicted.org/du8xe2.pdf}). Non-pre-registered analyses are labelled exploratory, and deviations are reported in Supplementary Information section~\ref{appendix:studyB_protocol}.

\emph{Participants.} We pre-registered a target of 750 complete responses after exclusions (150 per arm). Participants were US adults recruited on Prolific with quotas on political affiliation (40\% Democrat, 40\% Republican, 20\% Independent). In total, 851 participants began the survey. Two left before and 91 after random assignment. Dropout was higher in the two strategy arms (15\% versus 8\% in the other three arms; occurring mostly on the briefing pages). Bounds that allow for this selective dropout do not reverse the sign of either confirmatory contrast (Supplementary Information section~\ref{appendix:studyB_attrition}). Of the 758 participants who submitted the survey, we excluded nine (speeders, straight-liners, one record without an assigned arm and three completions through a leaked survey link), leaving 749 for analysis (155 in C0, 155 in C0b, 154 in C1, 140 in C2 and 145 in C3; 323 Democratic-leaning, 313 Republican-leaning and 113 Independent; Supplementary Information section~\ref{appendix:studyB_protocol}).

\emph{Materials and procedure.} Participants were randomly assigned to one of five between-subjects briefing arms that mirror the language-model experiment: no briefing (C0), decoupling only (C0b), accuracy prompt (C1), strategy knowledge (C2) and full briefing (C3), which adds the decoupling instruction to the strategy knowledge. After consenting and reading the task instructions, participants read their assigned briefing. The C2 and C3 briefings presented the 13 strategy groups with their corpus prevalence and main co-occurrence patterns (Fig.~\ref{fig:cialdini_pct} and~\ref{fig:cooc_lift}, with the corresponding textual descriptions), followed by three practice claims in which participants marked the strategies present and received feedback; C3 then added the decoupling instruction. The accuracy prompt and decoupling texts are reproduced in Supplementary Table~\ref{tab:briefing_texts}. Because the strategy briefings took longer to read, their effects combine content with reading time.

After one unscored practice claim, each participant rated 20 claims (7 true, 13 false) in random order. The claims came from one of four fixed batches, assigned at random and drawn from an 80-claim random sample of the 280 claims used with the language model (\secref{Briefing a language-model evaluator}). For each claim, participants rated its believability on the same 1--7 scale, explained their rating in free text and reported how familiar they had been with its topic before the study (1--5). They then completed a cognitive reflection test and demographic questions, and were shown the verdict of every claim they had rated. Details of the arms, claim sampling and task are in Supplementary Information section~\ref{appendix:studyB_protocol}, and the true claims are listed in Supplementary Information section~\ref{appendix:true-claims}.

\emph{Analysis.} The pre-registered primary outcome is each participant's discernment gap, the difference between a participant's mean believability ratings of true and false claims. Three tests were confirmatory and Holm-corrected. The first two asked whether the full briefing widened the discernment gap compared with no briefing and compared with the decoupling instruction alone. The third asked whether strategy knowledge without the decoupling instruction lowered belief in true claims. We also pre-registered secondary analyses of discernment measured by AUC and of how often participants' textual explanations named a taught strategy. We used a mixed model to test whether the effect depended on claim familiarity, also pre-registered. Differences by political leaning were pre-registered as exploratory. Exploratory analyses are not corrected for multiple comparisons. Supplementary Information section~\ref{appendix:studyB_results} and Supplementary Table~\ref{tab:studyB_tests} give the hypotheses, model specifications and full results.

\subsection{Adversarial test}

The five generator-side conditions added strategy knowledge in steps, from zero-shot generation (A1) through the 48-strategy taxonomy (A2) and the 13 strategy groups with their prevalence (A3) to corpus-wide (A4) and topic-specific (A5) co-occurrence patterns (Supplementary Information section~\ref{appendix:conditions}). For each of 20 seeding events anchored to developments after the evaluator's October 2023 cutoff (Supplementary Table~\ref{tab:events}; 12 concern recurring topics on which the model reported pre-cutoff knowledge) and each condition, GPT-4o (\texttt{gpt-4o-2024-08-06}, temperature 0.6) generated 20 claims (2,000 claims in total); eight of the 20 events passed the stricter novelty probe used in the discernment study. A 100-persona evaluator panel, drawn from the same 2,058 persona pool, rated all 2,000 claims under the four evaluator-side briefings (C1, C2, C3, C0b), in batches of 20 claims per call. The same personas' no-briefing ratings of the same claims were collected in a separate run, also in batches of 20 (Supplementary Information section~\ref{subsec:persona_eval}). We measured each briefed rating as a shift ($\Delta$) from the same persona's no-briefing rating of the same claim (Supplementary Information section~\ref{appendix:adversarial_full}). As a cross-family replication, a Gemini 2.5 Flash evaluator rated a random subsample of 500 claims under no briefing, the accuracy prompt, the full briefing and the decoupling instruction alone, using the 50 personas from the discernment experiment (Supplementary Information section~\ref{appendix:adversarial_crossmodel}).

\subsection{Data availability}
Because the materials used to study persuasion in misinformation have dual-use potential, we do not publicly release the fully labelled claim-level corpus, the strategy-knowledge briefings, the strategy-guided generation prompts and scripts, or the 2,000-claim generation set used in the adversarial test. Releasing them could facilitate misuse, such as building or fine-tuning models for deceptive persuasion. These materials may be made available from the authors upon reasonable request. Requests will be evaluated on an individual basis, considering the researcher's credentials, purpose and commitment to mitigating dual-use risks. Aggregated results, the 280 experimental claims, the language-model ratings, de-identified human-subject data and the survey instrument (without the strategy-knowledge briefing pages) will be deposited in a public repository upon acceptance. The underlying fact-checked claims are available from the ClaimReview project under its own terms. The pre-registration is available on AsPredicted (\url{https://aspredicted.org/du8xe2.pdf}).

\subsection{Code availability}
Analysis code, annotation prompts, and model and parameter settings will be deposited in a public repository upon acceptance. The strategy-guided generation prompts and scripts, as well as the evaluator briefing prompts with strategy knowledge, are restricted for the same dual-use reasons and may be made available from the authors upon reasonable request (\emph{Data availability}).

\printbibliography[notcategory=mainrefs,title={Methods references}] %

\section*{Author contributions}
T.G. and Q.M. conceived the study. T.G. analysed the ClaimReview data. Y.X. and Q.M. performed the open coding of strategies and the micro-topic visualization. T.G. and T.Y. classified the mega-topics and consolidated the open codes into the 48-strategy taxonomy, which T.G. then mapped onto 13 strategy groups under three theoretical frameworks. T.G. and Q.M. designed and analysed the experiments. T.G. and Q.M. wrote the paper.

\newpage

\appendix
\clearpage
\begin{refsection} %
\setcounter{secnumdepth}{3} %
\setcounter{section}{0}\setcounter{figure}{0}\setcounter{table}{0}\setcounter{page}{1}
\renewcommand{\thesection}{S\arabic{section}}
\renewcommand{\thesubsection}{\thesection.\arabic{subsection}}
\renewcommand{\thesubsubsection}{\thesubsection.\arabic{subsubsection}}
\renewcommand{\thefigure}{S\arabic{figure}}
\renewcommand{\thetable}{S\arabic{table}}
\renewcommand{\figurename}{Fig.}
\captionsetup[figure]{labelsep=period}\captionsetup[table]{labelsep=period} %
\renewcommand{\contentsname}{Content}
\singlespacing
\begin{center}
{\Large Supplementary Information for}\\[1em]
{\LARGE\textbf{A misinformation playbook sharpens judgement without breeding cynicism}}\\[1.2em]
{\large Tong Guo$^{*}$, Teng Ye, Yutong Xie \& Qiaozhu Mei$^{*}$}\\[0.6em]
{\small $^{*}$Corresponding authors: tong.guo@duke.edu, qmei@umich.edu}
\end{center}
\vspace{1em}
\addtocontents{toc}{\protect\setcounter{tocdepth}{2}}
\tableofcontents
\clearpage

\section{Overview of Analyses}\label{appendix:estimands}

\begin{table}[htp]
\centering
\footnotesize
\setlength{\tabcolsep}{4pt}
\caption{Estimands and inferential units of the paper's five analyses.}
\label{tab:estimands}
\begin{tabular}{p{2.5cm}p{1.9cm}p{2.7cm}p{2.2cm}p{2.7cm}}
\toprule
\textbf{Analysis} & \textbf{Unit} & \textbf{Contrast} & \textbf{Outcome} & \textbf{N; inference} \\
\midrule
Prevalence, co-occurrence and network (\secref{Mapping the persuasion playbook}) & fact-checked claim & descriptive & strategy labels & 214{,}646 claims; pairwise lifts against a maximum-entropy baseline (BH-adjusted); signed network \\
\addlinespace[2pt]
Channel association (Supplementary Information section~\ref{appendix:lpm_avg}) & fact-checked claim & strategy indicators; topic, language, time fixed effects & social-media origination & 214{,}505 claims; LPM with 95\% CIs \\
\addlinespace[2pt]
Machine discrimination (\secref{Briefing a language-model evaluator}) & persona $\times$ claim rating & briefing conditions, within-persona & believability (1--7); per-persona ROC-AUC & 50 personas $\times$ 280 claims $\times$ 4 conditions (+ C0 floor); paired per-persona tests \\
\addlinespace[2pt]
Human discernment (\secref{Briefing human evaluators}) & participant & five randomized arms, between-subjects & true$-$false discernment gap & 749 participants (20 claims each); person-level Welch tests, pre-registered \\
\addlinespace[2pt]
Adversarial reversal (\secref{The adversarial stress test}) & persona $\times$ claim $\times$ condition & briefed versus no briefing, within persona-claim & $\Delta$ believability & 100 personas $\times$ 2{,}000 claims; paired shifts; within-model \\
\bottomrule
\end{tabular}
\end{table}

\section{Corpus Overview}\label{appendix:corpus_overview}

ClaimReview is a structured-data schema developed by the Duke Reporters' Lab with Google, Jigsaw and Schema.org. Fact-checking organizations use it to encode the metadata of their fact-checks so that search engines, social platforms and misinformation monitors can read them\citep{duke_reporters_lab_claimreview}. Each entry records the claim evaluated, the original claimant, appearance dates and contexts, the fact-checking organization, and the verdict. The database holds over 250,000 evaluated claims from hundreds of independent fact-checking outlets worldwide, including signatories to the International Fact-Checking Network (IFCN) Code of Principles such as PolitiFact, FactCheck.org, Full Fact and Maldita, with multilingual coverage dating back to 2015.

Our analysis sample comprises 214,646 fact-checked claims with effective fact-check dates between January 2019 and April 2024. We map the fact-checkers' original ratings onto three categories: \emph{False} (``False,'' ``Misleading,'' ``Mostly False,'' and equivalents), \emph{True} (``True,'' ``Mostly True,'' and equivalents), and \emph{Not Clear} (``Mixture,'' ``Half True,'' ``Missing Context,'' and equivalents). Of the 214,646 claims, 76.3\% are rated False, 19.4\% Not Clear, and 4.3\% True.

The sample spans 51 languages. English accounts for the largest share (31.6\%), followed by Arabic (13.7\%), Spanish (10.6\%), Hindi (7.5\%), Portuguese (6.3\%), Italian (4.3\%), Turkish (3.1\%), German (3.0\%), and Polish (2.7\%); the remaining 42 languages together account for 17.2\%.

\section{Characterizing Megatopics} \label{appendix:megatopics}
We recovered topics with a three-step LLM-assisted procedure. LLMs perform well at text comprehension and annotation\citep{gilardi2023chatgpt}, and in the evaluation of \citet{pham2023topicgpt} this style of topic modelling outperformed both latent Dirichlet allocation (LDA)\citep{blei2003latent} and the embedding-based BERTopic\citep{grootendorst2022bertopic}.

For each candidate number of topics $K$, we prompted GPT-4o to identify topics in a batch of 5,000 randomly sampled claims, repeating the procedure 15 times to cover the distribution.

We then prompted GPT-4o to merge the per-batch topic sets and remove duplicates, giving a final list of $K$ topics.

Finally, GPT-4o assigned each claim to at most one topic from the final $K$-topic set; claims that fit none were left in an ``other'' residual. Of the 214,646 claims, 196,306 (91.5\%) received one of the eight megatopics and 18,340 (8.5\%) fell in the residual, which is excluded from the per-topic breakdowns.

\begin{figure}[h]
    \centering
    \includegraphics[width=0.4\linewidth]{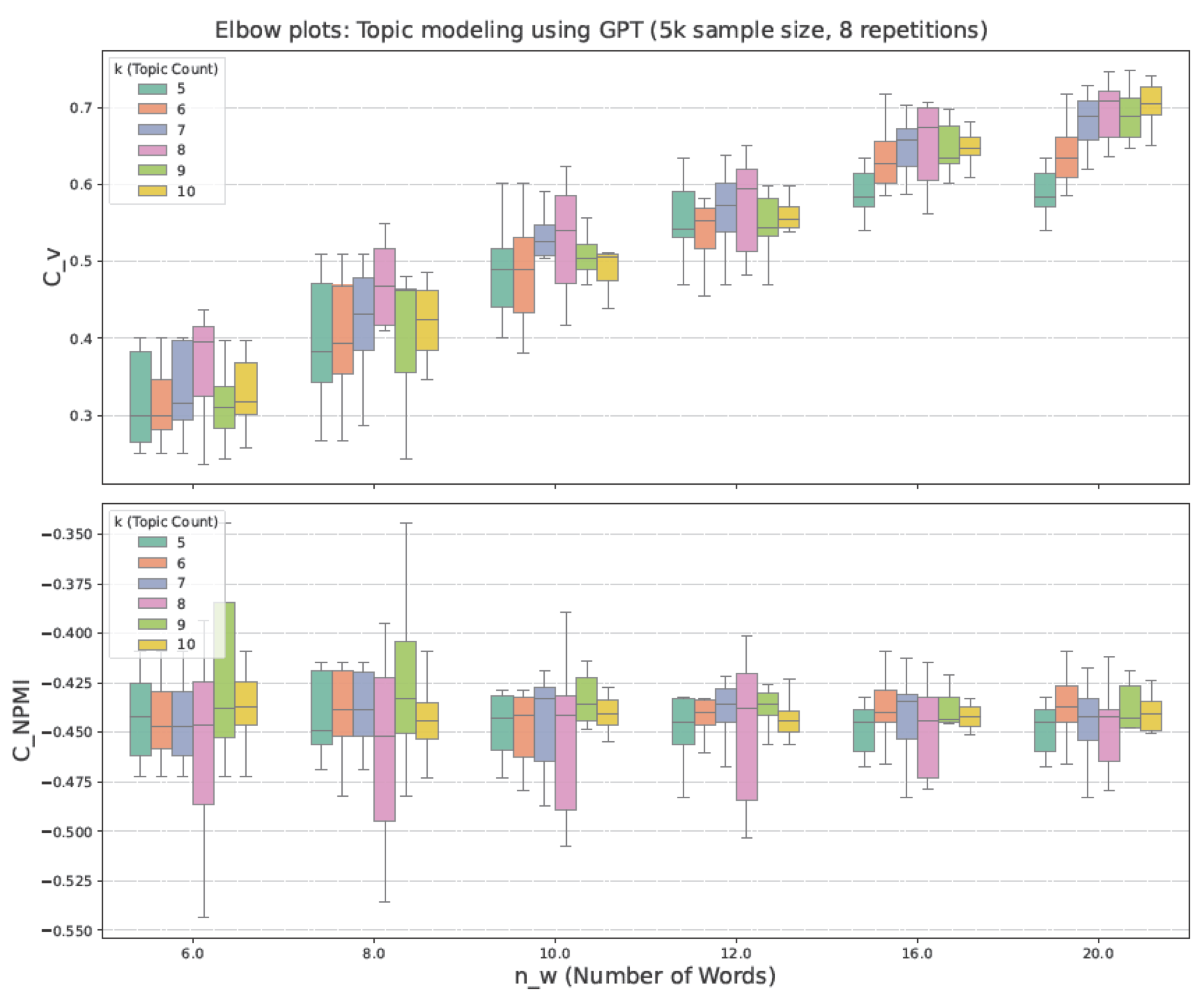}
    \caption{Topic coherence (CV and NPMI) as a function of the number of topics $K$. The curves bend at $K = 8$.}
    \label{fig:elbowplot}
\end{figure}

To choose the number of topics, we computed two coherence measures (CV and normalized pointwise mutual information, NPMI) across values of $K$ and selected $K = 8$, where the curves bend (Fig.~\ref{fig:elbowplot}).

The micro-topic map in Fig.~\ref{fig:topiccloud} is drawn for English-language claims only, because embeddings of claims in different languages would cluster partly by language rather than topic. It covers 66{,}130 of the 67{,}866 English claims in the analysis sample (97.4\%). The mega-topic shares in its legend are computed on all 214{,}646 claims.

\newpage
\section{48-Strategy Taxonomy}\label{appendix:48strat}
Table~\ref{tab:48strat} lists the 48 strategies with their definitions and prevalence, from most to least common. \emph{Emotional Appeal} and \emph{Appeal to Authority} each appear in more than half of all claims, and the five most common strategies (\emph{Emotional Appeal}, \emph{Appeal to Authority}, \emph{Appeal to Fear}, \emph{False Attribution} and \emph{Cherry-picking}) account for most labelled instances.

{\small
\begin{longtable}{p{3.2cm} p{9.5cm} r}
\caption{The 48-strategy taxonomy with corpus-wide prevalence. $N = 214{,}646$ fact-checked claims, January 2019--April 2024. Each claim receives up to five strategy labels; percentages therefore sum to more than 100\%.}
\label{tab:48strat} \\
\toprule
\textbf{Strategy} & \textbf{Definition} & \textbf{\%} \\
\midrule
\endfirsthead
\toprule
\textbf{Strategy} & \textbf{Definition} & \textbf{\%} \\
\midrule
\endhead
\midrule
\multicolumn{3}{r}{\footnotesize\textit{Continued on next page}} \\
\endfoot
\bottomrule
\multicolumn{3}{p{12.7cm}}{\footnotesize \textit{Note}: Strategies are listed in decreasing order of prevalence.} \\
\endlastfoot
\emph{Emotional Appeal} & Using emotionally charged language or scenarios to elicit a strong emotional reaction from the audience, which can cloud their judgement or critical thinking; Exaggerating or dramatizing an event to invoke strong emotional reactions such as shock or outrage & 64.7\% \\
\addlinespace[2pt]
\emph{Appeal to Authority} & Referencing a figure of authority or someone with perceived power to bolster the credibility of the claim & 50.2\% \\
\addlinespace[2pt]
\emph{Appeal to Fear} & Leveraging fear to influence an audience's reaction & 30.0\% \\
\addlinespace[2pt]
\emph{False Attribution} & Citing a reputable source or authority to lend credibility to false or misleading information & 23.7\% \\
\addlinespace[2pt]
\emph{Cherry-picking} & Selecting specific data or instances that support a particular position while ignoring relevant contradicting information & 20.4\% \\
\addlinespace[2pt]
\emph{Lack of Context} & Providing information without necessary background or situational details that are essential for an accurate understanding & 11.4\% \\
\addlinespace[2pt]
\emph{Unverified Claims} & Statements or assertions made without providing supporting evidence that can be independently confirmed or validated & 11.3\% \\
\addlinespace[2pt]
\emph{Exaggeration} & The representation of something as being larger, better, or worse than it really is & 11.2\% \\
\addlinespace[2pt]
\emph{Bandwagon Effect} & Suggesting that something is true or desirable because it is believed or done by many people, including the concept of Common Belief Bias & 11.2\% \\
\addlinespace[2pt]
\emph{Vivid Imagery} & Describing scenes in a vivid, detailed manner to create a strong mental image for the audience & 11.0\% \\
\addlinespace[2pt]
\emph{Anecdotal Evidence} & Using personal stories or isolated examples instead of sound reasoning or compelling evidence to support a claim & 10.6\% \\
\addlinespace[2pt]
\emph{Vagueness} & Using non-specific language that lacks detail, making the statement difficult to verify or refute & 7.1\% \\
\addlinespace[2pt]
\emph{Oversimplification} & A logical fallacy that involves reducing complex issues to overly simple terms or explanations, often omitting crucial details and leading to incomplete or distorted understanding & 5.5\% \\
\addlinespace[2pt]
\emph{Conspiracy Theory} & A belief that events or situations are secretly manipulated behind the scenes by powerful forces with negative intent, often without sufficient evidence & 5.5\% \\
\addlinespace[2pt]
\emph{Misrepresentation} & Presenting information in a misleading manner or distorting facts to support a particular viewpoint or agenda & 5.3\% \\
\addlinespace[2pt]
\emph{Association Fallacy} & Drawing an unjustified conclusion about an idea, person, or event based on its connection to another, often used to transfer qualities or implications without logical justification & 4.5\% \\
\addlinespace[2pt]
\emph{Specificity} & Providing detailed and specific information to create the impression of thorough research and accuracy & 4.3\% \\
\addlinespace[2pt]
\emph{False Causality} & Incorrectly suggesting that because two events occur together or in sequence, one causes the other & 3.8\% \\
\addlinespace[2pt]
\emph{Misleading Statistics} & The use of statistical data in a misleading way, which can involve incomplete statistics, wrong interpretation, or presenting numbers without proper context & 3.4\% \\
\addlinespace[2pt]
\emph{Confirmation Bias} & Focusing on information that confirms one's preexisting beliefs and ignoring information that challenges those beliefs & 3.4\% \\
\addlinespace[2pt]
\emph{Overgeneralization} & Making a broad or generalized claim based on limited evidence & 3.3\% \\
\addlinespace[2pt]
\emph{Urgency} & Creating a sense of immediacy or scarcity to prompt quick action without thorough consideration & 3.1\% \\
\addlinespace[2pt]
\emph{Misleading Headline} & Using a headline that is sensationalist, exaggerated, or unrelated to the actual content to attract attention and clicks & 2.6\% \\
\addlinespace[2pt]
\emph{Appeal to Nationalism} & Leveraging patriotic sentiment or pride in one's country to persuade the audience, often framing the information in a way that suggests national honor or security is at stake & 2.5\% \\
\addlinespace[2pt]
\emph{Appeal to Novelty} & Presenting information or objects that are new or unique to capture attention and create excitement & 2.2\% \\
\addlinespace[2pt]
\emph{Scapegoating} & Assigning blame to a person or group without considering the complete situation, often oversimplifying complex issues & 2.1\% \\
\addlinespace[2pt]
\emph{Character Assassination} & Attempting to discredit a person by presenting misleading or false information about their character or actions & 2.0\% \\
\addlinespace[2pt]
\emph{False Dichotomy} & Presenting two alternative states as the only possibilities when in fact more options are available & 1.9\% \\
\addlinespace[2pt]
\emph{Appeal to Tradition} & Arguing that something is better or correct simply because it is traditional or has been done for a long time & 1.6\% \\
\addlinespace[2pt]
\emph{Framing} & Presenting information in a way that influences the perception and interpretation of the audience & 1.5\% \\
\addlinespace[2pt]
\emph{False Equivalence} & Presenting two unequal things as if they are equivalent, suggesting a false balance between them & 1.4\% \\
\addlinespace[2pt]
\emph{Scientific Jargon} & Use of technical terminology to give an impression of credibility and authority & 1.3\% \\
\addlinespace[2pt]
\emph{Appeal to Identity} & This tactic seeks to align the information with the identity, values, or beliefs of a particular group to make it more persuasive to members of that group & 1.3\% \\
\addlinespace[2pt]
\emph{Fabrication} & The creation of false information with the intention to deceive & 1.2\% \\
\addlinespace[2pt]
\emph{Strawman Argument} & Distorting or oversimplifying an opponent's argument to make it easier to refute & 1.1\% \\
\addlinespace[2pt]
\emph{Historical Revisionism} & Distorting or altering historical facts and events to fit a particular agenda or perspective & 1.0\% \\
\addlinespace[2pt]
\emph{Appeal to Nature} & The belief that natural or organic remedies are inherently better or healthier than synthetic alternatives & 0.9\% \\
\addlinespace[2pt]
\emph{Appeal to Ridicule} & Using mockery or derision to frame an opposing argument or viewpoint as foolish or worthy of scorn, thereby undermining its credibility without addressing its merits & 0.8\% \\
\addlinespace[2pt]
\emph{Us vs.\ Them} & Creating a divide between two groups to incite opposition and solidarity within the in-group & 0.6\% \\
\addlinespace[2pt]
\emph{Loaded Question} & A question that contains a controversial or unjustified assumption, designed to manipulate the logical structure of inquiry and response & 0.5\% \\
\addlinespace[2pt]
\emph{Appeal to Prejudice} & Using loaded or emotive language to appeal to the biases or prejudices of the audience against particular groups & 0.5\% \\
\addlinespace[2pt]
\emph{Call to Action} & Encouraging readers to take immediate action, often creating a sense of urgency & 0.5\% \\
\addlinespace[2pt]
\emph{Stereotyping} & Generalizing or oversimplifying about a particular group of people to reinforce certain biases or preconceptions & 0.4\% \\
\addlinespace[2pt]
\emph{Appeal to Suspicion} & Using language that suggests hidden motives or conspiracies without concrete evidence, encouraging doubt and alternative interpretations & 0.4\% \\
\addlinespace[2pt]
\emph{Appeal to Ignorance} & Arguing that a proposition is true because it has not yet been proven false & 0.4\% \\
\addlinespace[2pt]
\emph{Polarization} & Dividing an audience into opposing groups, often through the use of provocative or controversial statements & 0.4\% \\
\addlinespace[2pt]
\emph{Repetition} & The act of repeating the same message or piece of misinformation to exploit the tendency to believe information after multiple exposures & 0.4\% \\
\addlinespace[2pt]
\emph{Partial Truth} & Providing some elements of truth mixed with falsehoods or misleading information to create a distorted version of reality & 0.4\% \\
\end{longtable}
}

\newpage
\section{Theoretical Mappings}\label{appendix:cialdini}

Six of Cialdini's seven principles of social influence\citep{cialdini2021influence} appear in the playbook (Table~\ref{tab:membership}). Each draws on a well-documented psychological tendency; for example, \emph{Commitment \& Consistency} relates to confirmation bias\citep{nickerson1998confirmation,rabin1999first} and \emph{Unity} to motivated reasoning\citep{kunda1990case,taber2006motivated}. \emph{Reciprocity} does not appear, probably because it unfolds over an exchange between people, whereas our corpus consists of single claims rather than conversations.

Table~\ref{tab:membership} shows how the 48 strategies map onto the 13 groups and a residual ``other'' set. The groups are a theoretical layer that we placed over the inductively derived strategies (\emph{Methods}); the table notes explain the least obvious assignments.

\begin{table}[htp]
\centering
\small
\begin{threeparttable}
\caption{Canonical 48-to-13 membership mapping.}
\label{tab:membership}
\begin{tabular}{llp{6.8cm}}
\toprule
\textbf{Group} & \textbf{Framework} & \textbf{Constituent strategies} \\
\midrule
Authority & Cialdini & Appeal to Authority; False Attribution; Scientific Jargon \\
Commitment \& Consistency & Cialdini & Cherry-picking; Partial Truth; False Dichotomy; Loaded Question; Confirmation Bias; Appeal to Tradition \\
Liking & Cialdini & Appeal to Ridicule; Association Fallacy \\
Scarcity & Cialdini & Appeal to Novelty; Urgency \\
Social Proof & Cialdini & Anecdotal Evidence; Bandwagon Effect \\
Unity & Cialdini & Appeal to Identity; Us vs.\ Them; Scapegoating; Appeal to Nationalism; Polarization; Appeal to Prejudice \\
Emotional Appeal & Affective & Emotional Appeal \\
Appeal to Fear & Affective & Appeal to Fear \\
Vivid Imagery & Affective & Vivid Imagery \\
Conspiracy Theory & Affective & Conspiracy Theory \\
Hide Information & Information & Lack of Context; Vagueness \\
Manipulate Information & Information & Exaggeration; Misrepresentation; Specificity; Oversimplification; Overgeneralization; Framing; Misleading Headline; Misleading Statistics \\
Fabricate Information & Information & Unverified Claims; Fabrication \\
Other & (unmapped) & False Causality; Character Assassination; False Equivalence; Historical Revisionism; Strawman Argument; Appeal to Nature; Stereotyping; Call to Action; Appeal to Suspicion; Appeal to Ignorance; Repetition \\
\bottomrule
\end{tabular}
\begin{tablenotes}
\small
\item \textit{Notes}: 37 of the 48 strategies map to the 13 groups (21 Cialdini, 4 affective, 12 information); the remaining 11 form the residual set used by the alignment analysis of Supplementary Information section~\ref{appendix:strategy_alignment}. \emph{Conspiracy Theory} is placed in the affective family because it persuades through threatening narratives rather than through evidence (\emph{Methods}); \emph{Appeal to Ridicule} and \emph{Association Fallacy} are placed under Liking because both transfer liking or dislike of a person or group onto the claim; \emph{Cherry-picking} and \emph{Confirmation Bias} are placed under Commitment \& Consistency because both exploit the audience's wish to stay consistent with prior beliefs.
\end{tablenotes}
\end{threeparttable}
\end{table}

\newpage
\section{Comparison with NLP Persuasion-Technique Taxonomies}\label{appendix:semeval_map}

Table~\ref{tab:semeval_map} maps the 23 persuasion techniques of the SemEval-2023 Task~3 taxonomy\citep{piskorski2023semeval}, which consolidates the 18-technique propaganda taxonomy of Da San Martino et al.\citep{dasanmartino2019finegrained} and its SemEval-2020 revision\citep{dasanmartino2020semeval}, onto their closest counterparts in our 48-strategy taxonomy (Supplementary Information section~\ref{appendix:48strat}). Twenty of the 23 techniques (13 direct, 7 partial) have a counterpart among our strategies. The reverse does not hold: many of our strategies have no SemEval counterpart, mostly in the information-manipulation family (\emph{Fabrication}, \emph{Unverified Claims}, \emph{Cherry-picking}, \emph{Lack of Context}, \emph{Misrepresentation}, \emph{Misleading Statistics}, \emph{Partial Truth}, \emph{Misleading Headline}, \emph{False Attribution}, \emph{Historical Revisionism}) and in evidence-simulation tactics (\emph{Specificity}, \emph{Scientific Jargon}, \emph{Anecdotal Evidence}, \emph{Vivid Imagery}, \emph{Conspiracy Theory}). The difference reflects what each corpus contains: propaganda-technique taxonomies describe rhetoric in news content, which need not be false, whereas fact-checked misinformation often also degrades or fakes evidence.

\begin{table}[h!]
\centering
\footnotesize
\begin{threeparttable}
\caption{Mapping the SemEval-2023 Task 3 persuasion-technique taxonomy\citep{piskorski2023semeval} onto the 48-strategy taxonomy. ``Direct'' = same construct under a near-identical definition; ``Partial'' = overlapping but not coextensive constructs; ``---'' = no counterpart in our taxonomy.}
\label{tab:semeval_map}
\begin{tabular}{p{5.2cm} p{5.6cm} l}
\toprule
\textbf{SemEval-2023 technique} & \textbf{Closest strategy in our taxonomy} & \textbf{Match} \\
\midrule
\multicolumn{3}{l}{\emph{Attack on Reputation}} \\
Name Calling or Labelling & Character Assassination; Stereotyping & Partial \\
Guilt by Association & Association Fallacy & Direct \\
Casting Doubt & Appeal to Suspicion & Direct \\
Appeal to Hypocrisy & --- & --- \\
Questioning the Reputation & Character Assassination & Partial \\
\addlinespace
\multicolumn{3}{l}{\emph{Justification}} \\
Flag Waving & Appeal to Nationalism & Direct \\
Appeal to Authority & Appeal to Authority & Direct \\
Appeal to Popularity & Bandwagon Effect & Direct \\
Appeal to Values & Appeal to Tradition; Appeal to Identity & Partial \\
Appeal to Fear, Prejudice & Appeal to Fear; Appeal to Prejudice & Direct \\
\addlinespace
\multicolumn{3}{l}{\emph{Distraction}} \\
Strawman & Strawman Argument & Direct \\
Red Herring & --- & --- \\
Whataboutism & False Equivalence & Partial \\
\addlinespace
\multicolumn{3}{l}{\emph{Simplification}} \\
Causal Oversimplification & False Causality; Oversimplification & Direct \\
False Dilemma or No Choice & False Dichotomy & Direct \\
Consequential Oversimplification & Oversimplification; Overgeneralization & Partial \\
\addlinespace
\multicolumn{3}{l}{\emph{Call}} \\
Slogans & Call to Action & Partial \\
Conversation Killer & --- & --- \\
Appeal to Time & Urgency & Direct \\
\addlinespace
\multicolumn{3}{l}{\emph{Manipulative Wording}} \\
Loaded Language & Emotional Appeal & Partial \\
Obfuscation, Vagueness, Confusion & Vagueness & Direct \\
Exaggeration or Minimisation & Exaggeration & Direct \\
Repetition & Repetition & Direct \\
\bottomrule
\end{tabular}
\begin{tablenotes}
\small
\item \textit{Notes}: SemEval-2023 techniques are grouped under the taxonomy's six coarse categories (italicized). Our strategies are listed by their Table~\ref{tab:48strat} names. The mapping is our judgement, based on published definitions; ``Partial'' entries differ in scope (e.g., \emph{Loaded Language} concerns wording whereas \emph{Emotional Appeal} also covers scenario-level emotional framing).
\end{tablenotes}
\end{threeparttable}
\end{table}

\newpage
\section{Signed Strategy Network: Null Model and Bloc Stability}\label{appendix:network}

Each claim is represented by a binary vector over the 13 strategy groups. Because a claim carries at most five labels, strategies compete for label slots, so comparing pairs with independence produces negative associations by construction: against independence, the mean pairwise $\log_2$ lift is $-0.40$. We therefore compare each pair's observed co-occurrence count with its expectation under a maximum-entropy null. In this null, every claim keeps its observed number of strategy groups, and groups are drawn with weights fitted (by iterative proportional fitting) so that each strategy's expected total equals its observed total. Expected pair counts and their variances are computed analytically; the null is the analytic equivalent of a swap randomization that preserves each claim's label count and each strategy's total. Against this null, the mean pairwise $\log_2$ lift is $-0.10$. A pair's edge weight is $\log_2[(\text{observed} + 0.5)/(\text{expected} + 0.5)]$; its $z$-statistic uses the null variance of the count, and two-sided $p$-values are Benjamini--Hochberg adjusted across the 78 pairs. The main network (Fig.~\ref{fig:network}) retains edges with $q < 0.001$ and $|\log_2 \text{lift}| \geq 0.3$: 13 positive and 20 negative associations.

We partitioned the signed network by maximizing the signed modularity of G\'omez et al.\citep{gomez2009analysis}, which rewards positive edges within blocs and negative edges between them. The best partition has two blocs ($Q = 0.355$), but the number of blocs is weakly determined: the best three-bloc partition scores almost the same ($Q = 0.353$), splitting bloc A into credibility strategies (\emph{Authority}, \emph{Social Proof}, \emph{Scarcity}) and vividness strategies (\emph{Liking}, \emph{Emotional Appeal}, \emph{Vivid Imagery}, \emph{Hide Information}); the core of bloc B (\emph{Conspiracy Theory}, \emph{Appeal to Fear}, \emph{Fabricate Information}, \emph{Unity}, \emph{Commitment \& Consistency}) is unchanged at every number of blocs from two to five under every rule we examined. In a network this small, $Q$ alone is not evidence of structure: placing the same edge weights at random positions gives a best $Q$ of 0.332 on average ($p = .20$ for the observed value), and shuffling signs and weights over the observed edges gives 0.330 ($p = .19$). The evidence for structure comes instead from the balance test below and from the composition of the blocs: bloc B carries 9 of the 13 positive edges and bloc A only 3 (\emph{Authority} and \emph{Hide Information} have none), and 17 of the 18 edges between blocs are negative. Bloc B also contains one strong negative association, between \emph{Conspiracy Theory} and \emph{Manipulate Information} ($-0.64$); \emph{Manipulate Information} belongs to bloc B through its positive associations with \emph{Appeal to Fear} ($+0.43$) and \emph{Commitment \& Consistency} ($+0.33$). The theoretical three-framework partition (Cialdini, affective, information) scores $Q = -0.118$ on the same network, and agreement between blocs and frameworks is no better than chance (adjusted Rand index $-0.11$). Of the network's 28 triangles, 20 (71\%) are structurally balanced, against 51\% on average when signs are shuffled among the retained edges ($p = .019$). The mean $\log_2$ lift is $-0.24$ for pairs within a framework and $-0.03$ for pairs across frameworks.

Table~\ref{tab:bloc_stability} reports, for each strategy, the share of eight edge-retention rules under which it falls in bloc A, the bloc containing \emph{Authority} (only rules under which the strategy keeps at least one edge are counted). Eleven of the 13 strategies sit in the same bloc under every such rule. \emph{Manipulate Information} sits in bloc B under six of its eight rules, and \emph{Liking} splits evenly (it falls in bloc B at the main threshold): its four significant edges all involve bloc-A strategies, one positive association (\emph{Vivid Imagery}, $+0.75$) and three negative associations (\emph{Authority}, \emph{Social Proof}, \emph{Scarcity}).

The C2 and C3 briefings use the 33 pairs with $|\log_2 \text{lift}| \geq 0.5$ against independence (Fig.~\ref{fig:cooc_lift}; 2 positive, 31 negative); the A4 generator received all 78 pairs. All 33 briefing pairs keep their sign against the maximum-entropy null; 22 also meet the network's retention rule, and the other 11, all negative, attenuate below it (to between $-0.12$ and $-0.29$).

\emph{Excluding conspiracy-labelled claims.} To check that the core does not depend on how \emph{Conspiracy Theory} is defined, we refit the maximum-entropy null and the network on the 202{,}759 claims (94.5\%) that carry no \emph{Conspiracy Theory} label (12 strategy groups). All 66 pairwise lifts keep their sign (correlation with the full-sample lifts $r = 1.00$). The core pairs persist: \emph{Unity}--\emph{Fabricate Information} $+0.64$ (full sample $+0.60$), \emph{Appeal to Fear}--\emph{Fabricate Information} $+0.48$ ($+0.51$) and \emph{Commitment \& Consistency}--\emph{Appeal to Fear} $+0.46$ ($+0.41$), all with BH-adjusted $q < 10^{-80}$. The best signed partition ($Q = 0.363$) recovers the same two groups: bloc A is unchanged, and \emph{Liking} joins the core group. Nine of 13 triangles are structurally balanced, against 50\% under sign shuffling ($p = .13$; the smaller network gives this test less power).

\begin{table}[htp]
\centering
\small
\begin{threeparttable}
\caption{Bloc membership and stability of the 13 strategy groups across eight edge-retention rules.}
\label{tab:bloc_stability}
\begin{tabular}{llcccc}
\toprule
\textbf{Strategy group} & \textbf{Framework} & \textbf{Prevalence} & \textbf{Rules} & \textbf{Share in A} & \textbf{Bloc} \\
\midrule
Authority & Cialdini & 0.652 & 8 & 1.00 & A \\
Emotional Appeal & Affective & 0.647 & 6 & 1.00 & A \\
Social Proof & Cialdini & 0.209 & 8 & 1.00 & A \\
Hide Information & Information & 0.181 & 8 & 1.00 & A \\
Vivid Imagery & Affective & 0.110 & 8 & 1.00 & A \\
Scarcity & Cialdini & 0.053 & 8 & 1.00 & A \\
\addlinespace
Manipulate Information & Information & 0.336 & 8 & 0.25 & B \\
Appeal to Fear & Affective & 0.300 & 8 & 0.00 & B \\
Commitment \& Consistency & Cialdini & 0.266 & 8 & 0.00 & B \\
Fabricate Information & Information & 0.124 & 8 & 0.00 & B \\
Unity & Cialdini & 0.071 & 8 & 0.00 & B \\
Conspiracy Theory & Affective & 0.055 & 8 & 0.00 & B \\
\addlinespace
Liking & Cialdini & 0.053 & 8 & 0.50 & bridge \\
\bottomrule
\end{tabular}
\begin{tablenotes}
\small
\item \textit{Notes}: Rules cross $q \in \{0.01, 0.001\}$ with $|\log_2 \text{lift}| \in \{0.1, 0.2, 0.3, 0.4\}$ against the maximum-entropy null; ``Rules'' counts the rules under which the strategy keeps at least one edge; ``Share in A'' is the share of those rules under which it falls in bloc A. Bloc A is the bloc containing \emph{Authority}; bloc B contains \emph{Conspiracy Theory}. $N = 214{,}646$ claims.
\end{tablenotes}
\end{threeparttable}
\end{table}

\newpage
\section{Variation in Persuasion Strategies Across Topics, Languages, and Channels}\label{appendix:variation}

Strategy use varies with a claim's topic, its language and the channel through which it spread (\secref{Mapping the persuasion playbook}).

\subsection{Strategy Distribution by Topic}\label{appendix:by_topic}

\begin{figure}[htp]
    \centering
    \includegraphics[width=\linewidth]{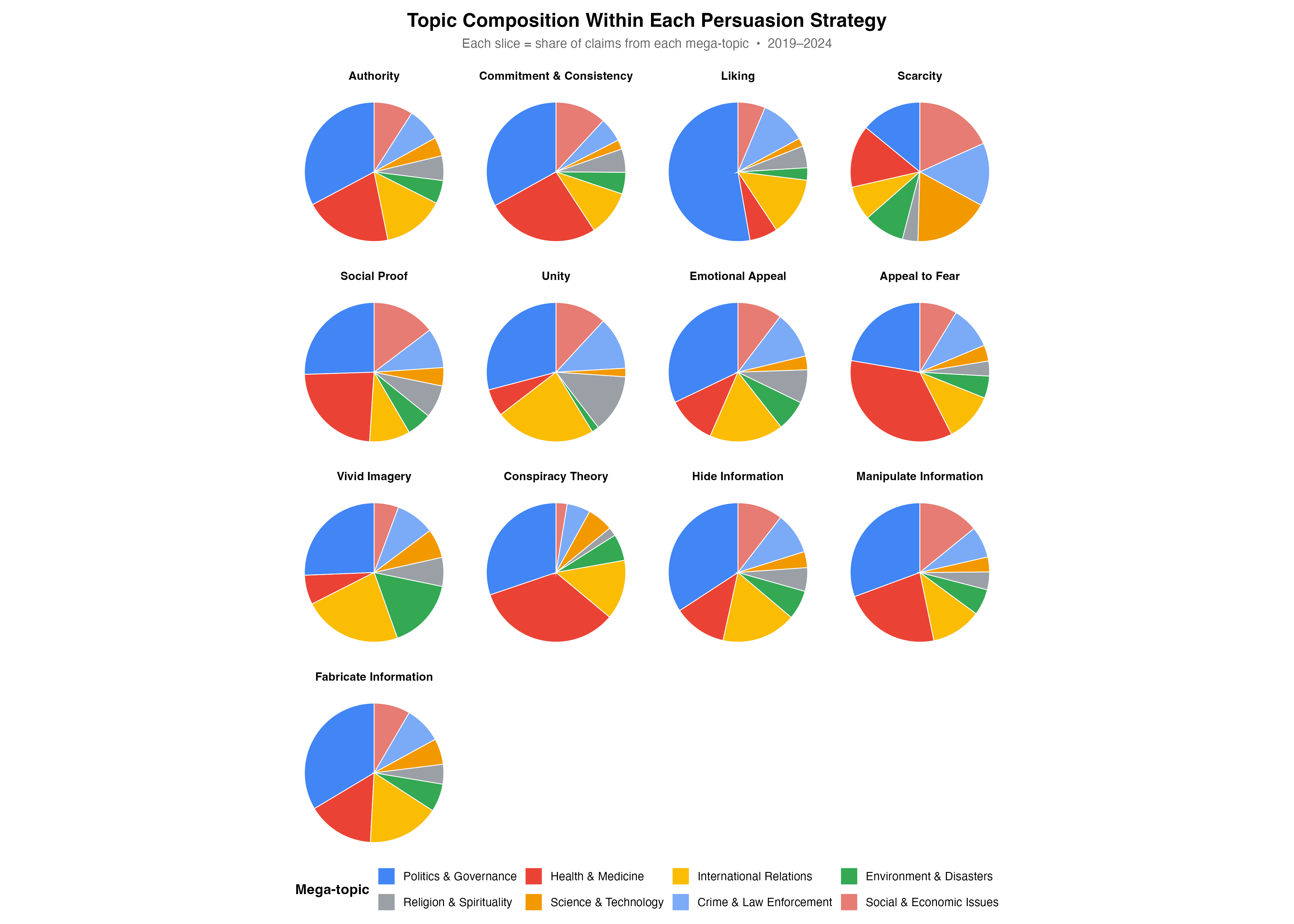}
    \caption{Topic composition of each strategy group. Each pie shows,
    among claims using a given strategy, the share from each
    mega-topic. Strategy groups follow the 13-group framework in
    Fig.~\ref{fig:cialdini_pct}.}
    \label{fig:by_topic}
\end{figure}

\emph{Authority} and \emph{Emotional Appeal} are among the most common strategies in every topic (54--70\% and 39--84\% of claims within a topic), but the rest of the strategy mix depends strongly on topic. Fig.~\ref{fig:by_topic} shows, for each strategy, the share of its claims that come from each mega-topic. \emph{Health \& Medicine} supplies 35.2\% of \emph{Appeal to Fear} claims and 33.7\% of \emph{Conspiracy Theory} claims, about twice its 17\% share of the corpus; \emph{Politics \& Governance} supplies 52.8\% of \emph{Liking} claims, against a 29\% corpus share; and \emph{Scarcity} is most common within \emph{Science \& Technology} claims (19.5\%, against 8\% or less in every other topic). Co-occurrence also varies by topic (the per-topic $\log_2$-lift matrices, which define generator condition A5, are available from the authors on request; \emph{Data availability}). The negative associations of \emph{Liking} and \emph{Unity} with \emph{Conspiracy Theory} turn positive within health, environmental and religious topics, and the links among \emph{Appeal to Fear}, \emph{Fabricate Information} and \emph{Conspiracy Theory} keep their sign in every topic but vary in strength. These patterns are consistent with producers adapting their rhetoric to each domain, although they do not establish it.

\subsection{Strategy Distribution by Language}\label{appendix:by_lg}

\begin{figure}[htp]
    \centering
    \includegraphics[width=\linewidth]{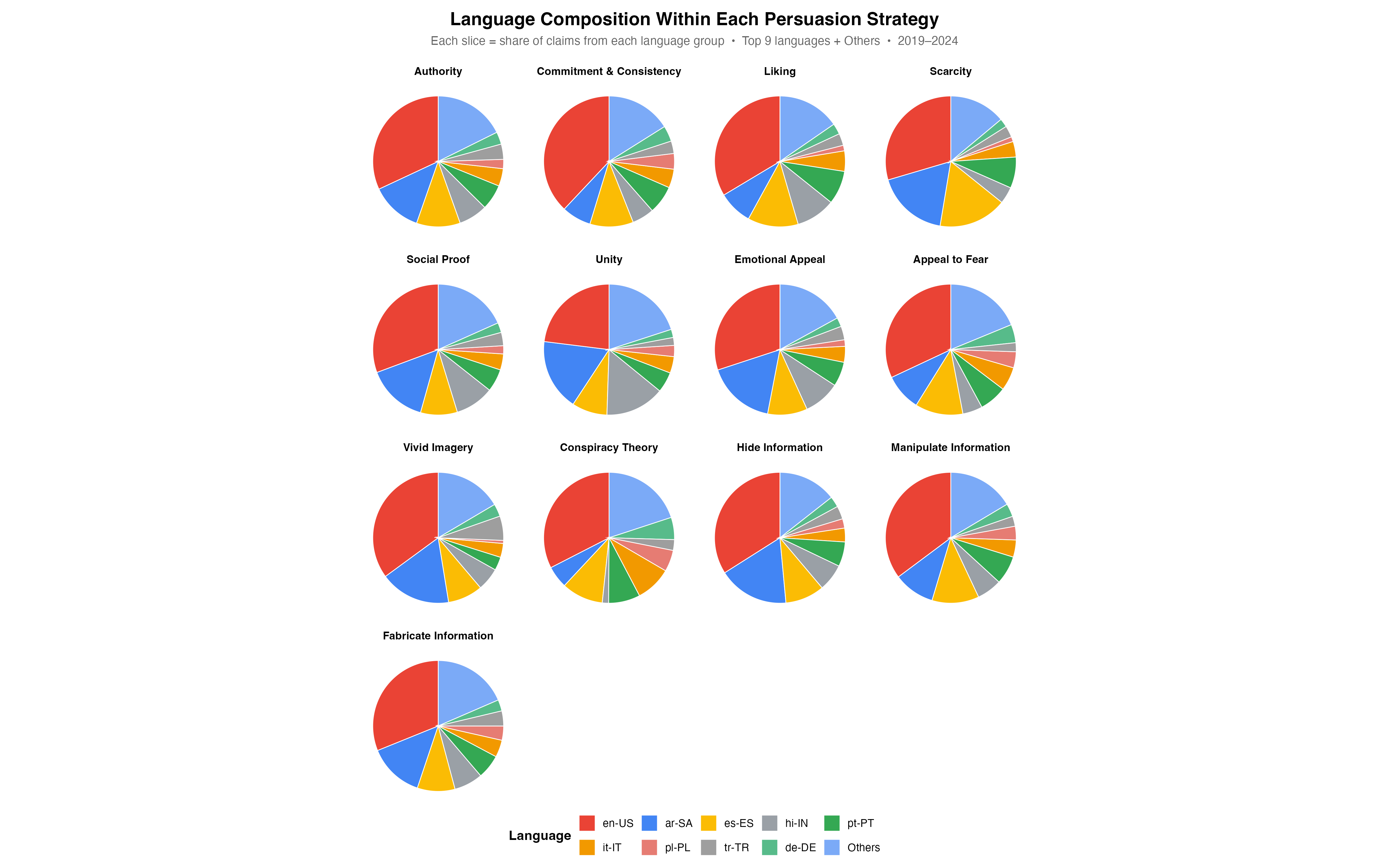}
    \caption{Language composition of each strategy group. Each pie
    shows, among claims using a given strategy, the share from
    each language; the nine most common languages and all others combined.}
    \label{fig:by_lg}
\end{figure}

Strategy use also varies across languages (Table~\ref{tab:by_lg_prev}; Fig.~\ref{fig:by_lg} shows the language composition of each strategy group). Among the 15 languages with at least 3{,}000 claims, \emph{Emotional Appeal} is most common in Arabic (80.2\% of claims), Bengali (78.8\%) and Hindi (77.5\%) and least common in Polish (39.4\%); \emph{Appeal to Fear} is most common in Slovenian (50.9\%), German (45.2\%) and Polish (44.8\%) and least common in Bengali, Hindi and Arabic (18--20\%); and \emph{Conspiracy Theory} appears in 20.0\% of Slovenian and 11.6\% of Italian claims but in 1.2\% of Hindi and Bengali claims. Co-occurrence also differs: \emph{Unity} and \emph{Conspiracy Theory} appear together less often than expected in Spanish (36 claims against 53 expected; $\log_2$ lift $-0.55$) but more than twice as often as expected in Arabic (99 against 43; $+1.20$). These differences are consistent with cross-national research showing that the content of disinformation reflects each country's political and media environment\citep{humprecht2019where,humprecht2020resilience}.

\begin{table}[htp]
\centering
\footnotesize
\caption{Share of claims (\%) using each strategy group, overall and in the eight most common languages (English 31.6\% of claims, Arabic 13.7\%, Spanish 10.6\%, Hindi 7.5\%, Portuguese 6.3\%, Italian 4.3\%, Turkish 3.1\%, German 3.0\%).}
\label{tab:by_lg_prev}
\begin{tabular}{lrrrrrrrrr}
\toprule
Strategy group & All & English & Arabic & Spanish & Hindi & Portuguese & Italian & Turkish & German \\
\midrule
Authority & 65.2 & 65.9 & 60.5 & 66.8 & 62.0 & 64.6 & 66.4 & 78.2 & 66.7 \\
Commitment \& Consistency & 26.6 & 32.0 & 14.2 & 27.1 & 19.3 & 29.3 & 28.9 & 26.2 & 34.6 \\
Social Proof & 20.9 & 20.3 & 22.9 & 18.1 & 26.8 & 18.5 & 19.5 & 21.9 & 17.1 \\
Unity & 7.1 & 5.2 & 9.2 & 5.9 & 13.9 & 5.7 & 6.8 & 4.4 & 4.8 \\
Scarcity & 5.3 & 4.9 & 6.9 & 8.4 & 2.9 & 6.4 & 4.8 & 4.8 & 3.7 \\
Liking & 5.3 & 5.6 & 3.3 & 6.2 & 6.8 & 6.9 & 6.3 & 5.0 & 4.7 \\
Emotional Appeal & 64.7 & 61.5 & 80.2 & 60.2 & 77.5 & 61.9 & 59.7 & 69.0 & 48.6 \\
Appeal to Fear & 30.0 & 30.3 & 20.1 & 33.6 & 19.5 & 32.1 & 41.0 & 21.2 & 45.2 \\
Vivid Imagery & 11.0 & 12.1 & 14.0 & 8.9 & 8.3 & 5.7 & 8.8 & 21.0 & 11.4 \\
Conspiracy Theory & 5.5 & 5.7 & 2.2 & 5.4 & 1.2 & 6.8 & 11.6 & 4.7 & 10.2 \\
Manipulate Information & 33.6 & 37.4 & 25.0 & 37.0 & 27.7 & 37.3 & 33.5 & 26.5 & 36.4 \\
Hide Information & 18.1 & 19.5 & 23.2 & 16.6 & 16.2 & 17.3 & 14.5 & 18.7 & 16.1 \\
Fabricate Information & 12.4 & 12.2 & 12.4 & 10.9 & 11.7 & 11.7 & 12.4 & 14.7 & 11.6 \\
\bottomrule
\end{tabular}
\end{table}

\emph{Cross-language similarity.} For each of the 15 languages, we refitted the maximum-entropy null within the language and computed the 78 pairwise $\log_2$ lifts. Each language's network correlates with the pooled network at a median $r = 0.83$ (lowest, Tamil: 0.63), and pairs of languages correlate at a median $r = 0.70$ (interquartile range 0.60--0.78), or 0.77 after correcting for sampling noise with split-half reliabilities (median 0.90). Among the 19 strongest pooled pairs ($|\log_2 \text{lift}| \geq 0.5$; 7 positive, 12 negative), the median share of languages matching the pooled sign is 100\% (minimum 73\%). The positive associations of \emph{Conspiracy Theory} with \emph{Appeal to Fear} and with \emph{Fabricate Information}, and the negative associations of \emph{Appeal to Fear} with \emph{Vivid Imagery} and of \emph{Social Proof} with \emph{Conspiracy Theory}, hold in all 15 languages. Similarity is lower within a single topic (median $r = 0.50$ across the nine languages with at least 1{,}500 \emph{Politics \& Governance} claims, not corrected for sampling noise), and each language's claims come from the organizations that fact-check in it, so the comparison is descriptive.

\subsection{Strategy Distribution by Channel}\label{appendix:by_channel}

We asked whether strategy use differs between social media and other outlets. In raw shares, emotional, social and identity-based strategies are more common on social media, and strategies that manipulate information are more common in longer-form outlets. Because these shares may reflect differences in topic, language and time, we regressed an indicator for social-media origin on the 13 strategy-group indicators in a linear probability model with mega-topic, language and year-month fixed effects (Supplementary Information section~\ref{appendix:lpm_avg}). Each coefficient is the change in the probability that a claim came from social media when it uses that strategy, holding topic, language and time fixed.

Adjusting for topic, language and time strengthens rather than weakens the raw pattern (Supplementary Table~\ref{tab:lpm_avg}). All four affective strategies and four of the six Cialdini strategies (\emph{Social Proof}, \emph{Unity}, \emph{Authority} and \emph{Liking}) are significantly more likely on social media. \emph{Hide Information}, \emph{Manipulate Information} and \emph{Commitment \& Consistency} are significantly less likely, and \emph{Fabricate Information} is negative but not significant. The concentration of affective strategies on social media fits evidence that false news there spreads farther than true news and draws more fear, disgust and surprise in replies\citep{vosoughi2018spread}. The signs are stable when the model is estimated within each mega-topic, although magnitudes vary (per-topic estimates are included with the deposited data). These are descriptive associations, not causal effects; they may reflect producers' choices, platform amplification or both.

\subsection{LPM Coefficient Estimates}\label{appendix:lpm_avg}

For the channel analysis in Supplementary Information section~\ref{appendix:variation}, we estimate the linear probability model
\begin{equation}
\label{eq:lpm}
Y_{it} = \alpha + \sum_{s=1}^{13} \beta_s \, \text{Strategy}^s_i
         + \sum_{k=1}^{K-1} \gamma_k \, \text{Topic}^k_i
         + \delta_t + \lambda_l + \varepsilon_{it},
\end{equation}
where $Y_{it}=1$ if claim $i$, fact-checked in year-month $t$, came from a social media URL; $\text{Strategy}^s_i$ and $\text{Topic}^k_i$ are strategy-group and mega-topic indicators; and $\delta_t$ and $\lambda_l$ are year-month and language fixed effects. $\hat{\beta}_s$ is the change in the probability of social-media origin associated with strategy $s$, holding topic, language and time fixed (Supplementary Table~\ref{tab:lpm_avg}).

\begin{table}[htp]
\centering
\caption{Linear probability model estimates (Equation~\ref{eq:lpm}): change in the probability that a claim came from social media, in percentage points, associated with each of the 13 strategy groups, with mega-topic, language and year-month fixed effects. Positive values indicate strategies more common on social media. Standard errors are heteroskedasticity-robust (HC1). The 141 claims that use none of the 13 strategy groups are excluded.}
\label{tab:lpm_avg}
\begin{tabular}{lcccc}
\hline
Strategy & Coefficient (pp) & SE & 95\% CI & $p$ \\
\hline
\multicolumn{5}{l}{\emph{Cialdini principles}} \\
\quad Authority & $+1.06$ & $0.18$ & $[+0.70, +1.42]$ & $<.001$ \\
\quad Commitment \& Consistency & $-1.64$ & $0.20$ & $[-2.04, -1.25]$ & $<.001$ \\
\quad Liking & $+0.88$ & $0.39$ & $[+0.12, +1.65]$ & $.023$ \\
\quad Scarcity & $+0.24$ & $0.36$ & $[-0.46, +0.95]$ & $.499$ \\
\quad Social Proof & $+2.39$ & $0.21$ & $[+1.97, +2.81]$ & $<.001$ \\
\quad Unity & $+1.40$ & $0.33$ & $[+0.75, +2.05]$ & $<.001$ \\
\multicolumn{5}{l}{\emph{Affective strategies}} \\
\quad Emotional Appeal & $+2.96$ & $0.19$ & $[+2.59, +3.33]$ & $<.001$ \\
\quad Appeal to Fear & $+0.68$ & $0.20$ & $[+0.29, +1.07]$ & $<.001$ \\
\quad Vivid Imagery & $+3.05$ & $0.27$ & $[+2.52, +3.58]$ & $<.001$ \\
\quad Conspiracy Theory & $+2.32$ & $0.38$ & $[+1.58, +3.06]$ & $<.001$ \\
\multicolumn{5}{l}{\emph{Information strategies}} \\
\quad Hide Information & $-0.49$ & $0.22$ & $[-0.92, -0.06]$ & $.025$ \\
\quad Manipulate Information & $-0.68$ & $0.18$ & $[-1.04, -0.32]$ & $<.001$ \\
\quad Fabricate Information & $-0.36$ & $0.25$ & $[-0.85, +0.13]$ & $.152$ \\
\hline
Mega-topic fixed effects & \multicolumn{4}{l}{Yes (9 categories, incl.\ unassigned)} \\
Language fixed effects & \multicolumn{4}{l}{Yes (51)} \\
Year-month fixed effects & \multicolumn{4}{l}{Yes (64)} \\
Mean of outcome & \multicolumn{4}{l}{30.2\% of claims from social media} \\
Within $R^2$ & \multicolumn{4}{l}{0.006} \\
Claims & \multicolumn{4}{l}{214,505} \\
\hline
\end{tabular}
\end{table}

\section{Evaluator and Generator Knowledge Conditions}\label{appendix:conditions}

This section specifies the four evaluator briefings (C1, C2, C3, C0b) used in the language-model experiment (\secref{Briefing a language-model evaluator}) and the adversarial test (\secref{The adversarial stress test}), the five generator conditions (A1--A5) used in the adversarial test, the generation protocol and the persona panels.

C1 is an accuracy prompt with no strategy content. C2 teaches the playbook's empirical regularities (\secref{Mapping the persuasion playbook}). C3 and C0b separate strategy knowledge from a short decoupling instruction:
\begin{itemize}
    \item[\textbf{C1}] \emph{accuracy prompt}: ``to the best of your knowledge, is this claim accurate?'', with no strategy content\citep{pennycook2021shifting}.
    \item[\textbf{C2}] \emph{strategy knowledge}: the 13 strategy groups with their corpus prevalence (Fig.~\ref{fig:cialdini_pct}) and the substantively meaningful strategy pairs ($\lvert\log_2\text{lift}\rvert \geq 0.5$) from the corpus-wide co-occurrence matrix (Fig.~\ref{fig:cooc_lift}).
    \item[\textbf{C3}] \emph{full briefing}: C2 augmented with an explicit instruction that ``the presence of a persuasion strategy describes \emph{how} a claim is communicated, not \emph{whether} it is true.''
    \item[\textbf{C0b}] \emph{decoupling only}: the same decoupling instruction without any strategy framework.
\end{itemize}
Under C2 and C3, the evaluator also names the strategies the claim uses before scoring. Table~\ref{tab:briefing_texts} gives the accuracy and decoupling texts verbatim.

\begin{table}[htp]
\centering
\small
\caption{Briefing texts that differ across arms, verbatim. Capitalization follows the prompts and survey screens as fielded. In the language-model prompts, each text was inserted into a common system message that frames the persona-rating task; in the human survey, each was shown on its own page before the claims.}
\label{tab:briefing_texts}
\begin{tabular}{@{}>{\raggedright\arraybackslash}p{2.0cm}>{\raggedright\arraybackslash}p{7.9cm}>{\raggedright\arraybackslash}p{5.4cm}@{}}
\toprule
\textbf{Arm} & \textbf{Language-model prompt} & \textbf{Human survey screen} \\
\midrule
C1 (accuracy prompt) & ``BEFORE SCORING, TAKE A MOMENT TO THINK TWICE. To the best of your knowledge, is the claim accurate? Do not use any external information or search; rely on your own assessment of whether the claim is accurate.'' & ``As you read each claim, take a moment to ask yourself: to the best of your knowledge, is this claim accurate?'' \\
\addlinespace
Decoupling instruction (C0b; C3, after the strategy knowledge) & ``BEFORE SCORING, APPLY THIS CRITICAL CHECK. The presence of a persuasion strategy describes HOW a claim is communicated, NOT WHETHER it is true. Many true, accurate claims also use these techniques: real experts and institutions are cited as authorities, genuine events evoke strong emotion, factual reports use vivid detail, and true statistics can be striking or surprising. Identifying a strategy is therefore NOT, by itself, evidence that the claim is false. Step back and assess the claim's factual plausibility independently of its rhetorical style: to the best of your knowledge, is the claim itself accurate? Do NOT lower a claim's believability merely because it uses a persuasion strategy if the underlying claim is likely true. Reserve skepticism for claims that are both strategy-laden AND factually implausible.'' & ``The presence of a persuasion strategy tells you how a claim is communicated, not whether it is true. Real, true claims can use emotional or authority-based framing too. Treat strategy cues as one signal among others, not as proof a claim is false.'' \\
\addlinespace
Strategy knowledge (C2, C3) & \multicolumn{2}{>{\raggedright\arraybackslash}p{13.6cm}@{}}{Not reproduced. Both versions present the 13 strategy groups with definitions, their corpus prevalence (Fig.~\ref{fig:cialdini_pct}) and the co-occurrence pairs with $|\log_2 \text{lift}| \geq 0.5$ (Fig.~\ref{fig:cooc_lift}); the language-model prompt then asks the model to name the strategies present before scoring, and the human briefing adds three practice claims with feedback. Full texts are available from the authors on request (\emph{Data availability}).} \\
\bottomrule
\end{tabular}
\end{table}

The five generation conditions (A1--A5) add strategy knowledge to the generator's prompt in steps:
\begin{itemize}
    \item[\textbf{A1}] \emph{zero-shot}: no strategy guidance; the generator receives only a brief description of the seeding event (mirrors C1).
    \item[\textbf{A2}] \emph{48-strategy taxonomy}, supplied as a generation reference.
    \item[\textbf{A3}] \emph{13 strategies + prevalence}; the generator is instructed to reflect the most common strategies in its output.
    \item[\textbf{A4}] \emph{+ corpus co-occurrence}: all 78 pairwise $\log_2$ lifts from the corpus-wide matrix, with no threshold; the generator is instructed to prefer positively associated strategy pairs and avoid negatively associated ones (mirrors C2, which shows only pairs with $\lvert\log_2\text{lift}\rvert \geq 0.5$).
    \item[\textbf{A5}] \emph{+ topic-specific co-occurrence}: all 78 pairwise $\log_2$ lifts from the seeding event's mega-topic matrix, with no threshold.
\end{itemize}

We selected 20 seeding events from developments after the evaluator's October 2023 training cutoff, spanning all eight mega-topics (Table~\ref{tab:events}). The two-stage probe (Supplementary Information section~\ref{appendix:basis_probe}) showed that the model had reliable general knowledge for 12 of these events, so only the remaining eight are treated as new to the evaluator. For each event and condition (A1--A5), two independent GPT-4o calls (temperature $0.6$) each requested 10 claims of 7--25 words with self-reported strategy labels, giving 20 claims per event and condition (2,000 in total). The 240 false claims in the discernment experiment (\secref{Briefing a language-model evaluator}) come from the eight new events, with 6 claims per event and generation condition (30 per event and 48 per condition; Supplementary Information section~\ref{appendix:basis_probe}). The 40 true claims come from major news outlets and official institutional sources (Supplementary Information section~\ref{appendix:true-claims}).

Personas come from a pool of 2,058 LLM-simulated personas\citep{toubia2025database} that vary in demographic, psychological, economic, personality and cognitive characteristics. In the discernment experiment, 50 randomly drawn personas each rated all 280 claims (240 false and 40 true), one claim per call, under the four briefings and, in a separate run, under no briefing. The adversarial test uses a panel of 100 personas from the same pool. All ratings use the 1--7 believability scale.

\newpage
\section{Seeding Events for Strategy-Guided Generation}\label{appendix:seeding-events}

Table~\ref{tab:events} lists the 20 seeding events used in the strategy-guided generation experiment (\secref{The adversarial stress test}), each anchored to a development after the evaluator's October 2023 training cutoff. The events span all eight mega-topics (\emph{Methods}); the mega-topic column determines which topic-specific co-occurrence matrix condition A5 uses. Twelve events concern recurring topics (for example, climate change, 5G and COVID-19) on which the model reported reliable pre-cutoff knowledge (Supplementary Information section~\ref{appendix:basis_probe}); only the eight events that passed this probe supply the claims for the discernment experiments.

{\footnotesize
\begin{longtable}{@{}r p{3.5cm} p{7.5cm} p{3cm}@{}}
\caption{The 20 seeding events used in the strategy-guided generation experiment, grouped by mega-topic.}
\label{tab:events} \\
\toprule
\textbf{\#} & \textbf{Event} & \textbf{Context} & \textbf{Mega-topic} \\
\midrule
\endfirsthead
\multicolumn{4}{l}{\emph{Table~\ref{tab:events} continued from previous page}} \\
\toprule
\textbf{\#} & \textbf{Event} & \textbf{Context} & \textbf{Mega-topic} \\
\midrule
\endhead
\midrule
\multicolumn{4}{r}{\emph{continued on next page}} \\
\endfoot
\bottomrule
\endlastfoot
1  & Trump Legal Trials and Political Fallout
   & Trump faced multiple trials, convictions, and rallies amid a heated U.S.\ political climate.
   & Politics \& Governance \\
\addlinespace
2  & Project 2025 and U.S.\ Government Restructuring
   & Project 2025 proposed sweeping federal reforms, prompting widespread concern over its potential effects on U.S.\ institutions and society.
   & Politics \& Governance \\
\addlinespace
3  & Global Election Fraud Allegations
   & Multiple countries reported claims of vote manipulation, ballot tampering, and systemic irregularities.
   & Politics \& Governance \\
\addlinespace
4  & Worldwide Protests and Civil Unrest
   & Massive protests erupted in London, Bangladesh, and elsewhere over war, economy, and elections.
   & Politics \& Governance \\
\addlinespace
5  & US and Global Immigration Policy
   & Immigration policies around the world triggered debate over national-security threats, crime, and terrorism.
   & Politics \& Governance \\
\addlinespace
6  & Iran--Israel Missile Escalation
   & Iran launched missiles at Israel in retaliation, raising fears of a broader regional war.
   & International Relations \& Conflicts \\
\addlinespace
7  & COVID-19 Pandemic Fallout and Resurgence
   & New variants and vaccine controversies kept COVID-19 in headlines across several countries.
   & Health \& Medicine \\
\addlinespace
8  & Advances in Diabetes Treatment
   & Breakthroughs in diabetes care introduced new drugs and interventions, improving outcomes for millions with the chronic condition.
   & Health \& Medicine \\
\addlinespace
9  & Francis Scott Key Bridge Collapse in Baltimore
   & A cargo ship struck the bridge on March 26, 2024, killing six workers and halting traffic; investigations revealed prior structural risks.
   & Natural Disasters \& Environmental Issues \\
\addlinespace
10 & California \& LA Wildfires
   & Repeated wildfires devastated large parts of California, prompting mass evacuations and a federal response.
   & Natural Disasters \& Environmental Issues \\
\addlinespace
11 & Climate Change and Global Warming
   & Public controversies around climate science, climate policy (meat bans, net-zero, COP), and climate-linked weather as a hoax or globalist scam.
   & Natural Disasters \& Environmental Issues \\
\addlinespace
12 & 2024 Paris Olympics
   & The Paris Olympics drew global attention for sports triumphs and political protests over global conflicts.
   & Social \& Economic Issues \\
\addlinespace
13 & 2024 World Economic Forum (Davos)
   & Officials gathered at the 54th WEF Annual Meeting to discuss economic uncertainty, climate change, AI governance, and geopolitical conflicts.
   & Social \& Economic Issues \\
\addlinespace
14 & US Cost of Living
   & US inflation, cost of living, unemployment, tax, and economic-downturn concerns triggered widespread anxiety.
   & Social \& Economic Issues \\
\addlinespace
15 & Jeffrey Epstein Files and Client List
   & Court documents linked to Jeffrey Epstein renewed claims about an elite pedophile network allegedly protected by intelligence services.
   & Crime \& Law Enforcement \\
\addlinespace
16 & Murder, Homicide, and Violent-Crime Cases
   & Local and viral murder, homicide, and assault stories (stabbings, shootings, and prison violence) circulated alongside disputed crime statistics.
   & Crime \& Law Enforcement \\
\addlinespace
17 & Pope Francis and the Vatican
   & Statements and image-based claims attributed to Pope Francis sparked controversy over politics, the Bible, climate, and an alleged alliance with the WEF.
   & Religion \& Spirituality \\
\addlinespace
18 & Islam and Muslim Communities
   & Posts targeting Muslim communities, mosques, and Sharia law spread across Europe, India, and the Americas, often framing Islam as a security or cultural threat.
   & Religion \& Spirituality \\
\addlinespace
19 & NASA and Space Exploration
   & NASA missions, asteroid warnings, lunar landings, and James Webb Space Telescope discoveries sparked viral claims, often distorting agency announcements.
   & Science \& Technology \\
\addlinespace
20 & 5G, EMF, and Cell-Tower Conspiracies
   & Claims that 5G networks, electromagnetic fields, and cell towers cause illness, fainting spells, or covert surveillance recurred across multiple languages.
   & Science \& Technology \\
\end{longtable}
}

\pagebreak

\section{True Claims for the Post-Cutoff Prebunking Evaluation}\label{appendix:true-claims}

Table~\ref{tab:true_claims} lists the 40 true claims (five per post-cutoff event) used in the prebunking evaluation (\secref{Briefing a language-model evaluator}), together with the source consulted to verify each.

{\footnotesize
\setlength{\Urlmuskip}{0mu plus 1mu\relax}
\def\UrlBreaks{\do\/\do\-\do\.\do\_\do\?\do\&\do\=\do\#\do\:\do\@\do\a\do\b\do\c\do\d\do\e\do\f\do\g\do\h\do\i\do\j\do\k\do\l\do\m\do\n\do\o\do\p\do\q\do\r\do\s\do\t\do\u\do\v\do\w\do\x\do\y\do\z\do\A\do\B\do\C\do\D\do\E\do\F\do\G\do\H\do\I\do\J\do\K\do\L\do\M\do\N\do\O\do\P\do\Q\do\R\do\S\do\T\do\U\do\V\do\W\do\X\do\Y\do\Z\do0\do1\do2\do3\do4\do5\do6\do7\do8\do9}
\begin{longtable}{@{}r p{2.0cm} p{6.7cm} p{3.7cm}@{}}
\caption{The 40 true post-cutoff claims and their sources, grouped by seeding event.}\\
\label{tab:true_claims} \\
\toprule
\textbf{\#} & \textbf{Event} & \textbf{Claim} & \textbf{Source} \\
\midrule
\endfirsthead
\multicolumn{4}{l}{\emph{Table~\ref{tab:true_claims} continued from previous page}} \\
\toprule
\textbf{\#} & \textbf{Event} & \textbf{Claim} & \textbf{Source} \\
\midrule
\endhead
\midrule
\multicolumn{4}{r}{\emph{continued on next page}} \\
\endfoot
\bottomrule
\endlastfoot
1 & 2024 Paris Olympics & The opening ceremony of the Paris 2024 Olympics took place on July 26, 2024, with thousands of athletes parading in boats along the Seine River. & \url{https://www.olympics.com/en/news/paris-2024-olympic-opening-ceremony-seine} \\
2 &  & The Paris 2024 Olympics was the first Olympic Games to hold its opening ceremony on a river rather than in a stadium. & \url{https://www.olympics.com/en/news/whats-new-paris-2024-opening-ceremony} \\
3 &  & The Paris 2024 Olympics opening parade followed the Seine River from east to west over a distance of approximately 6 kilometers. & \url{https://en.wikipedia.org/wiki/2024_Summer_Olympics_opening_ceremony} \\
4 &  & Around 300,000 spectators attended the Paris 2024 opening ceremony, including about 100,000 with paid tickets and 200,000 with free tickets. & \url{https://en.wikipedia.org/wiki/2024_Summer_Olympics_opening_ceremony} \\
5 &  & The Paris 2024 opening ceremony parade departed from the Pont d'Austerlitz beside the Jardin des Plantes. & \url{https://www.olympics.com/en/news/whats-new-paris-2024-opening-ceremony} \\
\addlinespace
6 & Francis Scott Key Bridge Collapse in Baltimore & The Francis Scott Key Bridge in Baltimore collapsed on March 26, 2024, after being struck by the container ship Dali. & \url{https://en.wikipedia.org/wiki/Francis_Scott_Key_Bridge_collapse} \\
7 &  & Six members of a maintenance crew working on the Francis Scott Key Bridge were killed when the bridge collapsed in March 2024. & \url{https://www.cbsnews.com/news/francis-scott-key-bridge-baltimore-collapse-container-ship/} \\
8 &  & The container ship Dali suffered a complete electrical blackout at approximately 1:24 a.m. before drifting into a pier of the Francis Scott Key Bridge. & \url{https://en.wikipedia.org/wiki/Francis_Scott_Key_Bridge_collapse} \\
9 &  & The NTSB determined that a loose signal wire in the Dali's electrical control center contributed to the blackouts that led to the bridge collapse. & \url{https://www.ntsb.gov/news/press-releases/Pages/NR20251118.aspx} \\
10 &  & The collapse of the Francis Scott Key Bridge blocked most shipping to and from the Port of Baltimore for approximately 11 weeks. & \url{https://en.wikipedia.org/wiki/Francis_Scott_Key_Bridge_collapse} \\
\addlinespace
11 & 2024 World Economic Forum (Davos) & The 2024 World Economic Forum Annual Meeting took place from January 15 to 19, 2024, in Davos-Klosters, Switzerland. & \url{https://www.weforum.org/press/2024/01/wef24-rebuilding-trust-world-economic-forum-annual-meeting-2024/} \\
12 &  & The theme of the 2024 World Economic Forum Annual Meeting was 'Rebuilding Trust.' & \url{https://www.weforum.org/meetings/world-economic-forum-annual-meeting-2024/about/} \\
13 &  & The 2024 World Economic Forum Annual Meeting was the 54th annual meeting and brought together more than 2,800 leaders from 120 countries. & \url{https://www.weforum.org/press/2024/01/annual-meeting-2024-rebuilding-trust-amid-uncertainty/} \\
14 &  & More than 60 heads of state and government attended the 2024 World Economic Forum Annual Meeting in Davos. & \url{https://www.weforum.org/press/2024/01/annual-meeting-2024-rebuilding-trust-amid-uncertainty/} \\
15 &  & The 2024 World Economic Forum hosted more than 450 sessions and workshops during its five-day program. & \url{https://www.weforum.org/press/2024/01/annual-meeting-2024-rebuilding-trust-amid-uncertainty/} \\
\addlinespace
16 & Global Election Fraud Allegations & Donald Trump won 312 electoral votes in the 2024 U.S. presidential election, while Kamala Harris received 226 electoral votes. & \url{https://en.wikipedia.org/wiki/2024_United_States_presidential_election} \\
17 &  & Congress certified Donald Trump's 2024 electoral college victory in a joint session on January 6, 2025, with no objections from any member. & \url{https://www.washingtonpost.com/politics/2025/01/06/trump-transition-electoral-college-certification/} \\
18 &  & The Electoral College voted on December 17, 2024, and all electors voted as pledged with no faithless electors recorded. & \url{https://en.wikipedia.org/wiki/2025_United_States_Electoral_College_vote_count} \\
19 &  & Vice President Kamala Harris presided over the January 6, 2025 joint session of Congress that certified her own electoral loss to Donald Trump. & \url{https://www.cbsnews.com/live-updates/congress-election-electoral-college-2025/} \\
20 &  & The January 2025 electoral vote count was the first conducted under the Electoral Count Reform Act, which raised the threshold for raising a valid objection to one-fifth of each chamber. & \url{https://www.npr.org/2025/01/06/nx-s1-5217818/january-6-electoral-count-reform-act} \\
\addlinespace
21 & Worldwide Protests and Civil Unrest & Bangladesh Prime Minister Sheikh Hasina resigned on August 5, 2024 after weeks of student-led protests, ending her 15-year tenure in office. & \url{https://en.wikipedia.org/wiki/July_Uprising} \\
22 &  & The July 2024 protests in Bangladesh began as a movement against a Supreme Court decision reinstating quotas that reserved 30\% of government jobs for descendants of independence-war fighters. & \url{https://en.wikipedia.org/wiki/July_Uprising} \\
23 &  & Sheikh Hasina fled Bangladesh to India on August 5, 2024, traveling by helicopter and later by a Bangladesh Air Force C-130 to Hindon Air Force base near Delhi. & \url{https://en.wikipedia.org/wiki/July_Uprising} \\
24 &  & At least 135 people, including 24 police officers, were reported killed during the July-August 2024 protests in Bangladesh. & \url{https://en.wikipedia.org/wiki/July_Uprising} \\
25 &  & The 2024 Bangladesh student protest movement that led to Sheikh Hasina's resignation was organized under the banner of 'Students Against Discrimination.' & \url{https://en.wikipedia.org/wiki/July_Uprising} \\
\addlinespace
26 & Iran--Israel Missile Escalation & Iran launched a drone and missile attack on Israel on April 13, 2024, in retaliation for an earlier Israeli strike on Iran's embassy compound in Damascus. & \url{https://en.wikipedia.org/wiki/April_2024_Iranian_strikes_on_Israel} \\
27 &  & Iran launched approximately 170 drones, 30 cruise missiles, and 120 ballistic missiles at Israel during the April 13, 2024 attack. & \url{https://www.aljazeera.com/news/2024/4/14/iran-attacks-israel-with-over-300-drones-missiles-what-you-need-to-know} \\
28 &  & The April 2024 Iranian attack on Israel was named 'Operation True Promise' by Iran's armed forces. & \url{https://en.wikipedia.org/wiki/April_2024_Iranian_strikes_on_Israel} \\
29 &  & Israeli and allied air defences intercepted approximately 99\% of the projectiles launched by Iran in the April 13, 2024 attack. & \url{https://www.iranintl.com/en/202404152333} \\
30 &  & The April 1, 2024 Israeli airstrike on the Iranian embassy compound in Damascus killed two senior Iranian generals, triggering Iran's retaliatory attack two weeks later. & \url{https://en.wikipedia.org/wiki/April_2024_Iranian_strikes_on_Israel} \\
\addlinespace
31 & Project 2025 and U.S.\ Government Restructuring & Project 2025 was established by the Heritage Foundation in 2022 to provide a personnel database and policy framework for a future conservative U.S. presidential administration. & \url{https://en.wikipedia.org/wiki/Project_2025} \\
32 &  & The Heritage Foundation published Project 2025's policy document, 'Mandate for Leadership 2025: The Conservative Promise,' which runs approximately 920 pages. & \url{https://en.wikipedia.org/wiki/Project_2025} \\
33 &  & Project 2025's advisory board includes more than 100 conservative organizations contributing to its policy framework. & \url{https://en.wikipedia.org/wiki/Project_2025} \\
34 &  & Project 2025 builds on the Heritage Foundation's original 'Mandate for Leadership,' first published in January 1981 as a guide for the incoming Reagan administration. & \url{https://en.wikipedia.org/wiki/Project_2025} \\
35 &  & Paul Dans served as the original director of Project 2025 from its founding in 2022. & \url{https://en.wikipedia.org/wiki/Project_2025} \\
\addlinespace
36 & Murder, Homicide, and Violent-Crime Cases & The FBI reported that national violent crime in the United States decreased by an estimated 4.5\% in 2024 compared to 2023. & \url{https://www.fbi.gov/news/press-releases/fbi-releases-2024-reported-crimes-in-the-nation-statistics} \\
37 &  & The 2024 violent crime rate in the United States was the lowest recorded since 1969, according to the FBI. & \url{https://www.americanprogress.org/article/nationwide-2024-crime-data-demonstrate-the-value-of-violence-prevention-and-local-law-enforcement/} \\
38 &  & Murder and non-negligent manslaughter in the United States decreased by an estimated 14.9\% in 2024 compared to 2023, according to FBI data. & \url{https://www.fbi.gov/news/press-releases/fbi-releases-2024-reported-crimes-in-the-nation-statistics} \\
39 &  & The FBI's 2024 crime statistics were drawn from reports by more than 16,000 law enforcement agencies covering 95.6\% of the U.S. population. & \url{https://www.fbi.gov/news/press-releases/fbi-releases-2024-reported-crimes-in-the-nation-statistics} \\
40 &  & Property crime in the United States declined by 8.1\% in 2024, reaching its lowest recorded rate since 1961, according to the FBI. & \url{https://www.cbsnews.com/news/fbi-crime-report-2025/} \\
\end{longtable}
}

\pagebreak

\section{Validating the Post-Cutoff Regime: Two-Stage Cutoff Probe}\label{appendix:basis_probe}

The post-cutoff design (\secref{Briefing a language-model evaluator}) requires that the evaluator lack claim-specific knowledge of the AI-generated false claims. We checked this in two stages with GPT-4o-mini (temperature 0, structured output). Both stages rely on the model's own reports; neither inspects its training data.

\emph{Stage 1: event-level probe.} For each of the 20 seeding events, we asked the model in what year the event occurred, whether it had reliable factual knowledge of it and what specific facts it could recall. The response recorded the year estimate, whether the model had reliable knowledge, whether the event postdated its training cutoff, any specific facts, and a confidence rating (1--5). Eight events were judged to postdate the cutoff, with no reliable knowledge, no specific facts and confidence 1 (in the names of Table~\ref{tab:events}: 2024 Paris Olympics; Francis Scott Key Bridge Collapse in Baltimore; 2024 World Economic Forum, Davos; Global Election Fraud Allegations; Worldwide Protests and Civil Unrest; Iran--Israel Missile Escalation; Project 2025 and U.S.\ Government Restructuring; Murder, Homicide, and Violent-Crime Cases). The other 12 returned reliable pre-cutoff knowledge with confidence 4, and we excluded them.

\emph{Stage 2: claim-level basis probe.} For each of the 800 AI-generated false claims about the eight retained events, the model gave a verdict (true, false or uncertain) and named its basis as one of four options: \texttt{factual\_recall} (specific facts from training), \texttt{general\_knowledge} (general topical knowledge sufficient to refute the claim, for example that the World Economic Forum does not implant microchips), \texttt{pattern\_recognition} (the claim matches a known conspiracy template) or \texttt{no\_basis} (no reliable basis). The prompt stated that recognizing a pattern is not a factual basis. The probe classified 398 claims as \texttt{no\_basis} (mean confidence 2.31 of 5), 370 as \texttt{general\_knowledge} (mean 4.36), 31 as \texttt{pattern\_recognition} and 1 as \texttt{factual\_recall}. The share of \texttt{no\_basis} claims varied by event, from 55--63\% for the Paris Olympics, Iran--Israel, Project 2025 and Baltimore bridge events to 17\% for the 2024 World Economic Forum, a topic surrounded by conspiracy narratives that predate the cutoff. The claim-level label serves as the moderator in the discernment analysis (Fig.~\ref{fig:post_cutoff_moderator}).

\emph{Claim sample.} From the 800 probed false claims we drew 240 for the main evaluation: six at random from each of the 40 event $\times$ generation-condition cells, without stratifying on the basis label. The sample is roughly balanced across generation conditions (19--28 \texttt{no\_basis} and 19--28 \texttt{general\_knowledge} claims among each condition's 48) but not across events (the World Economic Forum event contributes 3 \texttt{no\_basis} claims out of 30; Iran--Israel and Project 2025 contribute 18--19 each). The final claim set contains 108 \texttt{no\_basis}, 122 \texttt{general\_knowledge}, 9 \texttt{pattern\_recognition} and 1 \texttt{factual\_recall} false claims, plus the 40 true claims.

The probe and sampling code and the per-cell counts are in the public repository.

\section{Post-Cutoff Discrimination: Full Per-Condition Results}\label{appendix:discrimination_full}

\begin{figure}[htp]
    \centering
    \includegraphics[width=\linewidth]{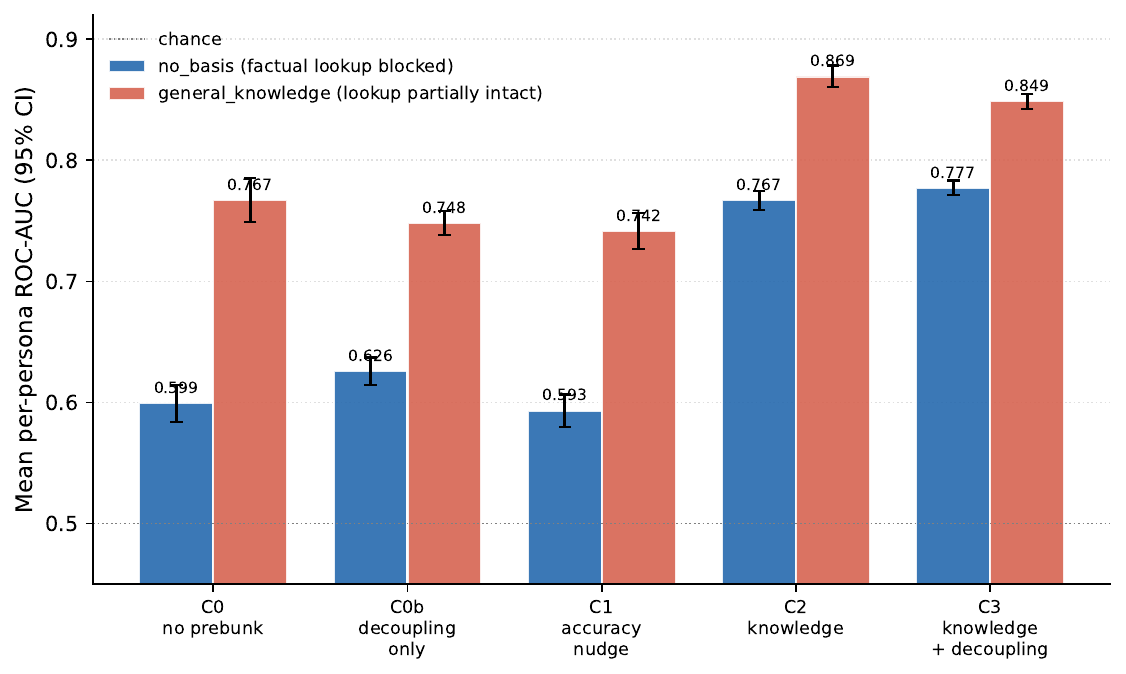}
    \caption{AUC version of Fig.~\ref{fig:post_cutoff_moderator}. Mean per-persona AUC ($N = 50$ personas, 95\% confidence intervals) by briefing and basis stratum. C2 and C3 score highest on both strata; C1 and C0b fall toward chance on the \texttt{no\_basis} stratum. Relative to no briefing (C0), the full briefing (C3) raises AUC by $+0.178$ on \texttt{no\_basis} claims and $+0.082$ on \texttt{general\_knowledge} claims (difference $+0.096$, $[0.083, 0.110]$; paired Wilcoxon $p = 1 \times 10^{-14}$). Relative to the decoupling instruction alone (C0b), which isolates strategy knowledge, the increase is $+0.151$ versus $+0.101$ (difference $+0.051$, $[0.040, 0.062]$; $p = 7 \times 10^{-12}$).}
    \label{fig:post_cutoff_moderator_auc}
\end{figure}

Table~\ref{tab:discrimination_full} reports, for each briefing, the discernment gap (the primary outcome: a persona's mean rating of true minus false claims, averaged over the 50 personas) and the per-persona AUC, on all 280 post-cutoff claims. The main text reports the paired contrasts on the gap (paired $t$-tests across personas); the full set, including the basis strata and the Gemini 2.5 Flash replication, is in the public repository. On AUC the two strategy briefings are tied ($-0.005$; paired Wilcoxon signed-rank test across personas, $p = .17$), and the full briefing exceeds the decoupling instruction alone by $0.126$ (higher for all 50 personas; $p = 2 \times 10^{-15}$). Fig.~\ref{fig:post_cutoff_moderator_auc} gives the AUC version of Fig.~\ref{fig:post_cutoff_moderator}.

\begin{table}[h!]
\centering
\caption{Per-condition discernment on the full post-cutoff claim set (GPT-4o-mini evaluator; 50 personas; 240 AI-generated false and 40 true claims). Gap (primary outcome) is the within-persona mean-true minus mean-false believability on the 1--7 scale; AUC is per-persona ROC-AUC. Both are means over personas with 95\% confidence intervals. C0 was collected in a separate run on the same claims and personas.}
\label{tab:discrimination_full}
\begin{tabular}{lcc}
\hline
Condition & Gap [95\% CI] & AUC [95\% CI] \\
\hline
C0 no briefing         & $+0.77$ [0.71, 0.84] & 0.682 [0.667, 0.698] \\
C1 accuracy prompt     & $+0.77$ [0.71, 0.83] & 0.666 [0.653, 0.678] \\
C0b decoupling only    & $+1.00$ [0.94, 1.06] & 0.686 [0.677, 0.695] \\
C2 strategy knowledge  & $+1.61$ [1.56, 1.65] & 0.818 [0.811, 0.825] \\
C3 full briefing       & $+2.06$ [2.03, 2.09] & 0.812 [0.807, 0.818] \\
\hline
\end{tabular}
\end{table}

\newpage
\section{Cross-Model Robustness of the Discrimination Result}\label{appendix:crossmodel}

To test whether the discernment result depends on one model, we repeated the experiment with Gemini 2.5 Flash, a model from another family of similar size and cost, using the same claims, personas and prompts, the four briefings and a no-briefing baseline. Because each model lacks a factual basis for a different set of claims, we re-ran the basis probe (Supplementary Information section~\ref{appendix:basis_probe}) for each model and scored each on its own \texttt{no\_basis} claims: 108 for GPT-4o-mini and 105 for Gemini 2.5 Flash, each against the 40 shared true claims and rated by 50 personas. Gemini 2.5 Flash's training cutoff (January 2025) postdates the 2024 seeding events, so for this model the design rests on the basis probe rather than on event timing. The similar \texttt{no\_basis} counts are unsurprising: the generated false claims never circulated, so a model can lack a basis for them even when it knows the underlying event. Table~\ref{tab:crossmodel} reports AUC and the discernment gap by briefing.

\begin{table}[htp]
\centering
\caption{Cross-model discernment on each evaluator's own \texttt{no\_basis} claims. AUC is pooled over all persona $\times$ claim ratings, so the C0 value for GPT-4o-mini (0.601) differs slightly from the per-persona mean in Fig.~\ref{fig:post_cutoff_moderator_auc}. Gap is the mean rating of true minus false claims within persona; True and False are the mean believability (1--7) of the 40 true claims and of the model's \texttt{no\_basis} false claims, averaged over 50 personas.}
\label{tab:crossmodel}
\begin{tabular}{lcccc}
\hline
Condition & AUC & Gap & True & False \\
\hline
\multicolumn{5}{l}{\emph{GPT-4o-mini} (108 \texttt{no\_basis} false claims)} \\
C0 no briefing (floor)    & 0.601 & $+0.28$ & 4.03 & 3.75 \\
C1 accuracy prompt        & 0.593 & $+0.24$ & 3.89 & 3.65 \\
C2 strategy knowledge     & 0.767 & $+1.30$ & 4.20 & 2.90 \\
C3 full briefing          & 0.777 & $+1.66$ & 4.69 & 3.03 \\
C0b decoupling only       & 0.626 & $+0.41$ & 4.19 & 3.78 \\
\hline
\multicolumn{5}{l}{\emph{Gemini 2.5 Flash} (105 \texttt{no\_basis} false claims)} \\
C0 no briefing (floor)    & 0.680 & $+0.88$ & 5.26 & 4.38 \\
C1 accuracy prompt        & 0.701 & $+1.16$ & 5.11 & 3.95 \\
C2 strategy knowledge     & 0.602 & $+0.44$ & 4.82 & 4.38 \\
C3 full briefing          & 0.812 & $+1.96$ & 5.40 & 3.44 \\
C0b decoupling only       & 0.742 & $+1.29$ & 5.45 & 4.17 \\
\hline
\end{tabular}
\end{table}

Two results hold in both models. First, the full briefing gives the highest AUC. For GPT-4o-mini it reaches 0.777, against 0.626 for the decoupling instruction alone (mean per-persona difference $+0.151$, paired Wilcoxon $p \approx 2 \times 10^{-15}$), although it is tied with strategy knowledge alone (0.767); for Gemini 2.5 Flash it reaches 0.812, exceeding the decoupling instruction alone by $0.075$ ($p = 9.8\times10^{-14}$) and the accuracy prompt by $0.115$ ($p = 1.8\times10^{-15}$). Second, the accuracy prompt adds little over no briefing. On AUC it scores 0.593 versus 0.601 for GPT-4o-mini and 0.701 versus 0.680 for Gemini 2.5 Flash; on the gap it leaves GPT-4o-mini unchanged ($-0.04$, $[-0.10, +0.02]$) and widens Gemini 2.5 Flash's modestly ($+0.28$, $[+0.21, +0.35]$), far less than the full briefing ($+1.08$, $[+1.01, +1.15]$).

\begin{figure}[htp]
    \centering
    \includegraphics[width=0.9\linewidth]{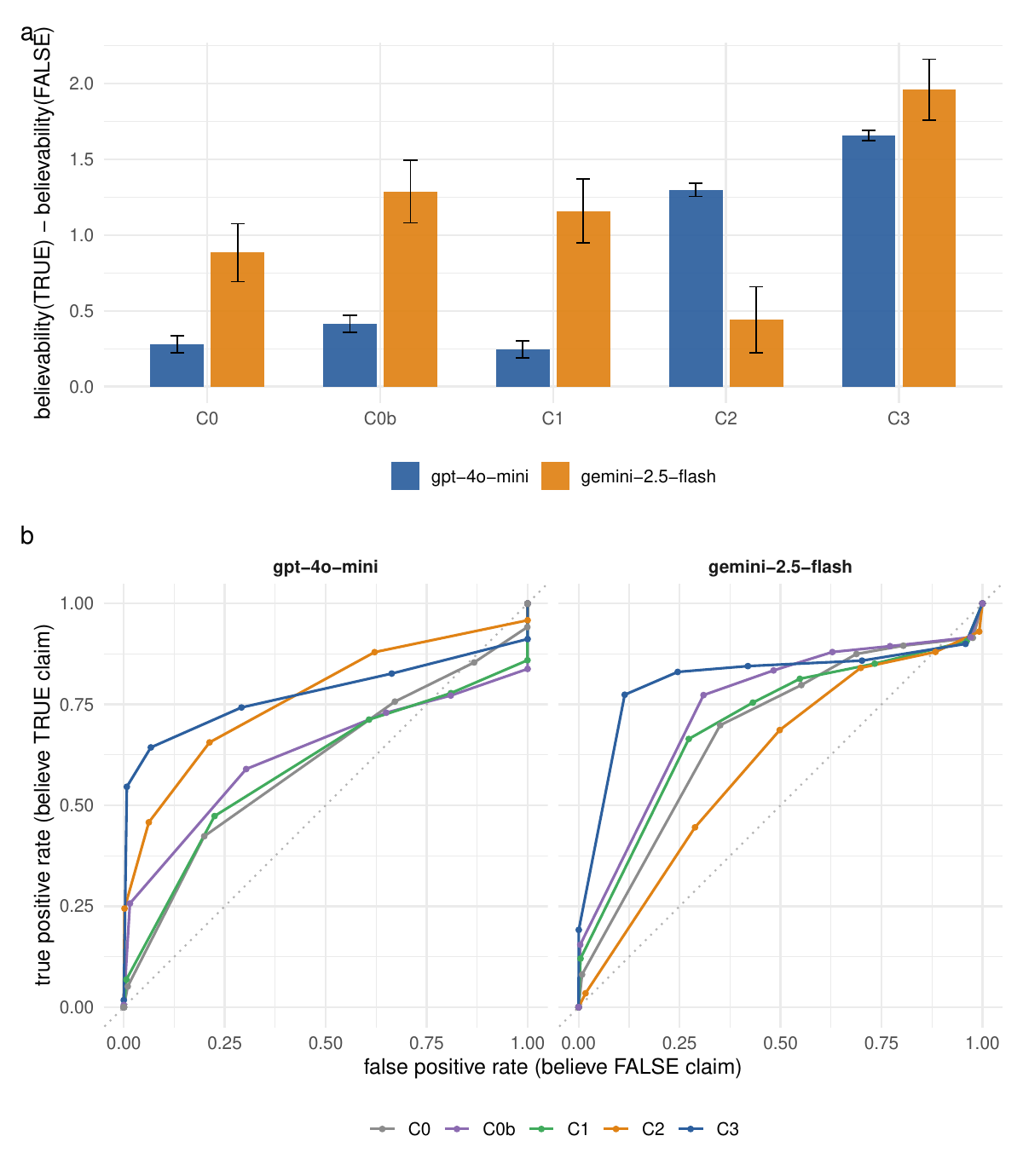}
    \caption{Cross-model discernment by briefing. (\emph{a}) Discernment gap (mean true minus mean false rating per persona) on each model's own \texttt{no\_basis} false claims against the 40 shared true claims (GPT-4o-mini, 108 false claims; Gemini 2.5 Flash, 105); error bars, 95\% confidence intervals over 50 personas. (\emph{b}) Pooled ROC curves by briefing, sweeping the 1--7 believability threshold: true positive rate, share of true claims rated at or above the threshold; false positive rate, the same share for \texttt{no\_basis} false claims; dotted diagonal, chance. The full briefing (C3) gives the widest gap in both models and the highest pooled AUC, tied with C2 for GPT-4o-mini. For Gemini 2.5 Flash, strategy knowledge without the decoupling instruction (C2) falls below the no-briefing level (C0), with its ROC curve near the diagonal; GPT-4o-mini shows no such drop.}
    \label{fig:crossmodel_gap}
\end{figure}

The models differ under strategy knowledge alone (C2). For GPT-4o-mini it ties the full briefing on AUC; for Gemini 2.5 Flash it falls below both the accuracy prompt ($-0.095$, $p = 1.1\times10^{-9}$) and the decoupling instruction alone ($-0.135$, $p = 1.8\times10^{-15}$) and gives the narrowest gap of the four briefings ($+0.44$). On the gap (Fig.~\ref{fig:crossmodel_gap}), the full briefing leads in both models, with 95\% confidence intervals that do not overlap those of any other briefing (GPT-4o-mini $[1.63, 1.69]$ versus C2 $[1.26, 1.34]$; Gemini 2.5 Flash $[1.76, 2.16]$ versus C0b $[1.08, 1.49]$). Gemini 2.5 Flash's drop below its own no-briefing level under C2 parallels the loss of belief in true claims that we pre-registered as a risk for people, in whom it was not detected (\secref{Briefing human evaluators}). The analysis code and per-model ratings are in the public repository.

\newpage
\section{Human Prebunking Experiment}
\subsection{Protocol and Disclosures}\label{appendix:studyB_protocol}

This section describes the protocol of the pre-registered human experiment (\secref{Briefing human evaluators}). The study was pre-registered on AsPredicted before launch (\url{https://aspredicted.org/du8xe2.pdf}).

\emph{Inclusion and ethics.} The study was approved by the Duke Campus Institutional Review Board (protocol 2025-0404). Before the study, participants read a consent page stating that they would read a short educational briefing about how misinformation is written, judge whether a set of news statements were true or false, and give reasons, and that the research concerned how people interpret AI-generated content. Participation was voluntary, and participants could withdraw at any time. At the end of the study, a debriefing page listed every claim the participant had rated together with whether it was true or false.

\emph{Arms.} Participants were randomly assigned, in equal proportions, to one of five briefings: \textbf{C0} (no briefing), \textbf{C0b} (the decoupling instruction only, with no strategy content), \textbf{C1} (accuracy prompt), \textbf{C2} (strategy knowledge: the 13 strategy groups with their prevalence in the corpus, the main co-occurrence patterns, and three practice claims with feedback) and \textbf{C3} (C2 followed by the decoupling instruction). C0b, C1, C2 and C3 are the human versions of the evaluator briefings in Supplementary Information section~\ref{appendix:conditions}, and C0 corresponds to the no-briefing condition of the language-model experiment (Supplementary Information section~\ref{appendix:crossmodel}). The full briefing texts are part of the survey instrument; the strategy-knowledge briefings are available from the authors on request (\emph{Data availability}).

\emph{Claims and task.} After one unscored practice claim, each participant rated 20 claims (7 true, 13 false) on the 1--7 believability scale, rated their familiarity with each claim's topic (1--5) and explained each rating in free text. Each participant saw one of four fixed batches drawn from an 80-claim random sample of the 280 claims used with the language model (\secref{Briefing a language-model evaluator}); the sampling procedure was fixed before launch. Batches were balanced only on the ratio of true to false claims. The 80-claim sample matched the full claim set on generation approach ($\chi^2$ goodness-of-fit $p = .59$) and mega-topic ($p = .41$) and covered all eight seeding events. Batches were assigned at random, and claim order was randomized within each batch.

\emph{Recruitment, quotas and pay.} Participants were US adults recruited on Prolific, with quotas on political affiliation (40\% Democrat, 40\% Republican, 20\% Independent) set through Prolific's political-affiliation screener. The analysed sample was 43\% Democratic-leaning, 42\% Republican-leaning and 15\% Independent. Participants were paid \$7.50 for a session expected to take 30--45 minutes; median completion time ranged from 26 to 33 minutes across arms. Arms were balanced on party ($p = .76$), gender, education, age group, cognitive reflection and familiarity (all $p \geq .13$).

\emph{Exclusions.} Of 852 participants who accepted the study on Prolific, five returned it without opening the survey. Together with three entries through a leaked survey link and one abandoned session without a Prolific ID, 851 sessions began the survey. Two ended before random assignment and 91 after it, leaving 758 submissions. We excluded the three leaked-link completions, whose Prolific IDs were fabricated and which fell outside the pre-registered population; speeders, who finished in less than one third of the median time (9.4 minutes; five); straight-liners, who gave the same believability rating to more than 90\% of claims (two); and one record with no assigned arm. Because some records met more than one criterion, nine submissions were excluded in total, leaving 749 participants.

\emph{Deviations from the pre-registration.} First, the pre-registered attention check (exclusion rule (b)) was never fielded, so the rule could not be applied; the other quality checks (straight-lining, speed and free-text effort) flagged few participants. Second, the three leaked-link completions were identified after data collection and excluded as ineligible. Third, H2b was pre-registered with two measures of the explanations for false claims, the rate of taught vocabulary and a blind LLM-judge score; we report only the vocabulary measure (Table~\ref{tab:studyB_tests}). Both of its contrasts ($p = 2 \times 10^{-9}$ and $1 \times 10^{-7}$) remain significant under any correction for multiple comparisons. In addition, one C1 participant entered a birth year (1988) in the age field, which we treated as missing. Analysis and figure-generation code is included in the public repository.

\subsection{Full Hypothesis Tests}\label{appendix:studyB_results}
Table~\ref{tab:studyB_arms} reports per-arm summary statistics, and Table~\ref{tab:studyB_tests} reports all pre-registered tests (\secref{Briefing human evaluators}). The three confirmatory tests (H1, H2a and H3a) were Holm-corrected, and equivalence was tested with two one-sided tests (TOST) against the pre-registered bounds of $d = \pm 0.3$.

\begin{table}[htp]
\centering
\caption{Per-arm summary statistics ($N = 749$). Gap is the per-participant discernment gap (mean believability of true minus false claims, 1--7 scale); AUC is the per-participant area under the ROC curve; bel(true) and bel(false) are mean believability of true and false claims; CRT/5 is the mean score on a five-item Cognitive Reflection Test; median min is the median completion time in minutes.}
\label{tab:studyB_arms}
\footnotesize
\begin{tabular}{lccccccc}
\hline
Arm & $n$ & Gap & AUC & bel(true) & bel(false) & CRT/5 & Median min \\
\hline
C0 no briefing          & 155 & 2.05 & .76 & 5.08 & 3.03 & 3.02 & 26.1 \\
C0b decoupling only     & 155 & 1.96 & .76 & 5.08 & 3.12 & 3.16 & 28.2 \\
C1 accuracy prompt      & 154 & 2.17 & .78 & 5.09 & 2.92 & 3.14 & 25.8 \\
C2 strategy knowledge   & 140 & 2.15 & .79 & 5.09 & 2.94 & 3.40 & 29.7 \\
C3 full briefing        & 145 & 2.28 & .80 & 5.08 & 2.80 & 3.32 & 32.8 \\
\hline
\end{tabular}
\end{table}

\begin{table}[htp]
\centering
\scriptsize
\setlength{\tabcolsep}{4pt}
\caption{All pre-registered tests. The confirmatory family (H1, H2a, H3a) is Holm-corrected. Equivalence tests (TOST) use bounds of $d = \pm 0.3$. Rows labelled ``H3a extension'' apply the same bound to true-claim believability in every arm, which was not itself pre-registered. The AUC contrasts are pre-registered secondary tests outside the confirmatory family; MWU, Mann--Whitney test. H2b is reported on the vocabulary measure only (Supplementary Information section~\ref{appendix:studyB_protocol}).}
\label{tab:studyB_tests}
\begin{tabular}{llcccl}
\hline
Test & Contrast (DV) & $\Delta$ [95\% CI] & $d$ & $p$ (Holm) & Verdict \\
\hline
H1 (confirmatory) & C3 $-$ C0 (gap) & $+0.230$ $[-0.12, +0.58]$ & 0.15 & .193 (.386) & not supported \\
H2a (confirmatory) & C3 $-$ C0b (gap) & $+0.317$ $[-0.03, +0.66]$ & 0.21 & .069 (.206) & not supported \\
H3a (confirmatory) & C2 $-$ C0 (true-claim belief) & $+0.014$ & 0.02 & .551 (.551) & no backfire \\
TOST (supporting) & C0b $\equiv$ C0 (gap) & --- & --- & $p_{lo}=.017$, $p_{hi}=.001$ & equivalent \\
TOST (predicted null) & C2 $\equiv$ C0 (gap) & --- & --- & $p_{lo}=.001$, $p_{hi}=.024$ & equivalent \\
TOST (H3a extension) & C3 $\equiv$ C0 (true-claim belief) & $+0.003$ $[-0.22, +0.22]$ & 0.00 & $p_{lo}=.004$, $p_{hi}=.005$ & equivalent \\
TOST (H3a extension) & C2 $\equiv$ C0 (true-claim belief) & $+0.014$ $[-0.21, +0.23]$ & 0.02 & $p_{lo}=.003$, $p_{hi}=.007$ & equivalent \\
TOST (H3a extension) & C1 $\equiv$ C0 (true-claim belief) & $+0.014$ $[-0.21, +0.24]$ & 0.01 & $p_{lo}=.003$, $p_{hi}=.006$ & equivalent \\
TOST (H3a extension) & C0b $\equiv$ C0 (true-claim belief) & $+0.001$ $[-0.22, +0.22]$ & 0.00 & $p_{lo}=.004$, $p_{hi}=.004$ & equivalent \\
Supporting & C3 $-$ C2 (gap) & $+0.12$ & --- & n.s. & not supported \\
Descriptive & C1 $-$ C0 (gap) & $+0.12$ & --- & n.s. & null \\
Secondary (AUC) & C3 $-$ C0 & $+0.042$ & --- & .047 (MWU .014) & nominal; unstable \\
Secondary (AUC) & C3 $-$ C0b & $+0.044$ & --- & .039 (MWU .017) & nominal; unstable \\
H2b (lexicon) & C3 vs.\ C0 (false expl.) & .041 vs.\ .013 & --- & $2 \times 10^{-9}$ & supported \\
H2b (lexicon) & C3 vs.\ C0b & .041 vs.\ .016 & --- & $1 \times 10^{-7}$ & supported \\
\hline
\end{tabular}
\end{table}

The two AUC contrasts are not robust: before the three leaked-link completions were excluded ($N = 752$), they had $p = .20$ and $p = .33$. Belief in true claims was nearly identical across arms (5.08 to 5.09; Table~\ref{tab:studyB_arms}), whereas belief in false claims was lowest under the full briefing (2.80, against 3.03 under no briefing). Because H3a is a one-sided test, failing to reject it shows only an absence of evidence. The equivalence tests go further: every arm was equivalent to no briefing on true-claim belief, with 90\% confidence intervals excluding shifts larger than $d = 0.19$ (C3), $0.21$ (C2), $0.20$ (C1) and $0.19$ (C0b).

We repeated these tests within familiarity strata (exploratory). On true claims that a participant rated unfamiliar (familiarity $\leq 2$; participants with at least two such ratings), equivalence was not established, because the estimates were positive and imprecise (C3 $+0.16$ [$-0.12$, $0.45$]; C2 $+0.24$ [$-0.03$, $0.51$]; C0b $+0.21$ [$-0.05$, $0.46$]; C1 $-0.12$ [$-0.38$, $0.14$]). The one-sided test that any decline is smaller than $0.3d$ passed for C3, C2 and C0b ($p < .001$) and was marginal for C1 ($p = .089$). On claims rated familiar (familiarity $\geq 3$), equivalence held for C3 ($-0.01$ [$-0.28$, $0.25$]; TOST $p = .013$), C0b ($p = .011$) and C1 ($p = .018$) and was marginal for C2 ($-0.11$ [$-0.38$, $0.15$]; $p = .071$). No arm significantly lowered true-claim belief in either stratum. Because familiarity was rated after the briefing, membership in these strata could differ by arm.

\subsection{Attrition}\label{appendix:studyB_attrition}

Arm assignment was balanced among entrants (169, 169, 165, 168 and 169; goodness-of-fit $p = .999$), so the unequal arm sizes in Table~\ref{tab:studyB_arms} reflect differential dropout. Here, entrants are the 749 analysed participants plus the 91 sessions that were randomly assigned but abandoned before submission (840 in total); the nine excluded submissions are in neither count. Most abandonment occurred on the briefing pages (74 sessions assigned to an arm with no claims rated; Table~\ref{tab:studyB_attrition}) and was concentrated in the strategy arms (25 in C2 and 17 in C3, against 10 or 11 in each other arm). Completion rates were .92 (C0), .92 (C0b), .93 (C1), .83 (C2) and .86 (C3) ($\chi^2(4) = 13.14$, $p = .011$); the two strategy arms together completed at 85\%, against 92\% in the other three (Fisher odds ratio $0.46$, $p = .0006$). Participants who completed the strategy arms may therefore be more patient or motivated than those in the other arms.

To bound the effect of this selective dropout on the two confirmatory contrasts, we computed Lee bounds\citep{lee2009training}. We assumed that assignment to a briefing can only discourage completion, so that every C3 completer (145 of 169 entrants) would also have completed in the comparison arm. The comparison arms, which retained more participants (C0 and C0b, each 155 of 169), were trimmed by the difference in retention (6.5\%, or ten of 155 completers), once from the top and once from the bottom of the outcome distribution. The resulting bounds exclude zero for both contrasts: C3~$-$~C0 gap $[+0.05,\,+0.43]$ around the observed $+0.23$; C3~$-$~C0b gap $[+0.12,\,+0.50]$ around $+0.32$; and per-participant AUC $[+0.02,\,+0.06]$ for both. Differential dropout alone therefore cannot reverse the sign of either contrast. The Imbens--Manski 95\% confidence intervals for these bounds\citep{imbens2004confidence} are $[-0.28,\,+0.76]$ and $[-0.20,\,+0.82]$ on the gap, consistent with the non-significant confirmatory tests. Tighter bounds that condition on demographics\citep{lee2009training} could not be computed, because 86 of the 91 participants who dropped out returned the study on Prolific, which hides their demographics. The bounds analysis (bootstrap within arm, 2{,}000 resamples) is included in the public repository.

\emph{Briefing length.} The strategy briefings also took longer (median completion 29.7 and 32.8 minutes in C2 and C3, against 25.8 to 28.2 in the other arms; Table~\ref{tab:studyB_arms}), so the strategy arms combine content with extra time and effort. The accuracy prompt, which encourages careful judgement without strategy content, showed no dependence on familiarity (\secref{Briefing human evaluators}). Separating content from time would require a placebo briefing of matched length.

\begin{table}[htp]
\centering
\caption{Stage at which the 91 randomly assigned sessions were abandoned. Entrants (840) are these 91 sessions plus the 749 analysed participants; two sessions that ended before random assignment are not counted. The briefing stage counts sessions assigned to an arm with no claims rated.}
\label{tab:studyB_attrition}
\footnotesize
\begin{tabular}{lrl}
\hline
Stage & $n$ & By arm \\
\hline
Briefing stage & 74 & C0 11, C0b 11, C1 10, C2 25, C3 17 \\
Mid-task & 16 & scattered \\
Post-task & 1 & --- \\
\hline
\end{tabular}
\end{table}

\subsection{Heterogeneity}\label{appendix:studyB_heterogeneity}

\emph{Claim familiarity (pre-registered secondary).} A mixed model of believability (believability $\sim$ condition $\times$ truth $\times$ low familiarity $+ (1|\text{participant}) + (1|\text{claim})$) showed a significant three-way interaction ($F = 5.13$, $p = .0004$). The full briefing's gain in discernment over no briefing was $+0.42$ on low-familiarity claims (familiarity $\leq 2$; Welch 95\% CI $[+0.03, +0.80]$, $p = .033$) and $+0.04$ on high-familiarity claims ($\geq 3$; $[-0.43, +0.50]$, $p = .88$). This matches one of the two directions allowed by the two-sided pre-registration, in which strategy knowledge substitutes for limited knowledge of the claim, and parallels the no-basis stratum in the language-model experiment (\secref{Briefing a language-model evaluator}). The same model restricted to C0 and C1 showed no such interaction ($F = 1.00$, $p = .32$): the accuracy prompt's gain over no briefing was small on low-familiarity claims (gap $1.73$ against $1.66$, $+0.07$) and slightly larger on familiar ones ($+0.14$). Compared directly with the accuracy prompt, the full briefing's gap on low-familiarity claims was $0.35$ larger ([$0.00$, $0.69$], Welch $p = .050$; exploratory, uncorrected). A significant interaction for one briefing and a null interaction for another do not by themselves show that the two briefings differ.

\emph{Familiarity was measured during the task.} Participants rated familiarity after the briefing, so the moderator could in principle have been affected by the briefings. Two results limit this concern: mean familiarity was balanced across arms (all $p \geq .13$; Supplementary Information section~\ref{appendix:studyB_protocol}), and familiarity ratings were largely specific to each participant and claim rather than to arms (see below). Measuring familiarity before the briefing would remove this concern in future studies.

\begin{figure}[htp]
    \centering
    \includegraphics[width=\linewidth]{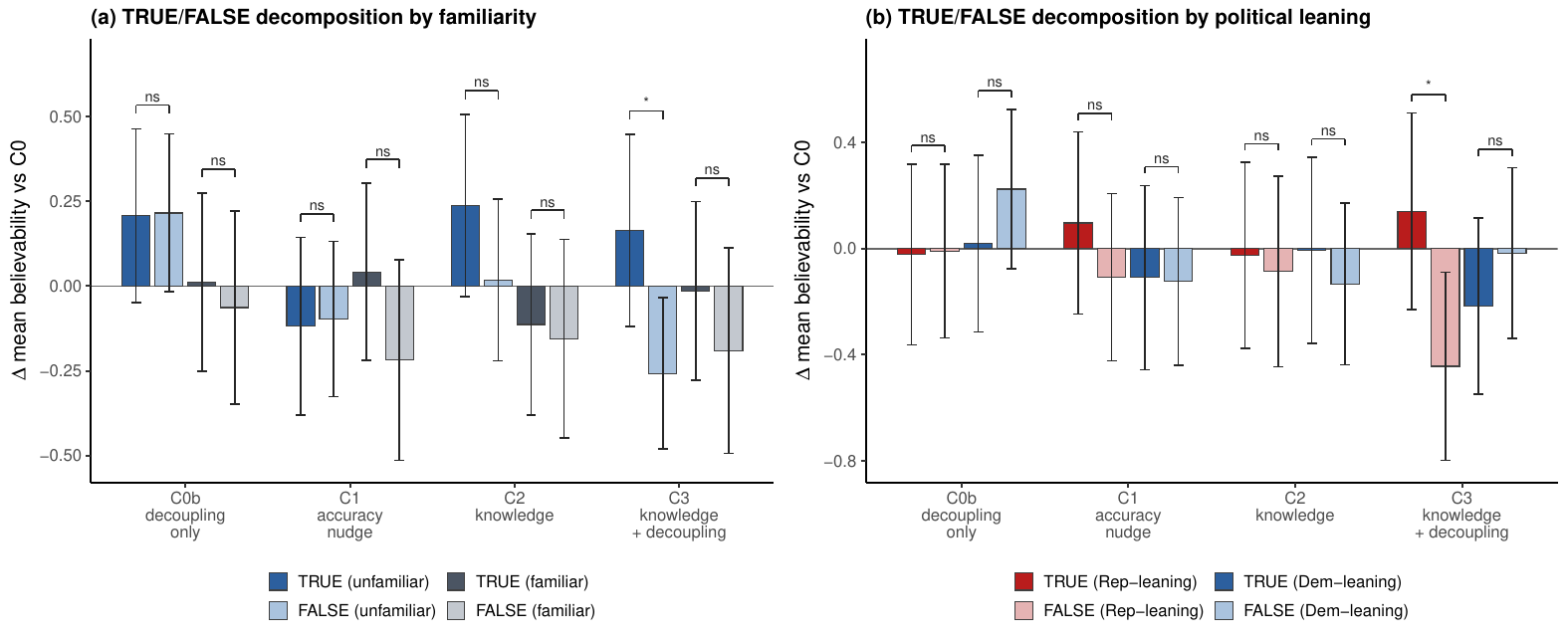}
    \caption{Belief in true and false claims behind the contrasts in Fig.~\ref{fig:studyB_human}. Bars show the change relative to no briefing (C0) in mean believability of true claims (solid) and false claims (shaded); error bars are Welch 95\% confidence intervals on participant-level means (at least two rated claims per cell). Brackets show the change in the discernment gap relative to C0 within each subgroup ($^{*}p<.05$, $^{**}p<.01$, $^{***}p<.001$; ns, not significant). (a) By familiarity (low $\leq 2$, high $\geq 3$): the full briefing's gain on unfamiliar claims came mainly from lower belief in false claims ($-0.26$), with belief in true claims unchanged. (b) By political leaning (exploratory; the condition $\times$ party interaction was not significant): the gain among Republican leaners was also on false claims (false $-0.44$; true $+0.14$, not significant), and changes among Democratic leaners were null on both sides.}
    \label{fig:studyB_decomposition}
\end{figure}

\emph{Political leaning (pre-registered as exploratory).} The condition $\times$ party interaction was not significant ($F = 0.83$, $p = .58$). Within parties, the full briefing's effect among Republican-leaning participants was $+0.58$ against C0 ($p = .041$) and $+0.60$ against C0b ($p = .041$); effects among Democratic leaners were near zero or negative, and Independents fell in between ($+0.49$ and $+0.48$, not significant). Without a briefing, the discernment gap was 2.44 among Democratic leaners and 1.80 among Republican leaners, and 2.02 and 1.20 on low-familiarity claims. The full briefing raised Republican leaners' low-familiarity gap from 1.20 to 1.80 ($+0.60$, against $+0.04$ on high-familiarity claims), compared with $+0.26$ (low) and $-0.23$ (high) among Democratic leaners. Familiarity did not differ by party (Republican-leaning 2.65, Democratic-leaning 2.60, Independent 2.61; $p = .61$). It was also largely specific to each participant and claim: in a variance-components model of the 14{,}980 familiarity ratings (familiarity $\sim 1 + (1\,|\,\text{claim}) + (1\,|\,\text{participant})$), the participant-by-claim residual accounted for 64\% of the variance, participants for 25\% and claims for 12\%, and two participants who rated the same batch ranked its claims by familiarity almost independently (mean Spearman $\rho = 0.15$). The party pattern therefore does not appear to reflect differences in familiarity. It is consistent with more room for improvement among participants with lower baseline discernment, although this was not tested as a mediator. These estimates are uncorrected and exploratory; we report them as differences in baseline discernment, with party as an observable proxy, not as a partisan effect. In both subgroup analyses, the estimated gain came mainly from lower belief in false claims, with belief in true claims about unchanged (Fig.~\ref{fig:studyB_decomposition}).

\emph{Other pre-registered exploratory moderators.} None showed reliable moderation: cognitive reflection (interaction $F = 0.04$, $p = .998$), claim-level baseline-discernment tertiles (all C3 $-$ C0 contrasts not significant), age, education, gender, media trust and familiarity with AI. Among roughly 20 subgroup contrasts, a few uncorrected results with $p < .05$ are expected by chance (for example, ages 35--54, C3 $-$ C0b $+0.53$, $p = .024$); none survives correction.

\pagebreak

\section{Adversarial Prebunking}
\subsection{Design and Per-Condition Effects}\label{appendix:adversarial_full}

\begin{figure}[htp]
    \centering
    \begin{subfigure}[b]{\linewidth}
        \centering
        \includegraphics[width=0.95\linewidth]{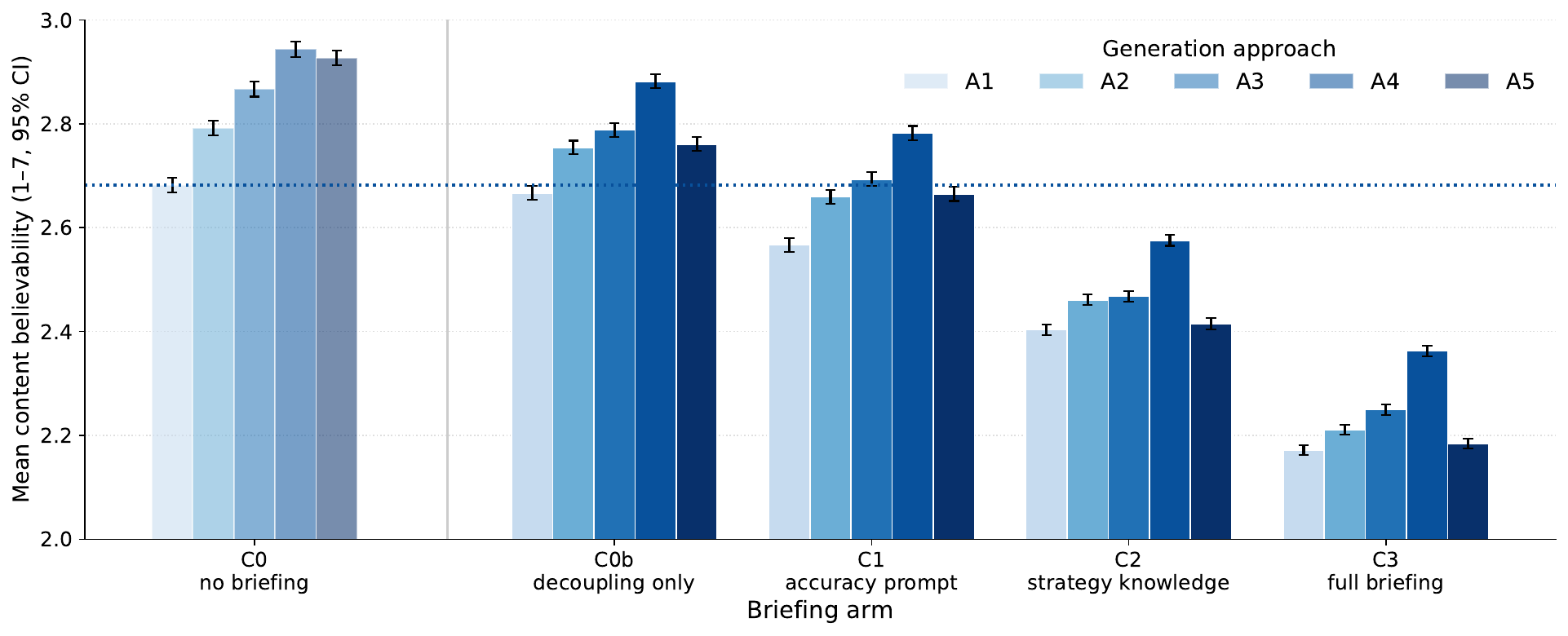}
        \caption{Mean content believability by briefing and generation approach
        (100 personas; no-briefing baseline (C0) at left; dotted line, A1 under no briefing).}
    \end{subfigure}
    \vspace{0.8em}
    \begin{subfigure}[b]{\linewidth}
        \centering
        \includegraphics[width=0.95\linewidth]{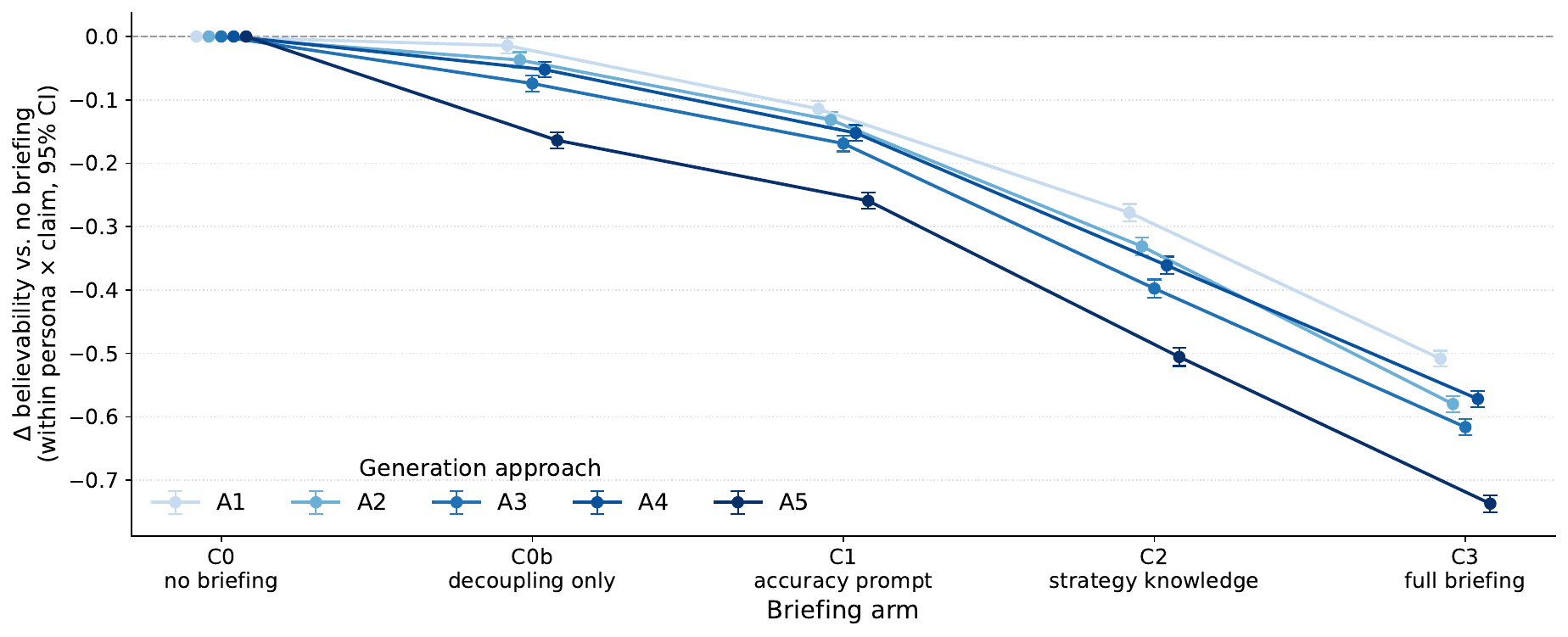}
        \caption{Shift in believability ($\Delta$) relative to no briefing for the same persona and claim, with 95\% confidence intervals. Every briefing lowers believability; the full briefing (C3) lowers it most and decoupling only (C0b) least.}
    \end{subfigure}
    \caption{Prebunking reverses the adversarial credibility advantage. A panel of 100 LLM-simulated personas rates 2,000 AI-generated claims under four briefings (C0b decoupling only; C1 accuracy prompt; C2 strategy knowledge; C3 full briefing; Supplementary Information section~\ref{appendix:conditions}). $\Delta$ = rating under the briefing $-$ rating under no briefing (C0) for the same persona and claim.}
    \label{fig:adversarial_prebunk}
\end{figure}

This section gives the design and per-condition results of the adversarial test (\secref{The adversarial stress test}, Fig.~\ref{fig:adversarial_prebunk}).

\emph{Design.} A panel of 100 personas from the 2,058-persona pool rated all 2,000 AI-generated claims (5 generation approaches $\times$ 20 seeding events $\times$ 20 claims) under the four evaluator briefings (C1, C2, C3, C0b; Supplementary Information section~\ref{appendix:conditions}), in batches of 20 claims per call. For each persona, claim and briefing, $\Delta$ is the briefed rating minus the same persona's no-briefing rating of the same claim. The no-briefing ratings were collected in a separate run with the same batching (Supplementary Information section~\ref{subsec:persona_eval}). Within this panel, no-briefing believability was $2.68$, $2.79$, $2.87$, $2.94$ and $2.93$ for A1--A5 and $3.13$ for the matched fact-checked claims; the full 2,058-persona means lie within $0.01$ of these. Pairing each rating with the same persona's baseline for the same claim removes differences between personas and between claims. All four briefings lowered believability for every generation approach. Because the test uses false claims only, it measures suppression of belief, not discernment, and it does not test authentic misinformation. Three further strategy-knowledge briefings were also run in this test: the 48-strategy taxonomy alone, the 13 strategy groups with their prevalence but no co-occurrence pairs, and the 13 groups with topic-specific instead of corpus-wide co-occurrence. They lowered believability by 0.25 to 0.38, comparable to C2 ($-0.37$); their ratings are included in the deposited data, and their prompts are available from the authors on request.

\emph{Per-condition effects.} Relative to no briefing, $\Delta$ was $-0.17$ under the accuracy prompt (C1), $-0.37$ under strategy knowledge (C2), $-0.60$ ($-21.2\%$) under the full briefing (C3) and $-0.07$ ($-2.4\%$) under decoupling only (C0b). The full briefing's reduction was about 60\% larger than that of strategy knowledge alone, whereas decoupling only had almost no effect. Reductions were largest for the approaches with the highest baseline believability, A4 and A5 (A5 fell by $0.74$ under C3). For A4, the generator's most believable approach, the full briefing erased the $+0.26$ knowledge gain and brought believability $0.32$ below the zero-shot baseline, a reduction 2.22 times the size of the gain, against 1.41 times under strategy knowledge alone (C2).

\subsection{Cross-Family Check of the Adversarial Reversal}\label{appendix:adversarial_crossmodel}
\begin{figure}[h!]
    \centering
    \includegraphics[width=0.68\linewidth]{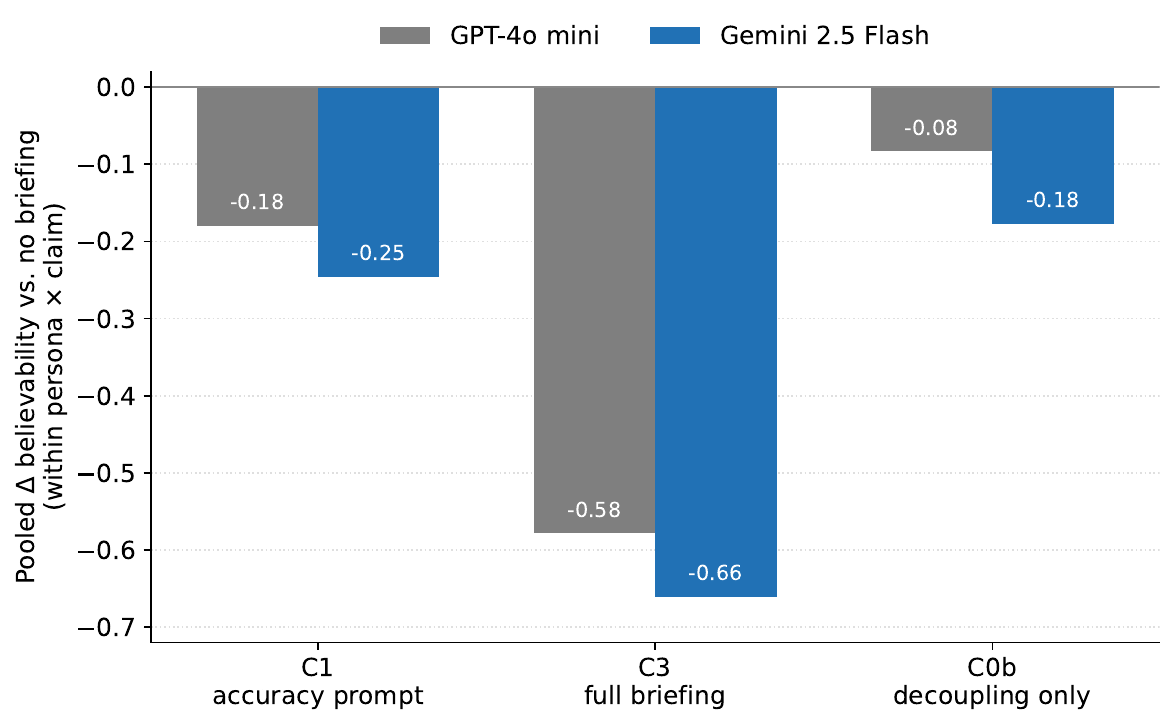}
    \caption{Cross-family replication of the adversarial reversal. Pooled shift in believability ($\Delta$) relative to no briefing under the accuracy prompt (C1), the full briefing (C3) and decoupling only (C0b), for GPT-4o-mini and Gemini 2.5 Flash on the same 50 personas and 500-claim random subsample (494 claims for GPT-4o-mini, whose no-briefing rating is missing for six). Both models show the same ordering, with the full briefing producing the largest reduction.}
    \label{fig:crossmodel_adversarial}
\end{figure}

Because the GPT panel shares a model family with the generator, we re-evaluated the generated claims with an evaluator from a different family, Gemini 2.5 Flash. It used the 50 personas from the cross-model discernment check (Supplementary Information section~\ref{appendix:crossmodel}), a 500-claim random subsample, and the same prompts, batching and parsing as the GPT panel. Both the personas and the claims are subsets of the GPT panel, so the GPT comparisons below use GPT ratings of the same personas and claims. Gemini rated the claims under no briefing, the accuracy prompt (C1), the full briefing (C3) and decoupling only (C0b); strategy knowledge alone (C2) was not rerun. We declared three criteria before the run: (S1) under no briefing, the knowledge-augmented generations (A3--A5) are more believable than A1--A2, as with GPT; (S2) $\Delta$ under C3 is negative and larger in magnitude on A3--A5 than under C1; and (S3) $\Delta$ is small under C1 and smallest under C0b, both well below C3. All three were met. Under no briefing, A3--A5 were more believable than A1--A2 (mean $3.67$ against $3.59$), and A4 was the most convincing approach ($3.81$ of $7$), as in the GPT panel (S1). Under C3, $\Delta$ was negative for every approach and largest for A3--A5 (A3 $-0.66$, A4 $-0.62$, A5 $-0.70$; pooled $-0.66$), against $-0.25$, $-0.19$ and $-0.26$ under C1; on A4, the full briefing brought believability $0.30$ below the zero-shot baseline. GPT on the same personas and claims gave a pooled shift of $-0.58$ under C3 (S2). Pooled shifts were $-0.25$ under C1 and $-0.18$ under C0b, both much smaller than under C3, as in the GPT panel on the same personas and claims ($-0.18$ and $-0.08$, against $-0.58$) (S3). Fig.~\ref{fig:crossmodel_adversarial} compares the two models.

\newpage
\section{Persona-Simulated Evaluation: Full Results}\label{subsec:persona_eval}
To measure perceived credibility at a scale the human experiment (\secref{Briefing human evaluators}) cannot reach, we ran a large persona-simulated evaluation. AI-generated messages are known to persuade people: LLM-authored messages shift policy attitudes\citep{bai2025llm}, personalization increases persuasion at scale\citep{matz2024potential}, and GPT-4 given basic personal information outperforms human debaters\citep{salvi2025conversational}. Here we ask how strategy knowledge changes the believability of AI-generated misinformation to simulated evaluators.

The evaluated claims are the full strategy-guided generation set (\secref{The adversarial stress test}): $2{,}000$ AI-generated claims, 20 for each combination of the five generator conditions A1--A5 (Supplementary Information section~\ref{appendix:conditions}) and the 20 post-cutoff seeding events (Table~\ref{tab:events}). Each claim is 7--25 words long and comes with the strategies the generator reported using. The panel also rated the $778$ unique fact-checked claims that serve as matched benchmarks: each generated claim is matched to its nearest fact-checked claim in embedding space (\texttt{text-embedding-3-small}), among claims in the same mega-topic about the same seeding event. In total, the $2{,}778$ claims span all eight mega-topics.

We used all $2{,}058$ LLM-simulated personas\citep{toubia2025database}, which vary in demographic, psychological, economic, personality and cognitive characteristics. Each persona rated all $2{,}778$ claims on three 1--7 scales: source trustworthiness, content believability and perceived likelihood of AI generation. Claims were presented in batches of 20 within a single API call, in shuffled order. %
This design yielded $5{,}714{,}201$ completed ratings (99.95\% of the $5{,}717{,}124$ persona--claim pairs).

\begin{figure}[htp]
    \centering
    \includegraphics[width=0.95\linewidth]{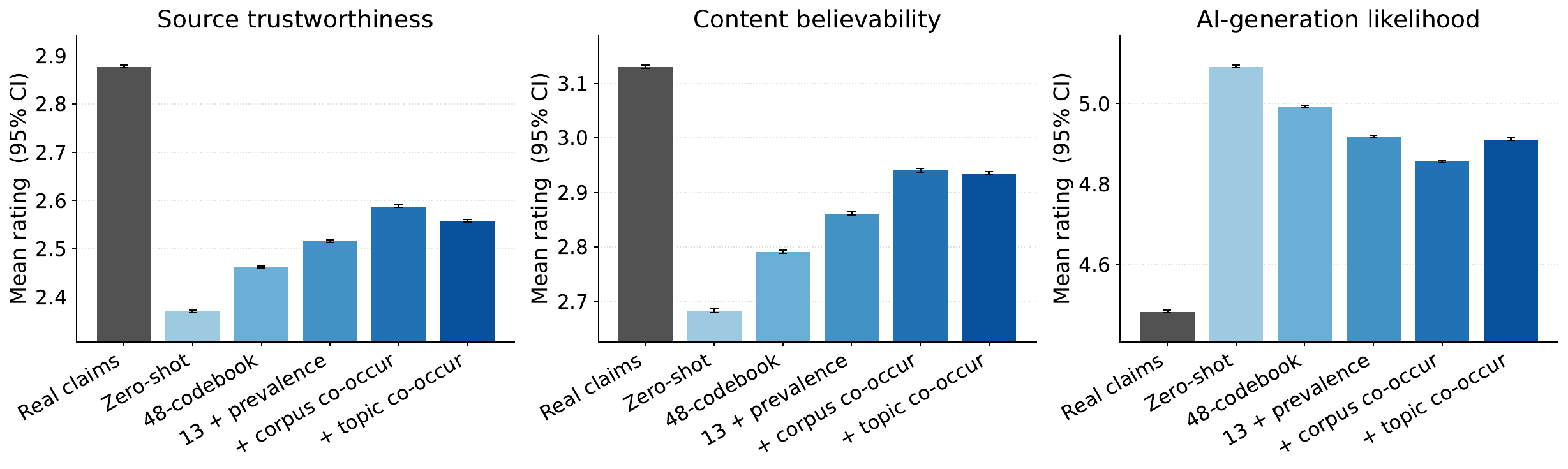}
    \caption{Mean source trustworthiness, content believability and AI-generation likelihood by condition, with 95\% confidence intervals. Real, fact-checked claims; A1, zero-shot; A2, 48-strategy taxonomy; A3, 13 strategy groups with corpus prevalence (Fig.~\ref{fig:cialdini_pct}); A4, A3 plus corpus-wide co-occurrence (Fig.~\ref{fig:cooc_lift}); A5, A3 plus topic-specific co-occurrence (per-topic matrices available from the authors on request). $N = 5{,}714{,}201$ ratings from $2{,}058$ personas on $2{,}778$ claims about 20 seeding events.}
    \label{fig:source_type}
\end{figure}

Fig.~\ref{fig:source_type} shows mean ratings by condition. Fact-checked claims received higher trustworthiness ($2.88$) and believability ($3.13$) ratings, and lower AI-generation likelihood ($4.48$), than claims from any of the five generator conditions. This is a difference in mean ratings; it does not show that the personas can tell individual fact-checked and AI-generated claims apart.

Among the generator conditions, ratings rose with each added layer of strategy knowledge. Trustworthiness rose from $2.37$ (A1) to $2.46$ (A2), $2.52$ (A3) and $2.59$ (A4); believability rose from $2.68$ to $2.79$, $2.86$ and $2.94$; and perceived AI-generation likelihood fell from $5.09$ to $4.99$, $4.92$ and $4.86$. Topic-specific co-occurrence (A5) did not improve on corpus-wide co-occurrence: it tied A4 on believability ($2.94$) and was slightly worse on trustworthiness ($2.56$) and AI-generation likelihood ($4.91$). %

The largest single step came from adding the 48-strategy taxonomy to zero-shot generation (A1 to A2: $+0.092$ trustworthiness, $+0.108$ believability, $-0.100$ AI-generation likelihood). The second largest came from adding corpus-wide co-occurrence to prevalence (A3 to A4: $+0.072$, $+0.079$ and $-0.062$), so knowing which strategies appear together added to knowing which are common. Even A4 claims remained below the matched fact-checked claims, by $0.29$ on trustworthiness, $0.19$ on believability and $0.38$ on AI-generation likelihood. Cosine similarity between each generated claim and its matched fact-checked claim was nearly the same across conditions (means $0.541$ to $0.554$; no contrast with A4 reached $p < .10$), so strategy knowledge did not bring generations closer to circulating misinformation in embedding space. Because each claim is matched to a fact-checked claim about the same event, some similarity is built into the design.

The three ratings are closely related. Across all ratings, trustworthiness and believability correlate at $r = 0.84$, and both correlate negatively with perceived AI-generation likelihood ($r = -0.82$ and $r = -0.78$). %

\subsection{Strategy Alignment Between Generated and Matched Fact-Checked Claims}\label{appendix:strategy_alignment}

This section details the strategy-alignment analysis reported in the main text (\secref{The adversarial stress test}). Each generated claim has two sets of strategy labels:
\begin{itemize}
    \item $A$: the strategies the generating model self-reported using during generation, parsed from the prompt's structured output;
    \item $B$: the strategy labels the annotation pipeline assigned to the matched nearest-neighbor fact-checked claim.
\end{itemize}
A1, A3, A4 and A5 report strategies in the 13-group framework (\secref{Mapping the persuasion playbook}); A2 reports them in the 48-strategy taxonomy. To compare all five conditions, we map A2's labels to the 13 groups using Table~\ref{tab:membership}, under which 37 of the 48 strategies belong to a group and the other 11 do not. After mapping, $92.8\%$ of A2 claims (371 of 400) have at least one group label. Of A2's $807$ strategy mentions, $68.9\%$ map to a group; most of the rest are \emph{Appeal to Suspicion} ($20.7\%$ of all A2 mentions). On average, generated claims carry about 2.0 group labels (A2, about 1.3) and matched fact-checked claims about 3.4.

Table~\ref{tab:strategy_alignment} reports four overlap measures across the 400 generated claims per condition. Jaccard ($\lvert A \cap B \rvert / \lvert A \cup B \rvert$) measures overall overlap; precision ($\lvert A \cap B \rvert / \lvert A \rvert$) is the share of the generator's strategies that the fact-checked claim also uses; recall ($\lvert A \cap B \rvert / \lvert B \rvert$) is the share of the fact-checked claim's strategies that the generator also used; and any-overlap indicates at least one shared strategy.

\begin{table}[htp]
\centering
\footnotesize
\caption{Strategy alignment between each generated claim and its matched fact-checked claim, by generation condition. Means ($\pm$ 95\% confidence-interval half-widths) over 400 pairs per condition; A2 labels are mapped to the 13 strategy groups. Bold marks the highest value for each measure.}
\label{tab:strategy_alignment}
\begin{tabular}{@{}l c c c c@{}}
\toprule
\textbf{Approach} & \textbf{Jaccard} & \textbf{Precision} & \textbf{Recall} & \textbf{Any-overlap} \\
\midrule
A1: Zero-shot                 & $0.191 \,(\pm 0.019)$ & $0.382 \,(\pm 0.033)$ & $0.232 \,(\pm 0.021)$ & $0.635 \,(\pm 0.047)$ \\
A2: 48-strategy taxonomy (mapped) & $0.154 \,(\pm 0.017)$ & $\mathbf{0.444 \,(\pm 0.045)}$ & $0.171 \,(\pm 0.019)$ & $0.547 \,(\pm 0.051)$ \\
A3: 13 + prevalence           & $0.218 \,(\pm 0.019)$ & $0.426 \,(\pm 0.032)$ & $\mathbf{0.266 \,(\pm 0.022)}$ & $\mathbf{0.705 \,(\pm 0.045)}$ \\
A4: + corpus co-occurrence    & $\mathbf{0.220 \,(\pm 0.020)}$ & $0.433 \,(\pm 0.035)$ & $0.258 \,(\pm 0.022)$ & $0.668 \,(\pm 0.046)$ \\
A5: + topic co-occurrence     & $0.208 \,(\pm 0.020)$ & $0.410 \,(\pm 0.035)$ & $0.244 \,(\pm 0.022)$ & $0.638 \,(\pm 0.047)$ \\
\bottomrule
\end{tabular}
\end{table}

Strategy knowledge improved alignment modestly. A3, A4 and A5 all scored above zero-shot generation (A1) on every measure, and the difference between A1 and A4 was marginally significant on Jaccard ($+0.028$, $p = .045$) and precision ($+0.051$, $p = .037$), with recall in the same direction ($p = .099$). Among A3, A4 and A5, alignment levelled off: A4 was nominally highest on Jaccard, precision and recall but did not differ from A5 on any measure ($p > .37$), and A3 was nominally highest on any-overlap ($p = .254$ against A4). A2 traded coverage for selectivity. Its precision was the highest of any condition ($0.444$; $p = .70$ against A4), but its recall was the lowest ($0.171$; $p < 10^{-7}$ against A4), and its Jaccard ($0.154$; $p < 10^{-4}$) and any-overlap ($0.547$; $p = .001$) were also lower, because fewer of its labels map to the 13 groups. Differences in alignment across conditions were smaller than the differences in believability (Supplementary Information section~\ref{subsec:persona_eval}), as overlap between strategy sets does not capture which strategies appear together.

\pagebreak
\clearpage
\printbibliography
\end{refsection}

\end{document}